\documentclass[review]{elsarticle}

\usepackage{lipsum}
\makeatletter
\def\ps@pprintTitle{%
 \let\@oddhead\@empty
 \let\@evenhead\@empty
 \def\@oddfoot{}%
 \let\@evenfoot\@oddfoot}
\makeatother

\usepackage{mypreamble}
\usepackage[a4paper, margin=1in]{geometry}
\usepackage{setspace}
\spacing{1.2}
\usepackage{xcolor}
\usepackage{soul}
\usepackage{mathtools}
\usepackage{nicefrac}
\usepackage{amsmath}
\usepackage{lipsum,times,graphicx,hyperref,cleveref}
\usepackage{multicol}
\usepackage{caption}
\usepackage{comment}
\newenvironment{nomenclature}[1][1em]
 {\begin{list}{}{%
   \settowidth{\labelwidth}{\textbf{#1}}%
   \setlength{\leftmargin}{\labelwidth}%
   \addtolength{\leftmargin}{\labelsep}%
 }}
 {\end{list}}

\newcommand{\nomenclatureentry}[2]{\item[#1] #2\par}

\begin{document}

    \begin{sloppypar}

    \begin{frontmatter}

\title{\textbf{Effect of the blast wave interaction on the flame heat release $\&$ droplet dynamics}}

\author[mymainaddress]{Gautham Vadlamudi}

\author[mymainaddress]{Balasundaram Mohan}

\author[mymainaddress]{Akhil Aravind}

\author[mymainaddress,mysecondaryaddress]{Saptarshi Basu\corref{mycorrespondingauthor}}
\cortext[mycorrespondingauthor]{Corresponding author: sbasu@iisc.ac.in}

\address[mymainaddress]{Department of Mechanical Engineering, Indian Institute of Science, Bangalore}
\address[mysecondaryaddress]{Interdisciplinary Center for Energy Research, Indian Institute of Science, Bangalore}

\begin{abstract}
The study comprehensively investigates the response of a combusting droplet during its interaction with high-
speed transient flow imposed by a coaxially propagating blast wave. The blast wave is generated using a miniature shock generator which facilitates wide Mach number range ($1.01<M_s<1.6$). The interaction of the shock flow occurs in two stages: 1) interaction of the temporally decaying velocity ($\rm v_s$) imposed by blast wave and 2) interaction with the induced flow ($\rm v_{ind}$). The flame base lifts off due to the imposed flow, and the advection of flame base towards flame tip results in flame extinction for $M_s > 1.06$. The timescale of the flame extinction is faster (interaction with $\rm v_s$) for $M_s>1.1$. The study investigates the effect on droplet regression, flame heat release rate and flame topological evolution during the interaction. The droplet regression rate gets enhanced after the interaction with blast wave for $M_s < 1.06$, while it slowed down due to complete extinction for $M_s > 1.06$. A momentary flame heat release rate (HRR) enhancement occurs during the interaction with shock flow, and this HRR enhancement is found to be more than 8 times the nominal unforced flame HRR for $M_s > 1.1$, where rapid flame extinction occurs due to faster interaction with $\rm v_s$ ($\sim O(10^{-1})ms$). HRR enhancement has been attributed to the fuel vapor accumulation during the interaction. Furthermore, for $M_s > 1.1$), compressible vortex interaction occurs with the droplet resulting in droplet atomization. The droplet shows a wide range of atomization response modes ranging from pure deformation, Rayleigh-Taylor piercing bag breakup, and shear-induced stripping. No significant effect of nanoparticle (NP) addition has been found on the flame dynamics due to the faster timescales. However, minimial effects of NP addition are observed during droplet breakup due to fluid property variation.
\end{abstract}


\end{frontmatter}

\section*{Novelty and Significance Statement}
{
This research is the first of its kind to examine a droplet flame response to blast waves. This study is a deep dive into the droplet flame topology, heat release rate during the interaction with a blast wave, extending upon our previous droplet flame shock interaction study which was limited to the hydrodynamic aspects of the shock - flame interaction. The insights from this study could be useful in practical high-speed combustion and propulsion systems, where flame interactions with pressure waves, such as decaying shocks and blast waves, are common. This study also provides insights into shock induced aero breakup of droplets in nanofuels during combustion. 
}

\section*{Author Contributions}
{
\textbf{Gautham Vadlamudi:} Conceptualisation, Experiments, Data processing, Data Analysis, Writing - Original Draft. \textbf{Balasundaram Mohan:} Conceptualisation, Experiments, Data Processing, Data Analysis, Writing - Review \& Editing. \textbf{Akhil Aravind:} Conceptualisation, Experiments, Data Analysis, Writing - Review \& Editing. \textbf{Saptarshi Basu:} Conceptualisation, Data Analysis, Writing - Review \& Editing, Fund acquisition.
}


    \section*{Nomenclature}

\begin{spacing}{1}

\begin{multicols}{2}
\begin{nomenclature}
  \nomenclatureentry{$M_{s}$}{Blast wave Mach Number}
  \nomenclatureentry{$R_{s}$}{Blast wave Radius}
  \nomenclatureentry{$d$}{Droplet diamater}
  \nomenclatureentry{$d_o$}{Initial droplet diamater}
  \nomenclatureentry{$\rm v_{s}$}{Blast imposed velocity variation}
  \nomenclatureentry{$p$}{Blast imposed pressure variation}
  \nomenclatureentry{$t$}{time from wire-explosion}
  \nomenclatureentry{$t_{d}$}{timescale of droplet lifetime ($t_d=0$ at ignition)}
  \nomenclatureentry{$t_{d,max}$}{Total droplet lifetime from ignition till extinction}
  \nomenclatureentry{$t_s$}{Instant corresponding to the incidence of blast with droplet flame from explosion ($t=0$)}
  \nomenclatureentry{$t_{amb}$}{Instant corresponding decay of $p$ to ambient levels}
  \nomenclatureentry{$\tau$}{Dimensionless time $t/t_{amb}$}
  \nomenclatureentry{$\tau_s$}{$\tau$ at the instance of shock incidence}  
  \nomenclatureentry{$\tau_{CVR}$}{$\tau$ at the begining of CVR interaction at the droplet}
  \nomenclatureentry{$t^*$}{time elapsed from $t_s$ normalized with inertial timescale of the droplet}
  \nomenclatureentry{$\rm v_{ind}$}{Bulk flow induced behind the blast wave}
  \nomenclatureentry{$h_{lft}$}{Flame base lift-off height}
  \nomenclatureentry{$\rm v_{imposed}$}{velocity scale imposed on the droplet flame due to shock-flow}
  \nomenclatureentry{$\rm v_{NC}$}{velocity scale corresponding to natural convection}
  \nomenclatureentry{$\rm v_{b,lft}$}{Flame base lift-off velocity}
  \nomenclatureentry{$\rm v_{s,Th}$}{Theoretically obtained velocity variation based on $M_s$}
  \nomenclatureentry{$t_{start}$}{Time, t from wire explosion where flame HRR starts to increase during interaction}
  \nomenclatureentry{$t_{dip}$}{Time, t from wire explosion where flame height becomes minimum}
  \nomenclatureentry{$t_{peak(HR)}$}{Time, t from wire explosion where flame HRR attains a maxima}
  \nomenclatureentry{$t_{Peak}$}{Time, t from wire explosion where flame height attains a maxima}
  \nomenclatureentry{$t_{Ext}$}{Time, t from wire explosion where flame extinction occurs}
  \nomenclatureentry{$\Dot{Q}$}{Heat release rate}
  \nomenclatureentry{$\Dot{Q}_o$}{Heat release rate of nominal unforced flame}
  \nomenclatureentry{$Re$}{Reynolds numer}
  \nomenclatureentry{$We$}{Weber number}  \nomenclatureentry{$We_{CVR}$}{Weber number of compressible vortex}
  \nomenclatureentry{$\epsilon_{\theta \theta}$}{Local strain rate at forward stagnation point}
  \nomenclatureentry{$\sigma$}{Surface tension of liquid} 
  \nomenclatureentry{$\rho_{l}$}{Density of the liquid fuel}
  \nomenclatureentry{$\rho_{a}$}{Density of ambient air}
  \nomenclatureentry{$r_{agg}$}{radius of nanoparticle aggregates inside droplet}
  \nomenclatureentry{$r_{bubble}$}{radius of bubbles inside the droplet due to internal boiling}
  \nomenclatureentry{$h_{fg}$}{Latent heat of vaporization}
\end{nomenclature}
\end{multicols}

\end{spacing}





    \newpage

\section{Introduction}

The ongoing progress in combustion systems for advanced propulsion applications has spurred extensive research into the fundamental dynamics of shock-waves with multiphase flows. One of the important applications of the shock-flame interactions is scramjet combustors which operate with in high-speed supersonic oxidizer intake \cite{yang_applications_1993}. The interactions with shock result in enhanced mixing, flame distortion, flame destabilization or blow-off, etc \cite{thomas_experimental_2001,yoshida_blowoff_2024}. Additionally, these investigations also contribute to the understanding of deflagration to detonation transition (DDT) \cite{oran_origins_2007}, which are crucial in futuristic propulsion systems. These studies also are relevant to the detonation-based engines like rotating detonation engines (RDE), which offer higher theoretical efficiencies compared to deflagration process \cite{kashdan2004new}. Since, liquid fuels compared to gaseous fuels, offer an intrinsic advantage of higher volumetric energy density desirable for propulsion applications \cite{patten2023exploration, anderson2005liquid,wei_effect_2017,ciccarelli_role_2010}. Thus, a fundamental investigation of the shock-droplet flame interaction  provides deeper insights into the interaction of the individual combusting fuel droplets (formed from spray atomisation) with the shock structures, which affects the global spray combustion characteristics in high-speed applications. 

In addition to the propulsion applications, understanding the shock - flame interactions becomes crucial for developing and enhancing large-scale fire suppression systems that utilize explosives or shock waves to extinguish flames in fire fighting applications. These systems effectively combat wildfires and oil fires by forcefully displacing flames from the fuel source \cite{akhmetov_extinguishing_1980,akhmetov_formation_2001,yoshida_blowoff_2024,chan_interactions_2016}. Thus, delving into the complexities of shock - flame interactions not only advances our understanding of propulsion systems but also significantly contributes to the enhancement of fire safety measures.

The shock wave or acoustic wave interaction with flame is associated with various flame instabilities that affect flame propagation and flame configuration in confined chambers \cite{maley_influence_2015,jiang_evolution_1999,khokhlov_interaction_1999}. Researchers showed that the shock interactions resulted in an enhancement in the flame heat release rate \cite{wei_effect_2017,thomas_experimental_2001}. The shock-flame interactions were shown to transform the circular flame bubble into toroidal shape due to vorticity generation and Richtmyer Meshkov (RM) instability \cite{picone_vorticity_1988,ju_vorticity_1998,dong_numerical_2008}. The results showed that the RM instability and vorticity generation contributed to the flame stability through enhanced mixing\cite{khokhlov_interaction_1999}. Pionering research by Markstein et al.\cite{markstein_shock-tube_1957} on shock-flame interaction also reported similar instability that results in the shape reversal of the flame as the shock swept across the flame, in premixed butane-air flames. Rudinger et al. also showed that the unburnt gases funnel into the product gases which eventually evolve into a vortex ring subsequently\cite{rudinger_shock_1958}. The experiments studying the shock interaction with a perforated plate by researchers showed that the increase in shock strength resulted in increased local gas temperature and pressure resulting in DDT and autoignition\cite{ciccarelli_role_2010,wei_effect_2017}. Researchers also showed that the hydrodynamic processes played more important role than the chemical proceses during the shock-flame interaction. 

In practical situations, the shockwaves typically arise when blast waves converge or focus due to the geometric features of the combustor, resulting in the formation of unsteady shocks, unlike the controlled shocks generated by shock tubes. A typical shock wave generated by a shock tube is a shock front (property discontinuity) followed by constant flow properties (steady shock), on the other hand a blast wave is a shock front with exponentially decaying flow properties (unsteady) \cite{apazidis_shock_nodate}. Fleche et al. \cite{la_fleche_dynamics_2018} investigated the dynamics of head-on interaction between blast waves and cellular flames in Hele-Shaw cells. The experiments were conducted in a circular combustor filled with fuel-air mixture, with ignitor and shock generator present at the center. The experiments allowed them to explore the response of cellular flames as the blast wave sweeps over the flame from burnt to unburnt side and vice versa. The direction of shock propagation resulted in different flame response such as pertubation growth and shape reversal depending on the pressure gradient induced by the RM instability. 

The current study investigates the interaction of a blast wave with a combusting droplet. Many researchers have investigated the droplet combustion dynamics under various external flow conditions, as the droplet combustion has been a prominent research area due to its relevance with spray combustion in multitude of applications. While droplet combustion studies cannot be directly applied to practical reacting sprays, they offer valuable insights into fundamental combustion processes such as flame stabilization, extinction limits, blow-off, and pollutant formation under controlled conditions \cite{williams1973combustion}. The relative motion between a droplet and the surrounding gas environment significantly influences vaporization characteristics and flame structure. It has been shown that the flame undergoes local extinction at the forward stagnation point before transitioning into the droplet wake due to external flow effects \cite{law1972kinetics,chen2012streamwise,vadlamudi2021insights,vadlamudi2024insights,pandey_self-tuning_2020,vadlamudi2023insights}. Such studies provide insights into local extinction phenomena relevant in spray combustion systems. Other researchers \cite{thirumalaikumaran2022insight,pandey_dynamic_2021,basu2016combustion} have conducted droplet combustion studies under various flow conditions such as vortex interaction, acoustic interaction, etc, to study the flame extinction criteria, flame shedding phenomena and flame heat release response to the external flow. These droplet combustion studies provide the insights into the flame stabilization criteria, flame evolution to external flow, flame topological response, etc, in a wide Reynolds number range of $0<Re<200$. However, very few literature are available that focus on the droplet flame interaction with a unsteady blast wave.     

In an unsteady blast wave propagation, it has been established that the initial blast wave propagation is characterized by a quasi-steady shock front with negligible fluctuations in shock Mach number ($M_s$), resulting in minimal property variations. However, eventually it exhibits non-linear effects, causing a departure from linear steady shock propagation characteristics. Subsequently, the blast wave gradually approaches a weak blast wave regime, and later behaves as an acoustic wave\cite{almustafa_fundamental_2023}. It has been shown that strong shock limit ($M_s>5$) the temporal evolution of shock radius follows a power law with exponent of 2/5, on the other hand, the weak blast wave limit ($M_s\rightarrow1$) shows a linear variation of shock radius with time. For a high-energy blast, a decaying blast wave initially follows a strong shock limit, which gradually transitions into intermediate phase, and it asymptotically approaches weak blast wave limit and acoustic limit in the far-field\cite{diaz_blast_2022}. The Mach number ($M_s$) in current experiments is in the range of $1.01 < M_s < 1.6$ which is in the intermediate transition regime, near the weak blast wave limit. 

Researchers like Taylor and Sedov \cite{taylor1950formation,sedov2018similarity} proposed a self-similar solution that is valid for strong shock regime ($1/M_s\rightarrow 0$), however a departure from this solution occurs for intermediate shock strengths. Researchers like Sakurai et al. and Oshima \cite{sakurai1956propagation,oshima1962blast} provided approximate linear velocity profile solution and quasi-similar solution accounting for the counterpressure effects that resulted in deviation. Other researchers such as Lee and Bach et al. \cite{lee1965propagation,bach1970analytical} showed that the power-law density profile whose exponent is obtained from mass conservation can be used to obtain the velocity and pressure profiles behind the blast wave. It has been shown that the approximate analytical model by Bach and Lee \cite{bach1970analytical} based on power-law density profile showed agreement with exact numerical solution by Goldstine et al.\cite{goldstine1955blast} even at lower shock Mach numbers ($M_s \rightarrow1$) alongside the strong shock limit ($1/M_s \rightarrow0$).  

As mentioned before, there is a dearth of literature on the interaction of the unsteady blast wave with a droplet flame. However, Aravind et al. \cite{aravind2023response,Aravind_2023phenomenology} 
investigated the interaction of an open blast wave with a premixed and non-premixed jet flames and showed different flame behavior regimes depending on the flame shedding, flame liftoff and extinction characteristics. This study also showed that the equivalence ratio and fuel flow velocity play an important role in the flame extinction and shedding response to the blast wave interaction. Moreover, our previous experimental study focused on the droplet flame interaction with a blast wave over the Mach number range of $1.01<M_s<1.6$\cite{vadlamudi2024insights}. A unique compact shock generation setup has been used in the previous experiments for shock generation. The study primarily focused on the comprehensive characterization of the shock flow imposed on the droplet flame as well as the hydrodynamic aspects of the flow-flame interaction observed in those experiments. The similar shock generation setup is used in the current study focusing on the flame heat release rate, flame topology and droplet breakup dynamics to further the understanding of the underlying phenomena.  

Our previous study on shock - droplet flame interaction \cite{vadlamudi2024insights} majorly focused on the hydrodynamic aspects of the flame dynamics in response to the interaction with the shock flow. Broadly two major flame behavior were observed during the interaction depending on the shock Mach number at the droplet ($M_s$). One is that slower and delayed flame response of the droplet flame during the interaction with the shock flow (for $M_s < 1.1$). The other is where the droplet flame shows rapid flame extinction at faster timescales (for $M_s>1.1$). 

The blast wave imposes a temporally decaying velocity profile ($\rm v_s$) at the droplet location as it passes by, which is followed by an induced flow ($\rm v_{ind}$). Using an approximate analytical blast wave formulation, the velocity imposed at the droplet location due to blast wave i.e., $\rm v_{s}$ is theoretically estimated based on the temporal variation of the experimentally obtained Mach number ($M_s$). It has been established that the flame-base liftoff rate during the first stage of interaction is the direct consequence of the imposed temporally decaying velocity $\rm v_{s}$ \cite{vadlamudi2024insights}. The flame base shape evolution due to the RM instability has also been addressed in our previous work. However, the previous work is limited to the hydrodynamic aspects of the flame response and did not provide any insights into the flame heat release rate, flame topology or droplet dynamics. Furthermore, the effect of the fuel type has also been not investigated which is a gap that will be addressed in the current study.   

In the current study, a detailed analysis on the flame heat release rates, droplet regression rates, droplet breakup phenomena in response to the shock flow imposition will be investigated. Two different fuels: n-Dodecane (low vapor pressure fuel) and ethanol (high vapor pressure fuel) will be used in the current experiments to study the effect of the fuel type on different phenomena. In the context of the climate crisis, the study of alternative fuels is also necessary. Nanoparticle laden fuels (Nanofuels) are one of the potential fuels which provide different advantages such as secondary atomization, net enhancement in fuel vaporization, increase of heat release rate, reduced ignition delay, etc. which can contribute to an improvement in combustion efficiency. Thus, two different concentrations of Alumina laden n-Dodecane Nanofuels will also be used in current study to investigate the effect of nanoparticle addition to fuel on the flame and droplet behavior during the interaction.  

In current study, initially the background of the previous experiments will be provided, which provides a general overview of the characteristics of the flow imposed on the combusting droplet during the experiments. Later, the general experimental observations of the flame response for dodecane fuel will be discussed focusing on the topological variation of the flame observed in OH* and CH* chemiluminescence. Alongside, the timescales involved and flame heat release rate (HRR) trends during the interaction will also be addressed. Subsequently, the two broad flame behaviors for Mach number ranges $M_s<1.1$ and $M_s>1.1$ are addressed. In these sections, the general trends of the flame heat release rate (HRR), timescales and droplet dynamics will be discussed. The effect of different fuel types, shock strength, etc will be discussed in detail for both the sub-regimes in $M_s<1.1$ regime. Similarly, the effect of the higher Mach number ($M_s>1.1$) shock interaction on the flame dynamics and the consequent droplet breakup phenomena will be investigated later on. The different stages of flame dynamics observed for $M_s>1.1$  are categorically discussed and the effect of different parameters will be highlighted. Finally, the different mechanisms involved in the droplet dynamics and breakup will be investigated, concluding with the effect of fuel type on droplet breakup. This study provides a comprehensive outlook at the interaction of a droplet flame and its response to a high-speed transient flow, in case of an isolated droplet combustion for different fuel types. such a study can be further extended to dilute spray limit conditions in spray combustion, which can potentially provide insights on spray combustion regime in a extreme environment.

    \section{Experimental Setup}
A specially designed shock generation apparatus used in our previous study \cite{vadlamudi2024insights} has been used for the current experiments to study the interaction of shock-flow with a droplet flame. The shock generation setup works on the principle of wire-explosion, where a high-voltage pulse is passed through a thin copper wire, which results in the generation of a blast wave due to Joule heating of the wire. This setup allows for the generation of blast wave in a wide range of shock Mach numbers from 1.01 to 1.6. Researchers like Liverts et al. \cite{liverts2015mitigation} and Sembian et al. \cite{sembian2016plane} give a detailed overview of the wire explosion technique. This wire-explosion technique provides smaller test facility size, better ease-of-operation and generation of a wide range of $M_s$ \cite{sembian2016plane}.

\begin{figure}[t!]
    \centering   \includegraphics[width=\textwidth,height=\textheight,keepaspectratio]{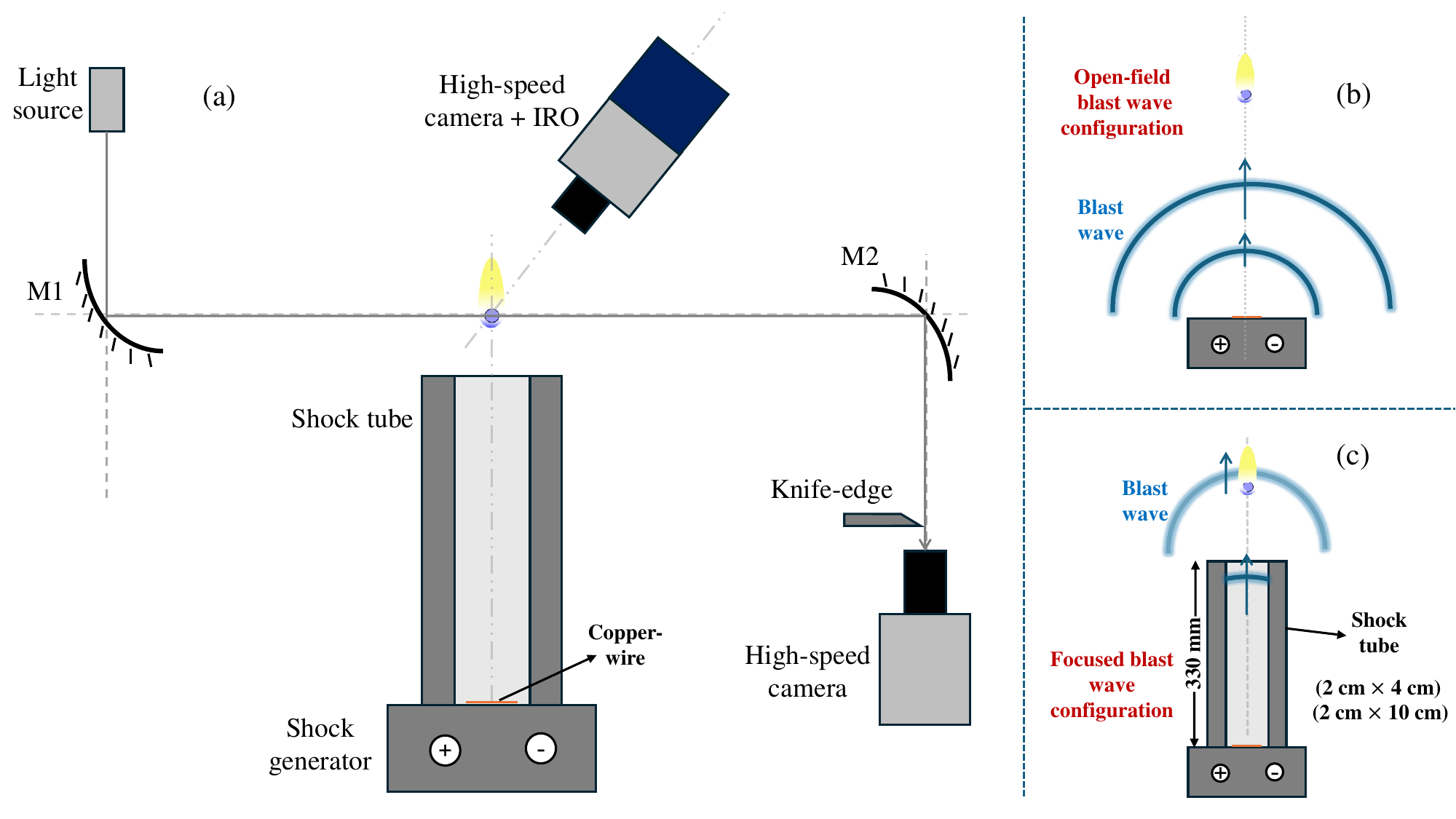}
    \caption{(a) Schematic of the experimental setup showing the droplet flame near the exit of the shock generation apparatus, and simultaneosus high-speed Schlieren and OH* chemiluminescence using high-speed IRO. Schematic of the shock generation apparatus in: (b) open-field blast wave configuration, (c) shocktube-focused blast wave configuration.}
    \label{fig1:IRO:Exp_setup}
\end{figure}

Fig. \ref{fig1:IRO:Exp_setup}a shows the schematic of the experimental setup consisting of a shock generation setup and flow visualization using high-speed imaging to study the shock-flame interaction in a combusting droplet, similar to our previous experiments \cite{vadlamudi2024insights}. The Fig. \ref{fig1:IRO:Exp_setup}a shows the shock generator consisting of an electrode chamber connected to high-voltage source. This system consists of 3kJ pulse power system (Zeonics Systech, India) is used to provide a high-voltage pulse across the copper wire (35 SWG), to generate the blast wave through wire-explosion technique. The center-to-center distance of the electrodes is 7.5cm. The droplet is placed at a specific distance of 365mm away from the copper wire.

When the system is triggered, the blast wave propagates radially outward from the origin (copper wire location) towards the droplet. Similar to previous experiments, the shock Mach number has been controlled by varying the charging voltage of the shock generator and the extent of shock focusing. Thus, the experiments were performed at different charging voltages in two configurations: Open blast wave, and focused blast wave (two shock focusing tubes with dimensions of $2 \times 4 \ cm^2$ and $2 \times 10\ cm^2$). \ref{fig1:IRO:Exp_setup}b,c show the schematic of the two configurations of the shock generation setup. In Open-field blast wave configuration the wire explosion occurs in the open, whereas in focused configuration, a shock tube is mounted on the base plate, which focuses and redirects the blast wave in single direction along the length of shock tube. Two different shock tubes with different crosssectional dimensions are used in the experiments for different levels of shock focusing (shown in \ref{fig1:IRO:Exp_setup}c).  The parametric space has been chosen based on the results of previous experiments corresponding to the Mach number ($M_s$) of the blast wave at the droplet location \cite{vadlamudi2024insights} as shown below:\\
5kV, 8kV: Open-field (Op) configuration ($M_s \sim 1.025, 1.063$)\\
5kV, 6kV: $2 \times 10\ cm^2$: Bigger (B) shock tube - Focused configuration ($M_s \sim 1.15, 1.2$0)\\
6kV, 7.5kV: $2 \times 4 \ cm^2$: Smaller (S) shock tube - Focused configuration ($M_s \sim 1.23, 1.36$)\\

 The length of both the shock tubes is 330 mm and the fuel droplet (diameter, $d \sim 2 mm$) is placed at 35mm from the shock tube opening (365mm from the copper wire) which is suspended at the tip of a quartz filament (0.4mm diameter) in pendant mode. A heater coil mounted on a linear solenoid is used to ignite the droplet. For data acquisition, simultaneous OH* chemiluminescence, high-speed Schlieren flow visualization and high-speed shadowgraphy of droplet has been used to record the flame dynamics and flow structures such as blast wave during the shock-droplet flame interaction. The high-speed Intensified Relay Optics (IRO) has been used along with LaVision SA5 high speed star camera has been used with UV lens and OH* filter ($\sim 310nm$) to perform OH* chemiluminescence imaging at 20,000 fps, with pixel resolution of $5.3187 px/mm$. OH* chemiluminescence imaging provides the indication of the heat release rate (HRR) variation of the flame. CH* chemiluminescence ($431nm$ band pass filter) has also been performed to obtain better insights into flame topology. Photron SA5 camera and Cavitar Cavilux smart UHS, $400 W$ power laser has been used for Schlieren flow imaging with acquisition rate of 40,000 fps and pixel resolution of $3.1701px/mm$. Photron MiniUX has been used along with Navitar lens with Veritas constellation 120E light source as backlighting to perform droplet shadowgraphy at 40,000 fps and pixel resolution of $34.82px/mm$. All the cameras, ignition mechanism (heater coil) and shock generation are synchronized using an arduino circuit and a BNC 745 T digital delay generator. A predefined time delays were set in the delay generator for triggering different components while performing the experiments. Firstly, the droplet is placed on the quartz filament and it's initial size is maintained at constant value using the zoomed-in shadowgraphy imaging. Further, the heater coil is released towards the droplet and retracted after sometime to achieve droplet ignition, and simultaneously the high-speed recording of the Shadowgraphy camera is triggered. This is the considered as the beginning of the droplet lifetime i.e., $t_d =0$, and the shock generator is triggered after a specific delay from $t_d=0$. Specific delays are set accordingly for the OH* chemiluminescence recording and high-speed Schlieren to record the flame imaging and flow, shock wave visualization during and after the interaction of the shock flow with the droplet-flame.
 
 In this study, the effect of the fuel on the flame dynamics is also investigated. Alongside the conventional fuels used, i.e., n-Dodecane and Ethanol, the effect of nanoparticle addition has been investigated in current experiments. The nanoparticle laden fuels were prepared at two different concentrations 0.5$\%$ and 5$\%$ (by weight) using n-Dodecane fuel, by adding Alumina nanoparticles ($Al_2O_3$) and the non-ionic surfactants: Tween 85 (Polyoxyethylene Sorbitan trioleate) and Span 60 (Sorbitan monostearate) to ensure steric stabilization. The droplet size has been maintained around $d \sim 2.00 \pm 0.3  mm$ for different runs, and additionally, the delay of shock incidence from the time of droplet ignition ($t_d = 0$) has also been varied in current experiments. This facilitated to study the effect on droplet regression and flame dynamics with respect to the instant of the beginning of the shock interaction ($t = t_s$), type of fuel and shock strength. It is to be noted that ‘$t_d$’ corresponds to droplet regression time, where, $t_d =0$ at the instant of ignition, and ‘t’ is the time corresponding to the blast wave propagation, where, $t = 0$ is the instant of the wire-explosion and $t = t_s$ is the instant at which blast wave reaches the droplet location.

    \section{Results and Discussions}
\begin{figure}[t!]
    \centering
    \includegraphics[width=1\textwidth,height=\textheight,keepaspectratio]{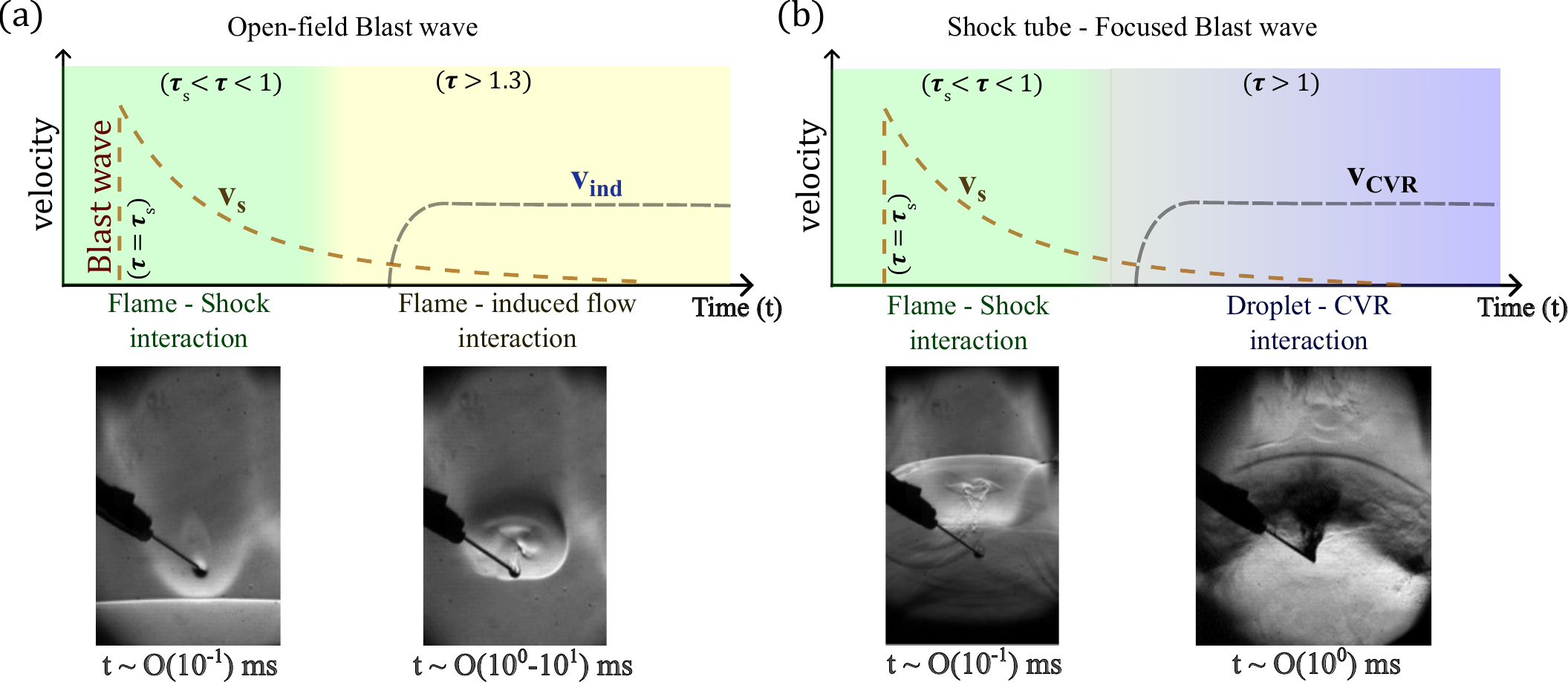}
    \caption{Schematic of the temporal variation of the velocity imposed at the droplet location as the blast wave passes by (a) for open-field blast wave configuration, (b) shocktube-focused blast wave configuration. Green background represents the interaction with decaying velocity profile ($\rm v_s$) behind the blast wave. Yellow background represent the interaction with the induced flow ($\rm v_{ind}$) and blue background indicate the compressible vortex (CVR) interaction with droplet. The corresponding representative Schlieren images of droplet flame during different interactions are shown below. $\tau_s$ represents the time instant of the beginning of blast wave interaction.}
    \label{fig1p5:background}
\end{figure}

In our previous experiments \cite{vadlamudi2024insights}, a similar shock generation setup has been used to conduct the shock - droplet flame interaction experiments to characterize the flow and study the hydrodynamic aspects of the interaction. It has been shown that the interaction of the shock flow occurs in two stages with a droplet flame, (see the schematic in Fig. \ref{fig1p5:background}a): 1) Firstly, when the blast wave passes by the droplet flame, it imposes a sudden discontinuity in the form of peak over-pressure followed by a temporally decaying velocity and pressure profiles at the droplet location. The temporally decaying velocity profile ($\rm v_s$), decays progressively to zero velocity at the droplet location around $t=1\ s$ from time of the wire-explosion (green region in Fig. \ref{fig1p5:background}). It has been shown that the droplet flame undergoes forward extinction and lifts off from the droplet due to this imposed flow ($\rm v_s$). This interaction with $\rm v_{s}$ occurs at faster timescales i.e., $\sim O(10^{-1})ms$. 2) Second stage of the interaction occurs due to the induced flow ($\rm v_{ind}$), characteristic to the current shock generation setup (green region in Fig. \ref{fig1p5:background}a). The induced flow ($\rm v_{ind}$) also interacts with the droplet flame resulting in flame stretching, shedding or extinction, depending on the magnitude of the flow velocity. This $\rm v_{ind}$ interaction occurs at slower timescales of $O(10^0-10^1)ms$ as the velocity scales of $\rm v_{ind}$ is significantly lower than that of $\rm v_{s}$. 

Thus, the flow imposed on the droplet flame in current experiments occurs in two modes. In open-field blast case, a blast wave is generated at the droplet due to wire-explosion and it propagates radially outward away from the copper wire, towards the droplet. On the other hand, the case of focused blast configuration (with shock tube), the blast wave generated due to wire-explosion is confined and is directed along the shock tube length forming a one-dimensional flow i.e., planar blast wave. This planar blast wave propagates further and exits the shock tube, forming a radially expanding cylindrical blast wave, as discussed in our previous study \cite{vadlamudi2024insights}. 

It has been established in our previous experiments \cite{vadlamudi2024insights} that the blast wave generated propagates away from the copper wire location, with a temporally decaying Mach number. The shock trajectory for the blast wave for this experimental setup i.e., the shock radius ($R_s$) variation with time (t, from explosion) is found to follow the correlation of $R_s \propto t^n$, with the exponent '$n$'$\sim 0.88-0.99$. This shows that the blast wave is in the transition between weak blast ($n \sim 1$) and strong blast limit ($n \sim 2/5$). The approximate analytical solution provided by Bach and Lee \cite{bach1970analytical} is shown to be valid for the current experimental setup in our previous study \cite{vadlamudi2024insights}. The approximate analytical model is based on the power-law density profile assumption, using which the velocity profile behind the blast wave can be obtained using the instantaneous shock Mach number ($M_s$). The further details of the model will be provided in the supplementary material. The model has been shown to be valid and in good agreement with the experimental measurements in our previous study \cite{vadlamudi2024insights}. The Bach and Lee model can be used to characterize the blast wave and estimate the velocity imposed by the propagating blast wave for different cases. 

 As the blast wave propagates and interacts with the droplet flame, the propagating blast wave imposes a temporally decaying velocity profile ($\rm v_s$) at the droplet location. This is accompanied with the decaying pressure profile behind the blast wave (estimated using the Bach and Lee model) which decreases continuously and drops below the ambient pressure. The timescale of the pressure decay to reach the ambient pressure is $t_{amb}$ and this timescale can be used to non-dimensionalize the time (t) during the shock interaction. Thus, the dimensionless time is given by $\tau = t/t_{amb}$ (see supplementary Figure S2). The instant at which the blast wave interaction with the droplet begins i.e., $t=t_s$ occurs at $\tau = \tau_s$. The velocity profile imposed by the propagating blast wave ($\rm v_s$) at a given location temporally decays and approaches zero for $\tau > 1$. The solid line in supplementary Fig. S5 shows the temporal variation of $\rm v_s$ for a specific Mach number. However, in the current shock generation setup, an induced flow ($\rm v_{ind}$) is also observed experimentally due to secondary effects such as entrainment even beyond $\tau > 1$. This results in the deviation of the local velocity from the blastwave-imposed decaying velocity profile ($\rm v_s$) at the droplet location during later stages ($t > 1.3 ms$ from the explosion), when the induced flow reaches the droplet location, i.e., $\tau > 1$. The temporally decaying velocity profile ($\rm v_s$) during first stage interaction (for $\tau_s < \tau < 1$) is shown in Fig. \ref{fig1p5:background} using green background, and yellow region represents the slower interaction with the induced flow ($\rm v_{ind}$) for $\tau > 1$. For shock tube-focused configuration, the compressible vortex (CVR) exiting the shock tube interacts with the droplet resulting in droplet breakup (see \ref{fig1p5:background}b), and this interaction is denoted by blue background.

\subsection{Droplet flame response during the interaction of the shock flow}\label{sec:Global}

Since, the current experiments are performed using the same shock generation setup, similar to the previous work \cite{vadlamudi2024insights}, the droplet flame interaction is observed to occur in two stages: Interaction with the velocity profile ($\rm v_s$) imposed by the blast wave and interaction with the induced flow ($\rm v_{ind}$) that interacts with the flame after some delay. In all the cases, during the interaction, the droplet flame undergoes local extinction at the forward stagnation point due to this externally imposed flow and the flame lift-off occurs towards the droplet wake. The current study focuses on the droplet regression rate, flame heat release rate and how the interaction with the shock-flow affects these characteristics by altering the flame topology.

\begin{figure}[t!]
    \centering
    \includegraphics[width=\textwidth,height=\textheight,keepaspectratio]{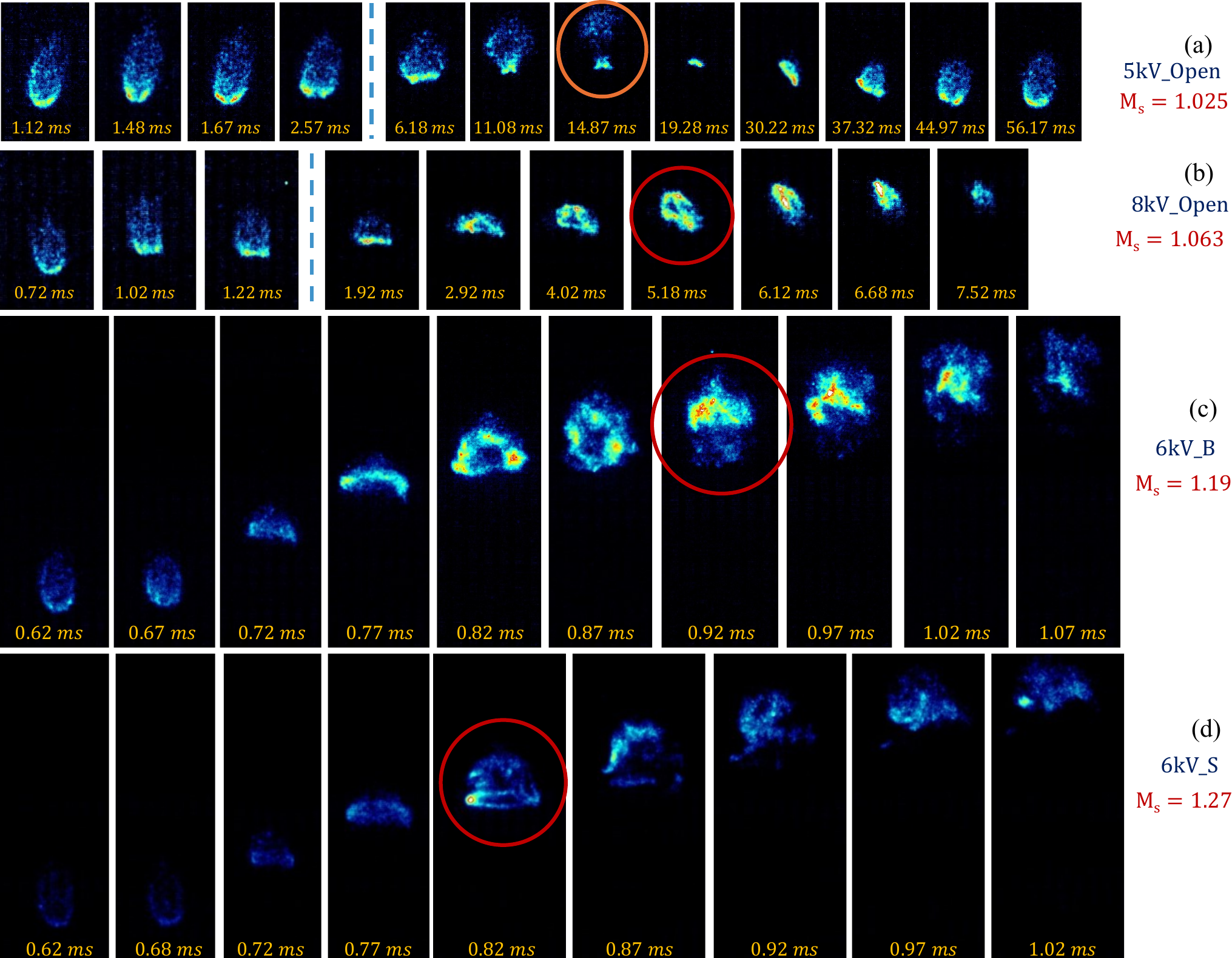}
    \caption{Time series of the OH* Chemiluminescence imaging depicting the overall interaction of the droplet flame with the shock flow with respect to time from explosion (t): For open-field configuration with charging voltage: (a) 5kV, (b) 8kV; (c) For bigger shock tube channel ($2 cm \times 10 cm$ crosssection) with charging voltages: 6kV; and (d) For the smaller shock tube channel ($2 cm \times 4 cm$ crosssection) with charging voltages: 6kV. The time shown in images is  from instant of wire-explosion which is considered to be time ($t=0$).}
    \label{fig2:IRO:Timeseries}
\end{figure}
Fig. \ref{fig2:IRO:Timeseries} shows the time series of the high-speed OH* Chemiluminescence imaging of the droplet flame during the interaction. Fig. \ref{fig2:IRO:Timeseries} (a,b) show the time series for interaction obtained in the open-field configuration experiments, whereas Fig. \ref{fig2:IRO:Timeseries} (c,d) show the time series of flame imaging for shock-focused configurations. It is to be noted that the different experimental configurations i.e., Open, Big section ($2 \times 10 cm^2$) and small section ($2 \times 4 cm^2$) are denoted using ‘Open’, ‘B’ and ‘S’ respectively, along with the charging voltage corresponding to the case (see Fig. \ref{fig2:IRO:Timeseries}). The blue vertical dotted lines shown in Fig. \ref{fig2:IRO:Timeseries} (a,b) indicate the time instant after which induced flow ($\rm v_{ind}$) interaction with the droplet flame occurs. The flame is observed to show a quick momentary liftoff (timescale $\sim O(10^{-1})ms$) during initial interaction with $\rm v_s$ (left side of the dotted line in Fig. \ref{fig2:IRO:Timeseries} a,b). Later, $\rm v_s$ decays temporally to zero which is followed by a receding phase. Subsequently, the flame starts to liftoff away from the droplet at a relatively slower rate, i.e., timescale $\sim O(10^0-10^1)ms$ (rightside of dotted line in Fig. \ref{fig2:IRO:Timeseries}a,b), due to the induced flow interaction ($\rm v_{ind}$). 

\begin{figure}[t!]
    \centering
    \includegraphics[width=\textwidth,height=\textheight,keepaspectratio]{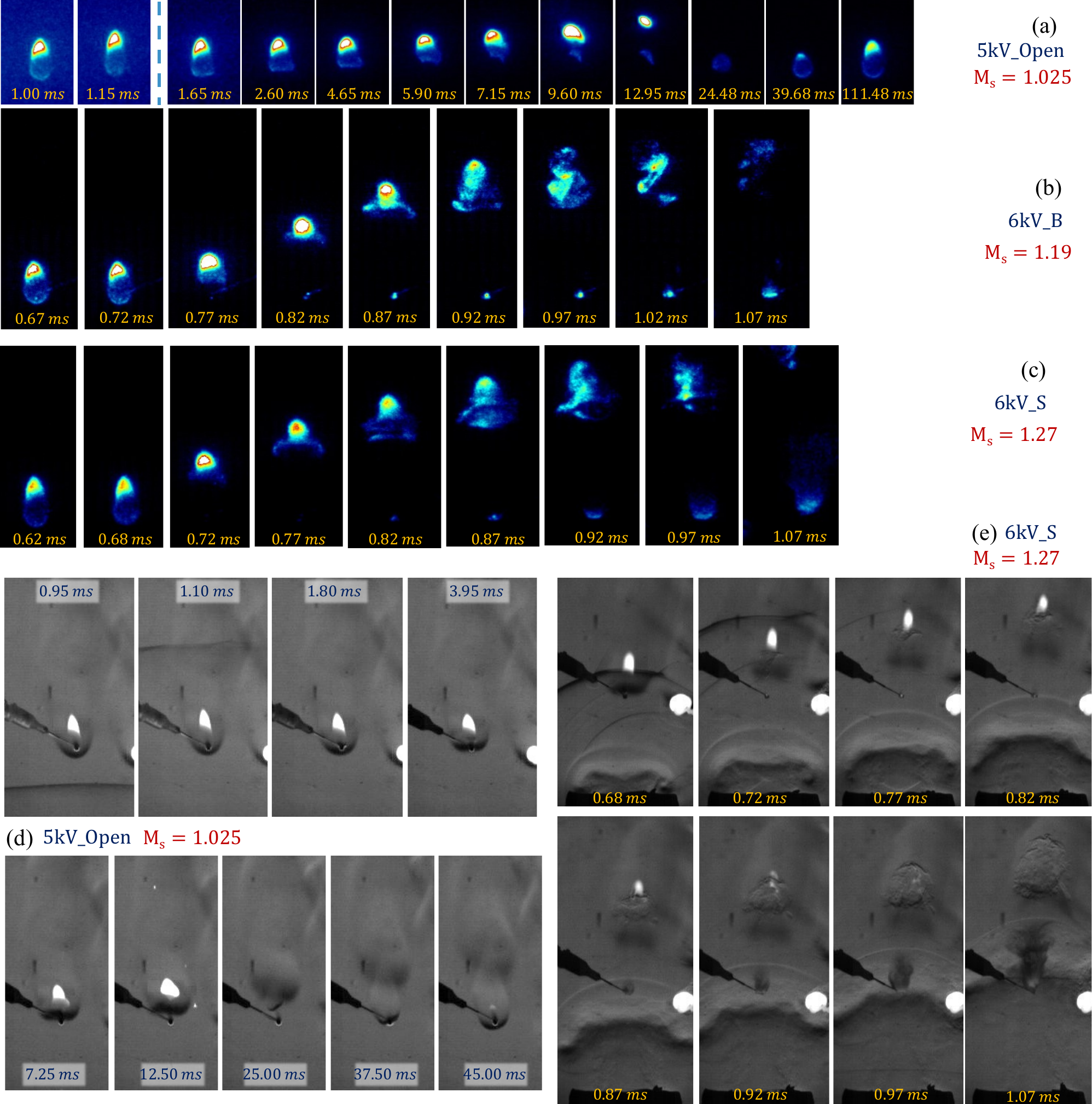}
    \caption{Time series (w.r.t. time from explosion ‘t’) of the CH* Chemiluminescence imaging depicting the overall interaction of the droplet flame with the shock flow for: (a) $5kV\_Open$; (b) $6kV\_B$; and (c) $6kV\_S$. Time series of high-speed Schlieren flow visualization depicting the overall interaction of the droplet flame with the shock flow for: (d) $5kV\_Open$ and (e) $6kV\_S$. The time shown in images is from instant of wire-explosion which is considered to be time ($t=0$).}
    \label{fig3:IRO:CH_Timeseries}
\end{figure}

The faster initial flame response during interaction with $\rm v_s$ becomes more pronounced with increase in shock strength ($M_s$), see Fig. \ref{fig2:IRO:Timeseries}a,b. Furthermore, the induced flow ($\rm v_s$) for $M_s \sim 1.025$ (Fig. \ref{fig2:IRO:Timeseries}a) is lower compared to $M_s \sim 1.063$ (Fig. \ref{fig2:IRO:Timeseries}b), resulting in significantly slower liftoff rate (timescale $\sim O(10^1)ms$) for $M_s \sim 1.025$, compared to $M_s\sim 1.063$ (timescale $\sim O(10^0)ms$). Thus, sufficient time is available for $M_s \sim 1.025$ case to facilitate vortex rollup, and circulation buildup that resulted in flame shedding, as shown in Fig. \ref{fig2:IRO:Timeseries}a (indicated with orange circle). This flame shedding occurred due to the interaction of the slower induced flow ($\rm v_{ind}$) with the droplet flame\cite{vadlamudi2024insights}. Later the flame reattaches to the droplet forming an enveloped flame around $t \sim 50ms$ (see Fig. \ref{fig2:IRO:Timeseries}a), thus regaining to the quasi-steady droplet flame condition after a time scale of $t \sim O(10^2)ms$ from wire explosion. Fig. \ref{fig3:IRO:CH_Timeseries}a and Fig. \ref{fig3:IRO:CH_Timeseries}d shows the time series of the corresponding high-speed CH* chemiluminescence and high-speed Schlieren imaging of the interaction of the droplet flame with blast wave for $M_s \sim 1.025$. 

The incident blast wave can be observed in first two images in Fig. \ref{fig3:IRO:CH_Timeseries}d, which imposes a temporally decaying velocity, $\rm v_s$ at the droplet location (for $t < 1.3ms$) and after this interaction, flame is observed to gradually liftoff again, after some delay (for $t > 1.80ms$ in Fig. \ref{fig3:IRO:CH_Timeseries}d), where induced flow ($\rm v_{ind}$) starts to interact with the droplet flame. The CH* chemiluminescence time series in Fig. \ref{fig3:IRO:CH_Timeseries}a also shows similar flame dynamics as discussed before, where initial faster flame response (for $t < 1.3ms$) and the flame shedding (for $t > 1.3ms$) are more clearly evident in Fig. \ref{fig3:IRO:CH_Timeseries}a because of the sooty bright yellow flame tip. A disturbance (undulation) on the flame boundary near the flame base is seen to travel downstream along the flame length from $t \sim 2.60ms$ to $t \sim 7.15ms$ (see Fig. \ref{fig3:IRO:CH_Timeseries}a), which finally results in flame shedding at $t \sim 9.60ms$. This corresponds to the vortex rollup and circulation buildup along the flame length \cite{vadlamudi2024insights}, and this shedding phenomenon is also clearly seen in Fig. \ref{fig3:IRO:CH_Timeseries}d (Schlieren time series) between $t \sim 7.25ms$ and $t \sim 12.50ms$. The hot plume around the droplet flame also clearly shows a vortical structure convecting downstream between $t \sim 7.25ms$ to $t \sim 37.50ms$, which corresponds to the vortex rollup that results in flame shedding. Subsequently, the flame reattachment at the droplet can be clearly observed between $t \sim 25ms$ and $t \sim 40ms$ in Fig. \ref{fig3:IRO:CH_Timeseries}d, where the lifted-off flame base moves towards the droplet to finally restore fully enveloped flame. Later, the bright yellow tip is observed to reappear around $t \sim 40ms$, as shown in Fig. \ref{fig3:IRO:CH_Timeseries}a,d.

As the Mach number is increased to $1.06 < M_s < 1.1$, flame reattachment is not observed. As shown in Fig. \ref{fig2:IRO:Timeseries}b, after the initial interaction with $\rm v_s$ (left side of dotted line), the flame base undergoes continuous lift-off resulting in full extinction when flame base reaches the flame tip, during the interaction with $\rm v_{ind}$ (right side of dotted line). The same can be observed in the time series shown in Fig. \ref{fig2:IRO:Timeseries}b where, as the flame base continuously swept towards the flame tip, it is accompanied by an increase in flame intensity (indicated by red circle in Fig. \ref{fig2:IRO:Timeseries}b). This is followed by full extinction of the flame. In all the flame dynamics discussed so far for the lower Mach number range ($M_s < 1.1$) the droplet flame sustains beyond the initial interaction with blast wave velocity ($\rm v_s$) and it subsequently interacts with $\rm v_{ind}$. 

However, when Mach number is increased further beyond $M_s > 1.1$ (achieved using shock focusing), complete flame extinction occurs which are observed to occur at faster rates (timescale $\sim O(10^{-1}) ms$) during the initial interaction stage with the blastwave-imposed velocity profile ($\rm v_s$). Thus, the flame does not sustain beyond the first interaction (with $\rm v_s$) to subsequently interact with the induced flow ($\rm v_{ind}$). This is shown in the OH* chemiluminescence time series in Fig. \ref{fig2:IRO:Timeseries}c,d ($6kV\_B$ $\&$ $6kV\_S$ cases), where the flame-base liftoff is observed to be significantly faster ($\sim O(10^{-1}) ms$). This rate of liftoff becomes faster with increase in Mach number from ($Ms \sim 1.19$ to $1.27$) as shown in Fig. \ref{fig2:IRO:Timeseries}c,d, before the imminent extinction or blowout for all the cases ($M_s > 1.1$). Interestingly, during this high-speed flame blowout, the flame intensity gets enhanced (circled in red, in Fig. \ref{fig2:IRO:Timeseries}c,d) before the imminent blowout. Fig. \ref{fig3:IRO:CH_Timeseries}(b,c) show the corresponding CH* chemiluminescence of $6kV\_B$ and $6kV\_S$ cases which shows the corresponding flame topology along with the bright flame tip. The droplet atomization due to the interaction with the compressible vortex (exiting the shock tube, see Fig. \ref{fig3:IRO:CH_Timeseries}e) is also visible in the CH* chemiluminescence imaging (bottom, near the droplet) as the atomized spray scatters the bright illumination of the blast (this illumination is captured by the high-speed camera in the CH* band spectral range, i.e., $\sim 431nm$). The same can also be observed in the high-speed Schlieren flow visualization time series shown for $6kV\_S$ case in Fig. \ref{fig3:IRO:CH_Timeseries}e. The Schlieren images in Fig. \ref{fig3:IRO:CH_Timeseries}e show the flame liftoff phenomena during the interaction with the blast wave (at $t \sim 0.68ms$ shown in Fig. \ref{fig3:IRO:CH_Timeseries}e), where the flame base location, advection, location of the hot gases is clearly visualized. After the blast wave interaction, the flame-base is swept downstream towards the flame tip, and it interacts with the flame tip around $t \sim 0.82ms$ and results in the disappearance of the bright yellow flame tip after $t \sim 0.92ms$. Subsequently, this results in complete extinction or blowout of the flame after $t \sim 1.07ms$ (see Fig. \ref{fig3:IRO:CH_Timeseries}c,e). 

\label{v_Th}It has been established in previous experiments \cite{vadlamudi2024insights} that the temporal variation of the flame-base advection velocity during the interaction with blast wave (for $\tau < 1$) is in good agreement with the temporally decaying velocity ($\rm v_s$) imposed by the propagating blast wave at the droplet location. Supplementary Fig. S5 shows that the temporal variation of the flame advection velocity (data points) and flow velocity imposed at the droplet location due to the blast wave for a sample shock-droplet flame interaction case (with $M_s \sim 1.3$ and the wire explosion is triggered with a 1.70-second delay after the droplet ignition). The temporal variation of the advection velocity of the flame-base ($\rm v_{flame base}$) is in good agreement with the temporal variation of velocity ($\rm v_{s,Th}$) imposed at the droplet location which is obtained theoretically from blast wave propagation using Bach and Lee formulation \cite{bach1970analytical} for $\tau < 1$. This implies that the flame lift-off and advection for $\tau < 1$ is the direct consequence of the temporally decaying velocity ($\rm v_s$) imposed at the droplet location by the propagating blast wave.

The compressible vortex (CVR) shown in Fig. \ref{fig3:IRO:CH_Timeseries}e is observed to only interact with the droplet, as the flame-extinction has already occurred before the vortex reaches the lifted flame. It can be clearly seen in Fig. \ref{fig2:IRO:Timeseries}d that the flame-base is convected downstream along the centerline, interacting with the bright flame tip from $t \sim 0.82ms$, exhibiting a dome-like topology (with central fuel jet convecting downstream towards flame tip), as seen in Fig. \ref{fig2:IRO:Timeseries}d and Fig. \ref{fig3:IRO:CH_Timeseries}c,e. This is accompanied by a significant enhancement in flame intensity momentarily, before the imminent blowout or extinction of the flame.

\begin{figure}[t!]
    \centering
    \includegraphics[width=\textwidth,height=\textheight,keepaspectratio]{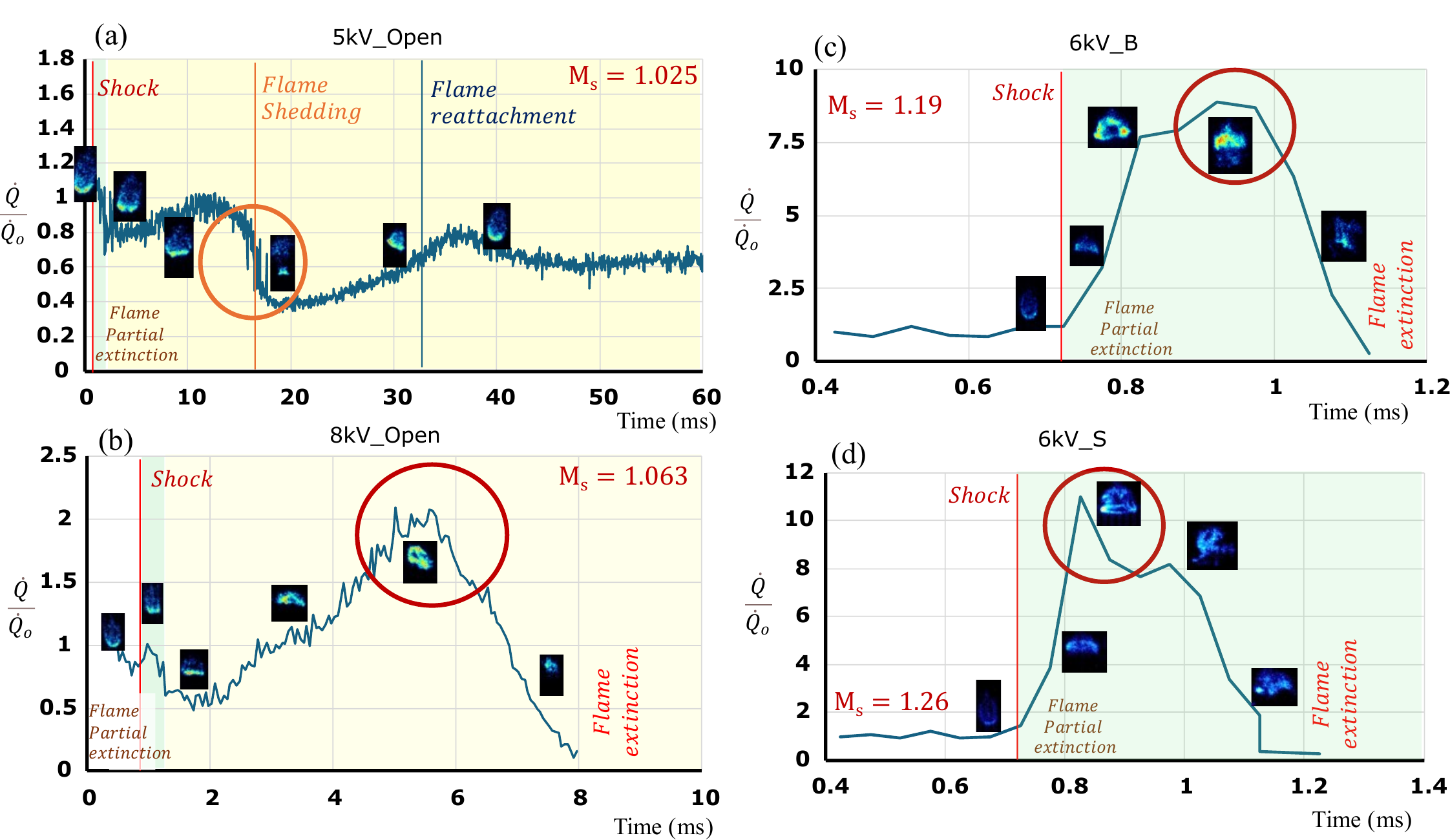}
    \caption{OH* Chemiluminesence flame intensity variation during interaction that is normalized using the unforced droplet flame intensity: Normalized flame heat release rate (HRR) i.e., ($\dot{Q}/\dot{Q}_o$) plotted against time from wire-explosion ($t$) for (a) $\rm 5kV\_Open$ ($M_s \sim 1.025$); (b) $\rm 8kV\_Open$ ($M_s \sim 1.063$). The blast wave incidence at the droplet location is denoted by solid vertical red line. Green region is the interaction with blastwave-imposed velocity profile ($\rm v_s$) and yellow region denotes the induced flow velocity ($\rm v_{ind}$) interaction; (c) $\rm 6kV\_B$ ($M_s \sim 1.19$) and (d) $\rm 6kV\_S$ ($M_s \sim 1.26$). OH* chemiluminescence images corresponding to the temporal variation of the normalized heat release plot are shown as sub-figures.}
    \label{fig4:IRO:HRR_Global}
\end{figure}

\subsubsection{Effect on flame heat release rate (HRR)}\label{sec:HRR_Effect}
Since the OH* Chemiluminescence is an indicator of flame heat release rate, the net integrated flame intensity variation is obtained from OH* chemiluminescence images following the procedure mentioned in experimental methodology section. As the primary focus of the current study is to investigate the effect of the interaction with the shock-flow with the droplet flame, the flame heat release rate variation ($\dot{Q}$) obtained is normalized using the average nominal flame heat release rate (unforced droplet flame) obtained from the average flame intensity ($\dot{Q}_o$) prior to the interaction. Thus, the temporal variation of the normalized flame heat release rate ($\dot{Q}/\dot{Q}_o$) is plotted in Fig. \ref{fig4:IRO:HRR_Global} for 5kV, 8kV cases in open-field configuration, i.e., $\rm 5kV\_Open$ ($M_s \sim 1.025$), $\rm 8kV\_Open$ ($M_s \sim 1.063$) and for $\rm 6kV\_B$ ($M_s \sim 1.19$), $\rm 6kV\_S$ ($M_s \sim 1.26$) cases in shocktube-focused configuration. 

The time instant of the beginning of shock interaction with droplet flame is indicated by solid red vertical line in the plots, and the blastwave-imposed velocity ($\rm v_s$) interaction, induced flow velocity interaction ($\rm v_{ind}$) are indicated by green and yellow regions respectively. Initially, the heat release rate (HRR) is observed to reduce during the interaction with the decaying velocity profile behind blast wave ($\rm v_s$) as the flame undergoes forward extinction at the forward stagnation point and continuously lifts-off downstream (for $\tau<1$), see green region in Fig. \ref{fig4:IRO:HRR_Global}. 

Subsequently, during the interaction with the induced flow ($\rm v_{ind}$) i.e., $\tau>1$, for lower Mach number ($M_s \sim 1.025$), the flame HRR increases momentarily showing a local peak (near the vertical orange line Fig. \ref{fig4:IRO:HRR_Global}a). This corresponds to the flame shedding that occurs due to the continuous rollup and detachment of vortex around the flame due to buoyancy-induced instability \cite{pandey_dynamic_2021} (as shown in Fig. \ref{fig2:IRO:Timeseries}a (orange circle) and Fig. \ref{fig3:IRO:CH_Timeseries}a,d). Consequently, the flame HRR drops drastically as the flame tip sheds away leaving back the lifted flame base, which later gradually recedes and reattaches at the droplet forming enveloped flame (see Fig. \ref{fig2:IRO:Timeseries}a, Fig. \ref{fig3:IRO:CH_Timeseries}a). This enveloping of the lifted flame after the interaction causes a rise in flame HRR as corresponding to the vertical blue line in Fig. \ref{fig4:IRO:HRR_Global}a. After enveloping, the flame undergoes few cycles of oscillations which gradually decay to restore the quasi-steady droplet flame similar to that present prior to the shock interaction. 

On the other hand, as Mach number is increased to $M_s \sim 1.063$, similar to $M_s \sim 1.025$ case, the flame HRR exhibits the initial spike in the green region ($\tau<1$). This initial spike in flame heat release rate trend (near the end of green region) corresponds to the momentary flame stretching that occurs during the interaction with $\rm v_s$. Subsequently, as the flame interacts with the induced flow ($\rm v_{ind}$), flame base lifts off and is continuously swept downstream towards the flame tip, resulting in a temporary reduction in both flame height (see Fig. \ref{fig2:IRO:Timeseries}b) and flame heat release rate (see Fig. \ref{fig4:IRO:HRR_Global}b in yellow region). Subsequently, it can be observed from experiments that both the flame-base and flame-tip begin to advect downstream (due to $\rm v_{ind}$), which results in the interaction of flame-base with the flame tip, as shown in the OH* Chemiluminescence timeseries images in Fig. \ref{fig2:IRO:Timeseries}b for $t > 2 ms$. This flame base interaction with flame tip results in flame extinction which is in agreement with our previous experiments \cite{vadlamudi2024insights}. However, interestingly, as the flame-base advects and reaches the flame tip, the flame becomes more brighter or intense just before it decreases again resulting in extinction (see Fig. \ref{fig2:IRO:Timeseries}b, $t \sim 5 ms$). It is interesting to note that this flame HRR enhancement in yellow region ($\tau>1$) is found to be almost twice that of the nominal HRR, which can be observed in the normalized HRR variation plot (see Fig. \ref{fig4:IRO:HRR_Global}b) in the form of a prominent local peak. This HRR enhancement can be attributed to the fuel accumulation near the flame tip as the flame base and fuel vapor gets transported downstream towards the flame tip due to $\rm v_{ind}$. This fuel vapor transport towards flame tip can result in local enhancement in the fuel concentration at the flame location, which can explain the enhancement in the flame heat release rate before the imminent extinction. After the momentary peak, the flame HRR finally decays to zero as the flame-extinction occurs consequently. This flame HRR enhancement is observed to occur in a timescale of $\sim O(10^0)ms$.

However, as the Mach number is increased beyond $M_s > 1.1$, the flame dynamics are completely altered (see Fig. \ref{fig2:IRO:Timeseries}c,d). At high Mach numbers, the droplet flame is observed to fully extinguish at faster rate before the timescale of $\tau<1$, i.e., in green region, which occurs at  timescales that are almost one order lower than $M_s<1.1$. This implies that the initial interaction with the decaying velocity profile ($\rm v_s$) imposed by the blast wave results in full extinction of the flame. Fig. \ref{fig4:IRO:HRR_Global}c,d show the normalized heat release rate (HRR) variation of the droplet flame during this interaction, where, the red vertical solid line indicates the instant at which shock wave begins to interact with the droplet flame and the time period of the interaction  with $\rm v_s$ is represented by green background. Similar to $1.06< M_s <1.1$, the HRR enhancement is also observed for high Mach numbers ($M_s>1.1$), however, interestingly, HRR enhancement is significantly higher (almost 8 times the nomimal unforced flame HRR), as shown in Fig. \ref{fig4:IRO:HRR_Global}c,d, which is a significant deviation from $M_s<1.1$ regime extinction. Thus, in the subsequent sections, the two Low ($M_s<1.1$) and High Mach number ($M_s>1.1$) regimes will be discussed in detail. 

\begin{figure}[b!]
	\centering
	\includegraphics[width=\textwidth,height=\textheight,keepaspectratio]{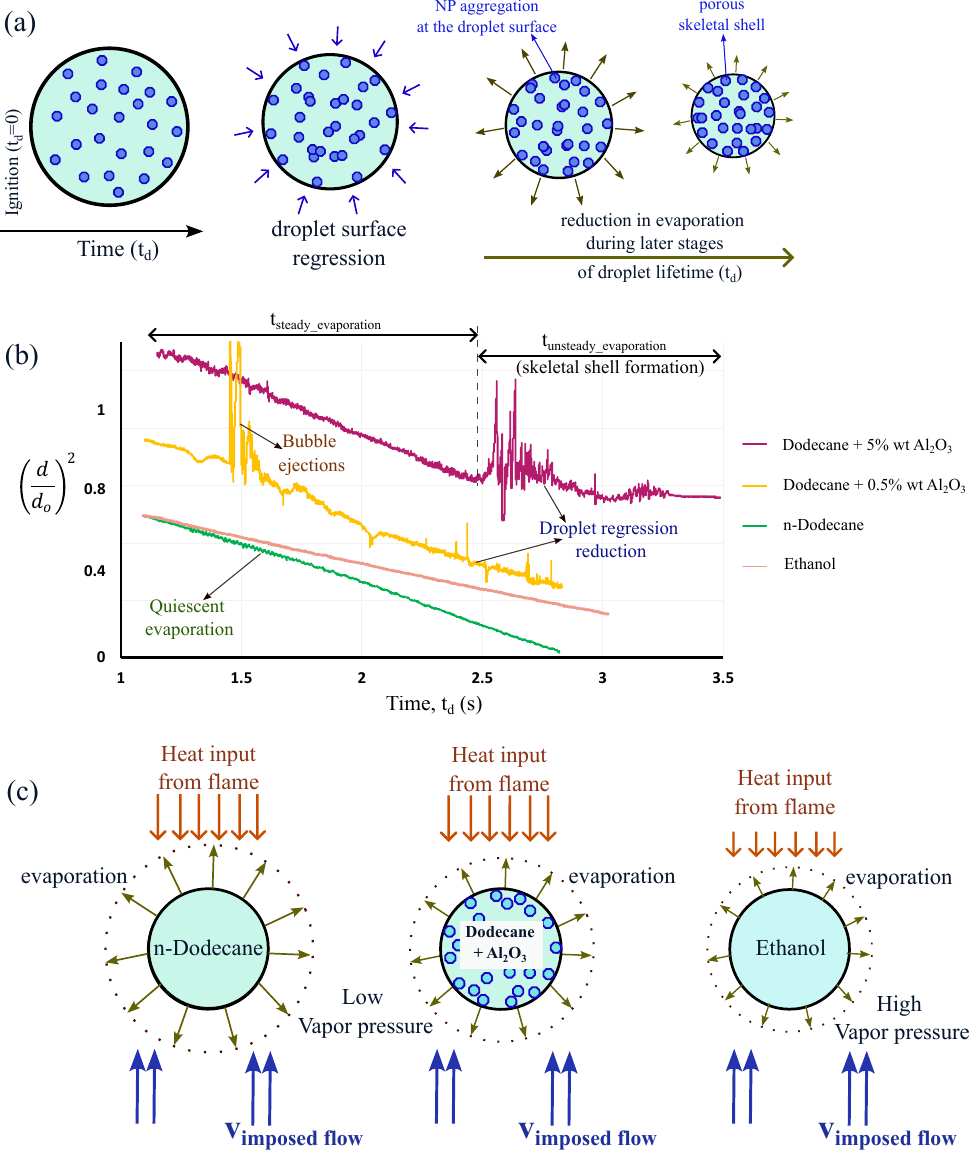}
	\caption{(a) Schematic of the temporal snapshots inside Nanofuel droplet during different stages of evaporation. (b) Temporal droplet regression of droplet combustion nanofuels for different particle loading rates (PLR) and for Ethanol plotted against droplet lifetime, $t_d$ (It is to be noted that the different plots are offset vertically for better visualization, and all the plots begin at $(d/d_o)^2 \sim 1$ at $t_d=0$). The peaks and oscillations in the droplet regression plot for "DD + $0.5\%$ wt $Al_2O_3$" case (yellow) correspond to bubble ejections. (c) Schematic showing the effect of heat input from the flame and fuel type.}
	\label{fig4p5:IRO:NP_Eth_effect}
\end{figure}

\subsection{Effect of type of the fuel}\label{sec:FuelEffect}
So far the general trend of the flame response to the shock-flow has been briefly discussed. Moreover, to further the understanding of different phenomena involved during the interaction of the shock-flow with droplet flame, the effect of different fuel types also needs to be investigated. 

Nanofuels are one of the promising option that offers various advantages over the conventional fuels. The combustion characteristics of conventional fuels are observed to be enhanced when the nanopowder of metal or metal oxide additive is dispersed in conventional liquid fuels \cite{gan2012combustion,choi1995enhancing}. One way of increasing the energy density of conventional fuels is the addition of the solid particles such as boron, aluminium, and carbon as "liquid-fuel extender"  \cite{gan2012combustion}. The solid additive is dispersed in the conventional fuels in the form of a nanopowder is shown to increase in surface area to volume ratio which enhances catalytic properties, high reactivity and significantly different thermo-physical properties \cite{pandey2019high}. It has also been reported that Nanofuels exhibit a heightened heat transfer coefficient of liquids with NP addition \cite{choi1995enhancing}.The nanoparticle (NP) addition also showed increased ignition probability\cite{tyagi2008increased}, enhaced burning rates\cite{tanvir2016droplet} and reduced soot formation\cite{rotavera2009effect}. Lenin et al. \cite{lenin2013performance} have reported decreased level of pollutants in diesel engine exhaust using nano-additives of manganese oxide (MnO) and copper oxide (CuO). Shaafi et al. \cite{shaafi2015effect} have provided a comprehensive review on performance and exhaust of CI engine fueled with nano-additive enhanced biodiesel. Sajith et al.\cite{sajith2010experimental} have reported an appreciable improvement in brake thermal efficiency of a single cylinder CI engine fueled with biodiesel laden with Ceria NPs. Thus, research on such nanofuels becomes relevant in the development of nexgen fuels that can contribute towards the smoother transition into the sustainable fuel sources of the future, thus addressing the global climate crisis. 

Fig. \ref{fig4p5:IRO:NP_Eth_effect}a shows the schematic of the droplet interior during the lifetime of nanofuel droplet combustion. The Fig. \ref{fig4p5:IRO:NP_Eth_effect}a shows that in nanofuel laden droplet combustion, as the fuel vapor evaporates at the droplet surface, droplet diameter regression occurs. The nanoparticles begin to aggregate in the droplet interior resulting in the formation of heterogeneous nucleation sites which are shown to increase the net vaporization rates through bubble ejections at lower NP particle loading rates (PLR). Furthermore, the temporally receding droplet surface results in the aggregation of the dispersed nanoparticles (NP) at the droplet surface, especially during the later stages of the droplet lifetime ($t_d > 0.6 t_{d,max}$), as shown in Fig. \ref{fig4p5:IRO:NP_Eth_effect}. In current experiments, $t_d = 0$ corresponds to the instant of ignition of the droplet. The aggregated NPs tends to form a gelatinous crust at the droplet surface entrapping liquid inside for higher NP particle loading rates (PLR), which results in the evaporation of fuel vapor through the skeletal network of NP aggregates analogous to flow through a porous medium as per Darcy's law \cite{pandey2019high,miglani2015coupled}. The capillary pressure at the porous skeletal structure ($\Delta P_{cap}$) ensures the flow of the solvent through the porous network and is given by
$\Delta P_{cap} \sim -2\sigma/r_m$ where, $\sigma$ is surface tension of solvent ($\sim 25mN/m$) and $r_m$ is the meniscus radius (order of NP particle size, $\sim 40nm$). Thus, in nanofuel droplet combustion, at the later stages of the droplet lifetime ($t_d > 0.6t_{d,max}$), the vaporization at the surface suppressed due to the porous skeletal structure formation, which can be observed in the droplet regression plots for different concentrations of nanofuels in Fig. \ref{fig4p5:IRO:NP_Eth_effect}b.

Thus, as shown in Fig. \ref{fig4p5:IRO:NP_Eth_effect}b, the droplet evaporation occurs in two stages. First stage can termed as the steady evaporation where, the droplet regression occurs at steady rate (constant slope). Furthermore, during the later point of droplet lifetime ($t_d$), the droplet regression slows down, (indicated by $t_{unsteady evaporation}$ in Fig. \ref{fig4p5:IRO:NP_Eth_effect}b, which is termed as the unsteady evaporation stage. This effect is observed to be more prominent with increase in the NP particle loading rate (PLR) as shown in Fig. \ref{fig4p5:IRO:NP_Eth_effect}b. While the shell formation (that result in reduction in droplet regression rate) is observed to be prominent for $5\%$ PLR of NP addition, the phenomena such as internal ebullition, bubble incipience, ejections (which result in relatively enhanced net vaporization rate) are prominent for $0.5\%$ PLR case as shown in Fig. \ref{fig4p5:IRO:NP_Eth_effect}b. It is known that the particle aggregation result in two competing phenomena of heterogeneous nucleation in the droplet interior and the shell formation at the droplet surface, which enhance and reduce the droplet regression rate respectively \cite{pandey2019high}. This is reflected in the experimental data shown in Fig. \ref{fig4p5:IRO:NP_Eth_effect}b, where shell formation (reduction in regression rate for $t_d > 0.6 t_{d,max}$) occurs at higher PLRs (solid purple line), whereas significant bubble ejections occur at intermediate PLRs (solid yellow line). For intermediate PLR (solid yellow line), the droplet regression plot showed relatively higher slope compared to pure Dodecane due to bubble ejection events for $t_d < 0.6 t_{d,max}$. However, during later stages of droplet lifetime ($t_d > 0.6 t_{d,max}$), the porous skeletal shell formation effects become dominant that slows down the droplet regression rate relative to the initial stages, which can be observed in Fig. \ref{fig4p5:IRO:NP_Eth_effect}b. This shell formation effect becomes more dominant with increase in NP PLR as shown in Fig. \ref{fig4p5:IRO:NP_Eth_effect}b, where the intensity of bubble ejection events reduce with PLR increase (from $0.5\%$ to $5\%$ by weight of NPs) for $t_d < 0.7 t_{d,max}$. Furthermore, the extent of reduction of the droplet regression rate becomes more prominent at high PLRs ($5\%$ wt) compared to intermediate PLRs ($0.5\%$ wt), which is evident from the flattening of solid purple line at the end of droplet lifetime ($t_d \sim 0.9 t_{d,max}$), as shown in Fig. \ref{fig4p5:IRO:NP_Eth_effect}b. However, significant droplet ejection events were also seen during this final stages of droplet lifetime ($t_d \sim 0.9t_{d,max}$), where a noticeable sol-gel transition is observed with an elastic shell formation. This is evident from the significantly higher amplitude of the oscillations in droplet regression plot for $5\%$wt PLR (high PLR) during final stages of droplet lifetime ($t_d \sim 0.9 t_{d,max}$). This may be attributed to the growth and ejection of the entrapped bubbles inside the shell or the bubble formation and ejection due to the heterogeneous nucleation of the small amount of entrapped liquid (fuel and/or surfactant) inside the shell. However, this final stage is out of scope of current study as the current study is limited to pendant droplet combustion where droplet size becomes similar diameter as that of the quartz rod holding it during these final stages ($t_d \sim 0.9 t_{d,max}$).

Alongside the effect of the particle addition to the base fuel, the effect of the vapor pressure of the fuel is also investigated in the current study. In contrast to the low vapor pressure i.e., n-Dodecane used in current study, experiments were also conducted using a high vapor pressure fuel such as Ethanol. Since, Ethanol is specifically chosen for current experiments as it is one of the prominent biofuels, which is also a primary fuel additive used in many applications. Experimentally, the Ethanol droplet flame is observed to show lower heat release rates (HRR) compared to n-Dodecane (Supplementary Fig. S3), due to lower heat of combustion of Ethanol ($\sim 1371 kJ/mol$) compared to n-Dodecane ($\sim 7900 kJ/mol$). Thus, even though Ethanol is more volatile than n-Dodecane due to its high vapor pressure, lower vaporization rate, i.e., slower droplet regression rate is observed in case of Ethanol due to lower heat input from the flame compared to n-Dodecane. This is also reflected in the droplet regression plot in Fig. \ref{fig4p5:IRO:NP_Eth_effect}b showing slower droplet regression for Ethanol compared to Dodecane, although both the pure fuels exhibited quiescent smooth, linear droplet regression mode due to the absence of NPs.

The schematic shown in Fig. \ref{fig4p5:IRO:NP_Eth_effect}c shows the evaporation at the droplet surface, heat input from the flame for different fuels (relative to each other). The evaporation rate at the droplet is highest for pure Dodecane which decreases with NP addition (due to skeletal shell formation), and is lower for Ethanol due to low HRR of the flame. Thus, the fuel type alters the fuel vaporization rate and HRR. Hence, in the current experiments, different fuels (n-Dodecane, low vapor pressure fuel; Ethanol, high vapor pressure fuel and nanoparticle laden fuels: $5\%$wt $\&$ $ 0.5\%$wt $Al_2O_3$ + Dodecane) were used. Since, the droplet composition varies with time during the nanofuel droplet combustion, the shock incidence time with respect to droplet lifetime ($t_d$) has also been varied experimentally along with the shock Mach number ($M_s$). The further sections will discuss the effects of the fuel-type, nanoparticle concentration and shock incidence delay on the flame and droplet dynamics for different $M_s$ regimes.

\subsection{Droplet flame response: Low Mach number regime ($M_s < 1.1$)} \label{sec:Ms<1.1_IRO}

For open blast wave cases ($M_s < 1.1$), the droplet interaction with the shock or induced flow does not show any breakup or atomization characteristics. However, due to the interaction of the blastwave, the droplet regression rate is observed to be altered. In an unforced droplet combustion, the droplet regression plot $(d/d_o)^2$ vs time shows an initial preheating time \cite{pandey2017combustion}, beyond which quasi-steady droplet combustion is observed where, the droplet regression plot $(d/d_o)^2$ vs time is observed to show a constant slope ($m_1$), which has been termed as $d^2$ law in the literature \cite{law1982recent}. This slope ($m_1$) is observed to be altered after the droplet flame interacts with the shock flow in current experiments. 

\begin{flushleft}\textbf{Regime-I: Partial extinction followed by reignition ($M_s < 1.06$)}\end{flushleft} \label{sec:Ms<1.06}
\par
\begin{figure}[t!]
    \includegraphics[width=\textwidth,height=1\textheight,keepaspectratio]{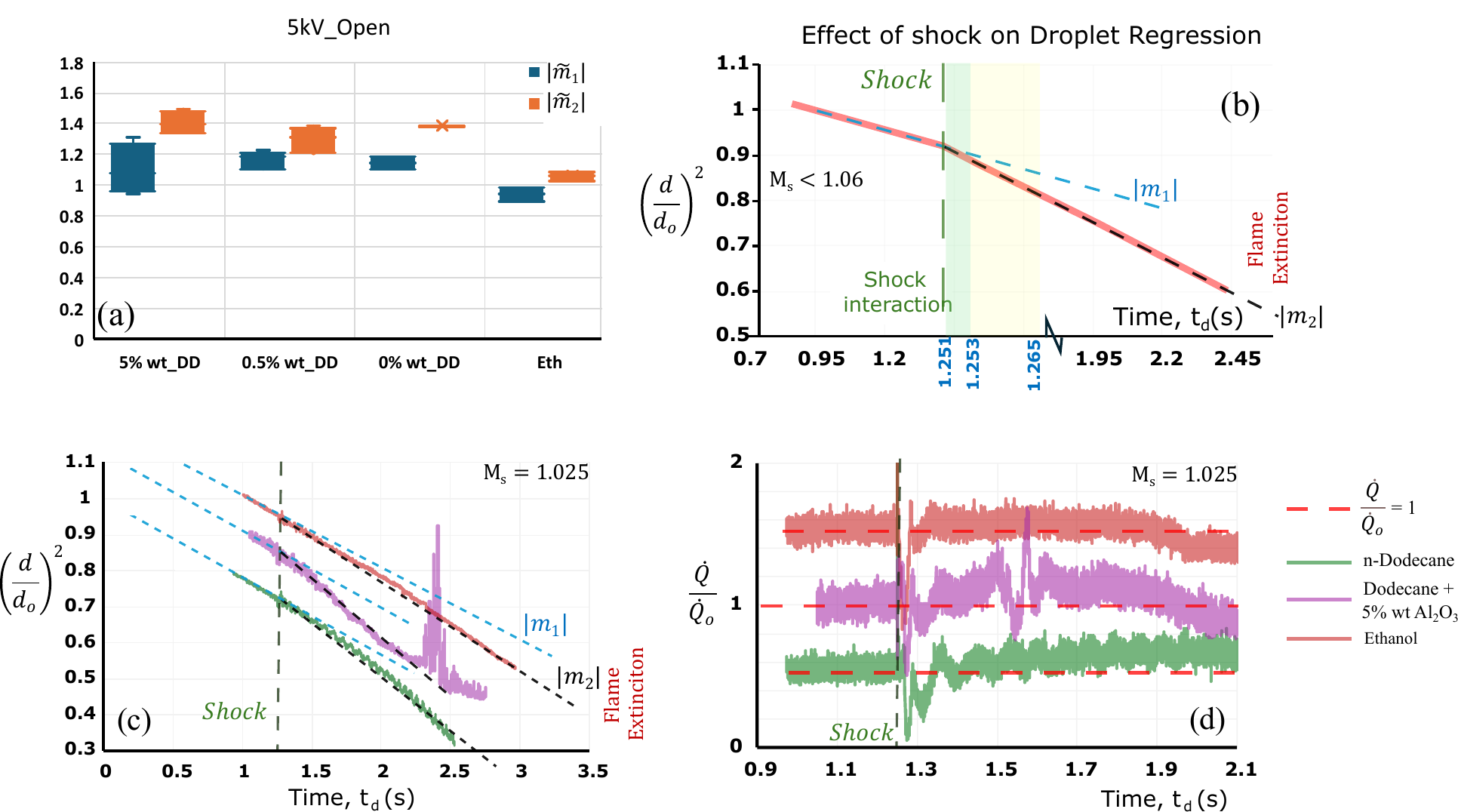}
    \caption{(a) Normalized slope ($ |m| $) variation of the droplet regression rate $(d/d_o)^2$ before and after interaction with the shock-flow for $M_s \sim 1.025$ (Regime-I); (b) Schematic of the droplet regression rate alteration due to interaction with the shock-flow for $5kV\_Open$ ($M_s \sim 1.025$); (c) Droplet regression rate slope ($ |m| $) variation before and after interaction with the shock-flow plotted against droplet life time ($t_d$) for  $5\%$ wt NP + DD, pure Dodecane and pure Ethanol cases with a shock delay of $\sim 1.25s$ for shock Mach number of $M_s \sim 1.025$. (d) Normalized heat release rate (HRR) of Droplet flame ($\dot{Q}/\dot{Q}_o$) for $5\%$ wt NP + DD, pure Dodecane and pure Ethanol cases with a shock delay of $\sim 1.25s$ for $M_s \sim 1.025$. The green vertical dotted line indicates the time instant of shock incidence, blue and black inclined dotted lines represent the droplet regression slope before ($|m_1|$) $\&$ after ($|m_2|$) the interaction with the shock-flow and red horizontal dotted line represents the unity value of normalized HRR ($\dot{Q}/\dot{Q}_o = 1$). The plots in (c $\&$ d) are offset vertically for better visualization, however all the plots in 'c' begin at $(d/d_o)^2=1$ at $t_d=0$ and all the plots in 'd' begin at $\dot{Q}/\dot{Q}_o = 1$ before shock interaction. The green and yellow backgrounds indicates the interaction with the blastwave-imposed velocity profile ($\rm v_s$) and induced flow ($\rm v_{ind}$) respectively.}
    \label{fig5:IRO:Ms<1.06}
\end{figure}

 For low Mach number regime ($M_s < 1.06$), the droplet flame is observed to reignite or reattach at the droplet after the liftoff due to the interaction with $\rm v_s$ and $\rm v_{ind}$ imposed by the shock-flow in current experiments. As mentioned before, in current experiments, the shock incidence time (referred to as ‘shock delay’) has been varied so that the blast wave interacts with droplet flame at different stages of droplet life-time ($t_d$). The plots of the droplet regression and flame HRR variation for different fuels with respect to the droplet life-time ($t_d$) are shown in Fig. \ref{fig5:IRO:Ms<1.06}c,d. The instant of shock incidence is indicated by a vertical green dotted line in the Fig. \ref{fig5:IRO:Ms<1.06}c plot. 

Fig. \ref{fig5:IRO:Ms<1.06}b shows the schematic of the general trend followed by the droplet regression $(d/d_o)^2$ which is plotted against the droplet lift-time ($t_d$), and the instant of beginning of shock interaction with droplet flame is denoted by vertical green dotted line. After the shock incidence, the droplet flame undergoes two stage interaction (as discussed in previous sections) with blastwave-imposed decaying velocity profile ‘$\rm v_s$’ (denoted by green background in Fig. \ref{fig5:IRO:Ms<1.06}b) and the subsequent induced flow velocity ‘$\rm v_{ind}$’ (denoted by yellow background in Fig. \ref{fig5:IRO:Ms<1.06}b). After the shock incidence, the first stage of interaction with $\rm v_s$ occurs within in a timescale of $\sim O(10^{-1})ms$ and subsequently the interaction with $\rm v_{ind}$ takes place in a timescale of $\sim O(10^0-10^1)ms$ from the shock incidence, both of which are denoted using green and yellow background respectively in Fig. \ref{fig5:IRO:Ms<1.06}b. After the droplet ignition, the initial preheating time is observed for around $\sim 200ms$ of time period after which quasi-steady droplet combustion is achieved. During this phase, the droplet regression follows $d^2$ law where the normalized square of droplet diameter $(d/d_o)^2$ decreases linearly with time ($t_d$). In current experiments, the different shock delays used are maintained to be significantly higher than the preheating time, to ensure the shock interaction with droplet flame occurs during the quasi-steady combustion regime. Thus, Fig. \ref{fig5:IRO:Ms<1.06}b shows the schematic of the initial unforced droplet regression i.e, linearly decreasing $(d/d_o)^2$ on the left side of the vertical green line (shock incidence), which gets altered after interaction with the shock. For low Mach number, $M_s \sim 1.025$, the original slope ($|m_1|$) of the droplet regression $(d/d_o)^2$ vs $t_d$ gets altered due to the interaction with the shock-flow to a newer value of slope from $|m_1|$ to $|m_2|$. The two slopes before and after the interaction with the shock-flow (i.e., $|m_1|$ and $|m_2|$) are depicted in the droplet regression schematic $(d/d_o)^2$ vs $t_d$ in Fig. \ref{fig5:IRO:Ms<1.06}b, using blue and black dotted lines respectively.

As different fuels are used in the current experiments, to eliminate the effect of initial droplet size ($d_o$) variation on the droplet lifetime, the time of droplet regression is normalized using the square of initial droplet diameter ($d_o$), to obtain normalized time, $t_d/(d_o)^2$. Hence, the normalized slope of droplet regression ($|\tilde{m}|$) is evaluated for different cases by plotting droplet regression as $(d/d_o)^2$ vs Normalized time [$t_d/(d_o)^2$]. Thus, the values of $|{\tilde{m}}_1 |$ and $|{\tilde{m}}_2 |$ (normalized slopes before and after interaction) for all the cases are plotted in the form of a box and whisker plot, to compare the alteration of the unforced droplet regression due to interaction with the shock-flow at low Mach number, $M_s \sim 1.025$, for different fuel types in Fig. \ref{fig5:IRO:Ms<1.06}a.

Fig. \ref{fig5:IRO:Ms<1.06}c show the droplet regression plots for different cases (based on fuel type). In these droplet regression plots i.e., $(d/d_o)^2$ vs $t_d$, the blue and black inclined dotted lines represent the droplet regression slope before $\&$ after the interaction with the shock-flow. The normalized flame heat release rates ($\dot{Q}/\dot{Q}_o$) are plotted in Fig. \ref{fig5:IRO:Ms<1.06}d corresponding to the different cases shown in the droplet regression plots. The red horizontal dotted line indicated in $\dot{Q}/\dot{Q}_o$ (Normalized heat release rate) vs $t_d$ plots represent the unity value of normalized HRR i.e., $\dot{Q}/\dot{Q}_o=1$. 

Interestingly, the low Mach number ($M_s \sim 1.025$) blast wave interaction results in the enhancement in the droplet regression rate, i.e., $|m_1| < |m_2|$. It is to be noted that the time scale of both the blast wave velocity ($\rm v_s$) interaction and induced flow ($\rm v_{ind}$) interaction $\sim O(10^{-1}-10^1)ms$ is insignificant compared to the overall droplet lift time $\sim O(10^3)ms$, which is also shown in Fig. \ref{fig5:IRO:Ms<1.06}b. Thus, the effect of the interaction is minimal on the overall droplet combustion characteristics. However, the interaction with the shock-flow resulted in the enhancement of the droplet regression rate (slope) for different cases, which can be attributed to the enhancement in the fuel vaporization and the vapor transport from the droplet to the flame during the interaction. The externally imposed flow velocity scale ($\rm v_{ind}$) corresponding to the $M_s \sim 1.025$ responsible for this is of the order $\sim 1m/s$ (obtained based on flame-base advection rate), whose Reynolds number ($Re)$ is in the range of $Re < 150$. Sharma et al. \cite{sharma2022evaporation} showed that the evaporation rate is enhanced under the externally imposed vortex for the Reynolds number range, which resulted in significant reduction in droplet lifetime. Thus, during the interaction of the shock-flow, the fuel vaporization at the droplet surface gets enhanced because of the imposed induced flow interaction ($\rm v_{ind}$). This fuel vapor gets transported into the flame as the flame-base lifts off and convects downstream, which results in a brighter flame, i.e., flame heat release rate (HRR) gets enhanced. This is evident from the Fig. \ref{fig5:IRO:Ms<1.06}d that the flame heat release rate (HRR) increased after the interaction with shock flow (above the red dotted line i.e., $\dot{Q}/\dot{Q}_o =1$). Hence, after the interaction of the shock-flow with the droplet flame, the fuel fed into the flame is enhanced, which increases the flame heat release rate (HRR). Therefore, a brighter flame with higher HRR is established after the reattachment of the lifted flame at the droplet post-interaction with the induced flow. This brighter flame with higher HRR results in higher fuel consumption rate, which can be hypothesized to contribute to the higher droplet regression rate after the interaction, i.e., $|m_2| > |m_1|$ for low Mach number case ($M_s \sim 1.025$).  Thus, the increase in the slope of droplet regression after the interaction of the shock-flow in current experiments can be attributed to the positive feedback coupling between HRR and fuel vaporization, where the fuel vaporization rate is increased as a result of the higher HRR post-interaction, which is caused by the initial fuel vaporization enhancement and transport of this fuel vapor towards the flame tip, during to the interaction with the induced flow ($\rm v_{ind}$).


\begin{flushleft}\textbf{Regime-II: Complete Flame Extinction  ($1.06<M_s < 1.1$)}\end{flushleft}

\begin{figure}[b!]
	\centering
	\includegraphics[width=1\textwidth,height=\textheight,keepaspectratio]{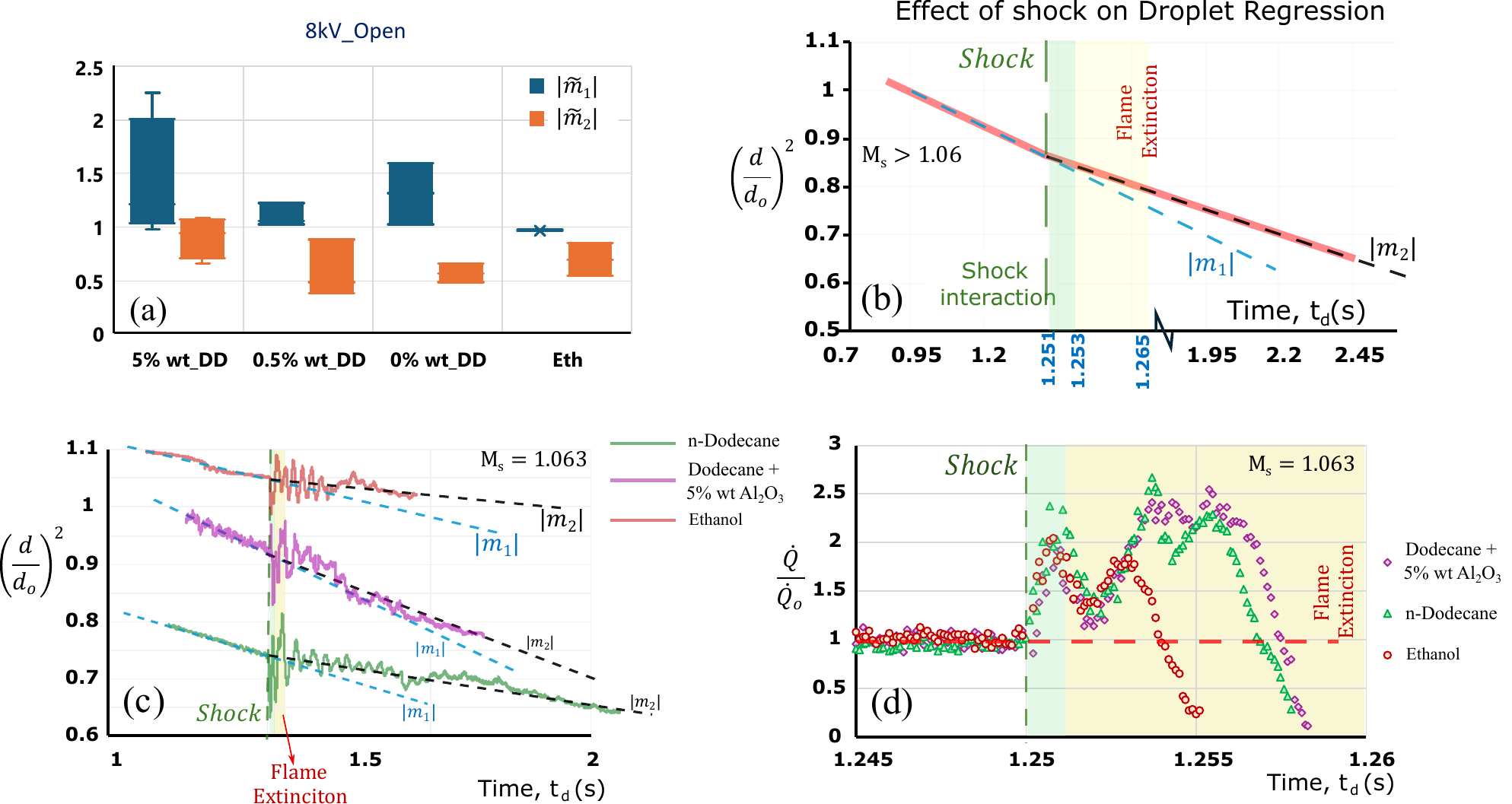}
	\caption{(a) Normalized slope variation ($|m|$) of the droplet regression rate $(d/d_o)^2$ before and after interaction with the shock-flow for $M_s \sim 1.063$ (Regime-II); (b) Schematic of the droplet regression rate alteration due to interaction with the shock-flow for $8kV\_Open$, $M_s \sim 1.063$ (Regime-II)); (c) Droplet regression rate slope variation before and after interaction with the shock-flow plotted against droplet life time ($t_d$) for $5\%$ wt NP + DD, pure Dodecane and pure Ethanol cases with a shock delay of $\sim 1.25s$ for $M_s \sim 1.063$; (d) The normalized flame HHR ($\dot{Q}/\dot{Q}_o$) variation before and after interaction with the shock-flow plotted against droplet life time ($t_d$) for $5\%$ wt NP + DD, pure Dodecane and pure Ethanol cases with a shock delay of $\sim 1.25s$ for $M_s \sim 1.063$.  
		The green vertical dotted line indicates the time instant of shock incidence, blue and black inclined dotted lines represent the droplet regression slope before $\&$ after the interaction with the shock-flow and red horizontal dotted line represents the unity value of normalized HRR. The plots in (c $\&$ d) are offset vertically for better visualization, however all the plots in 'c' begin at $(d/d_o)^2=1$ at $t_d=0$ and all the plots in 'd' begin at $\dot{Q}/\dot{Q}_o = 1$ before shock interaction. The green and yellow backgrounds indicates the interaction with the blastwave-imposed velocity profile ($\rm v_s$) and induced flow ($\rm v_{ind}$) respectively.}
	\label{fig6:IRO:Ms>1.06}
\end{figure}

\par Interestingly, as the Mach number is increased beyond $M_s > 1.06$, the droplet regression trend is observed to deviate from that of the previously observed experiments performed at lower Mach numbers ($M_s \sim 1.025$), that is regime-I. The flame undergoes complete extinction during the induced flow interaction in case of $M_s > 1.06$ (as discussed in the previous section). Fig. \ref{fig6:IRO:Ms>1.06}c and Fig. \ref{fig6:IRO:Ms>1.06}d show the droplet regression trend and the corresponding flame HRR trend for $M_s > 1.06$, for different fuel types: $5\%$ wt + DD, Pure Dodecane and pure Ethanol respectively, for different shock delays. Green vertical dotted lines shown in these plots represent the instant of shock incidence with droplet flame and the green, yellow background represent the two-stage interaction of the droplet flame with the shock flow i.e., with $\rm v_s$ and $\rm v_{ind}$ respectively. Similar to the previous section, the slope of the droplet regression rate before ($|m_1|$, blue dotted line) and after ($|m_2|$, black dotted line) the interaction with the shock flow is also shown in the droplet regression plots in Fig. \ref{fig6:IRO:Ms>1.06}c. The location of flame extinction is also indicated in each of the plots in Fig. \ref{fig6:IRO:Ms>1.06}.

Interestingly, the droplet regression slope is drastically reduced due to the interaction in regime-II, which is shown in the box whisker plot in Fig. \ref{fig6:IRO:Ms>1.06}a, where $|m_2|<|m_1|$ for all the cases. This can be attributed to the complete extinction of the flame that occurs for higher Mach numbers ($M_s > 1.06$) during the interaction with the induced flow ($\rm v_{ind}$). This extinction results in the removal of heat source that is necessary for the continual droplet evaporation, unlike $M_s < 1.06$ (regime-I) where the flame reattaches and acts as a continuous heat source for droplet vaporization even during later stages of droplet lifetime ($t_d> 0.6 t_{d,max}$). On the other hand, the enhancement in droplet vaporization observed in regime-I due to the externally imposed flow can also occur in regime-II. The velocity scale of the induced flow in regime-II ($\rm v_{ind}$) is found to be of the order of $\sim 5-7m/s$ which corresponds to $Re > 500$, which is far greater than the regime-I. However, as shown by Sharma et al. \cite{sharma2022evaporation}, the maximum enhancement in droplet evaporation occurs at $Re \sim 220$ beyond which there is no significant improvement in evaporation with further increase in the externally imposed flow. This suggests that the enhancement fuel vaporization should be similar in both regime-I and regime-II, however, the major contributor to the fuel vaporization i.e., flame as a heat source is absent in regime-II, which can explain the reduction in droplet regression rate for regime-II i.e., $M_s > 1.06$, as shown in Fig. \ref{fig6:IRO:Ms>1.06}a. This can also be observed in the droplet regression plots in Fig. \ref{fig6:IRO:Ms>1.06}c where the actual droplet regression slowly drifts upwards from the initial slope after the interaction of shock flow due to flame extinction. Fig. \ref{fig6:IRO:Ms>1.06}d shows the temporal variation of the normalized flame HRR for regime-II exhibiting flame HRR enhancement for Dodecane, Ethanol and nanofuels, before extinction. Fig. \ref{fig6:IRO:Ms>1.06}d also shows the two stage interaction of droplet flame with $\rm v_{s}$ (green background) and subsequently with $\rm v_{ind}$ (yellow background) which resulted in flame extinction. The plots in Fig. \ref{fig6:IRO:Ms>1.06}d also show that for all the fuels, the flame HRR shows an initial peak in the green region during the interaction with $\rm v_s$ before exhibiting another peak in yellow region ($\rm v_{ind}$ interaction). This flame HRR enhancements during the interaction is found to reach twice the value of base flame HRR (without interaction) for different fuels, which is attributed to fuel vapor accumulation near the flame tip, as discussed in \autoref{sec:HRR_Effect}.

\subsubsection{Droplet Regression model during the interaction ($M_s < 1.1$)}\label{sec_regressionmodel}
\par In addition to the detailed analysis provided based on experimental evidence, the general trend of droplet regression observed for $M_s < 1.06$ and $1.06 < M_s < 1.1$ regimes can also be explained using the droplet evaporation model given by McAllister et al. \cite{mcallister2011fundamentals}. A simplified energy analysis of spherical droplet is given below, showing the balance between the rate of loss of enthalpy through fuel mass evaporation and the net heat flux input to the total surface area of the droplet.
\begin{equation}\label{Eq1: nanofuel} 
    -\frac{d}{dt}\left[\rho_l\ \frac{4}{3}\pi\left(\frac{d}{2}\right)^3h_{fg}\right]\ \sim \pi d^2\ \cdot \ {q^{"}}_s
\end{equation}
Where, $\rho_l$,$d$, $h_{fg}$ and ${q^{"}}_s$ are the liquid fuel density, droplet diameter, latent heat of vaporization and heat flux input to the droplet per unit area. The heat flux towards the droplet (${q^{"}}_s$) can be obtained using heat conduction as
\begin{equation}\label{Eq2: nanofuel} 
    {q^{"}}_s \sim k\ \left.\frac{dT}{dr}\right|_s \sim k\ \frac{T_{amb} - T_d}{\delta}
\end{equation}
Where, $k$ is thermal conductivity and $\delta$ is the thickness of thermal layer surrounding the droplet which depends on the specific conditions, but it is proportional to the characteristic length scale i.e., droplet diameter ($d$), thus, we can write: $\delta \sim C_1d$ (‘$C_1$’ is a constant). Substituting the heat flux scaling in the energy balance equation, we get:
\begin{equation}\label{Eq3: nanofuel} 
    -\frac{d}{dt}\left[\rho_l\ \frac{4}{3}\pi\left(\frac{d}{2}\right)^3h_{fg}\right]\ \sim \pi d^2\ \cdot \ \left\{k\ \frac{T_a - T_d}{C_1d}\right\}
\end{equation}
This equation can be simplified to obtain:
\begin{equation}\label{Eq4: nanofuel} 
    2d\frac{d(d)}{dt}\ \sim -\frac{4k\left(T_a - T_d\right)}{\rho_l h_{fg}C_1}
\end{equation}
Re-writing the above equation gives Eq. \ref{Eq5: nanofuel} the traditional $d^2$ law that provides the temporal droplet diameter variation during droplet evaporation (shown in Eq. \ref{Eq6: nanofuel}).
\begin{equation}\label{Eq5: nanofuel} 
    \frac{d(d^2)}{dt}\ \sim -\beta_o
\end{equation}
    where $\beta_o=\frac{4k\left(T_a - T_d\right)}{\rho_l h_{fg}C_1}$

\begin{equation}\label{Eq6: nanofuel} 
    d^2\ \sim {d_o}^2-\beta_ot
\end{equation}

During the droplet combustion, the droplet regression initially undergoes a preheating phase where the droplet temperature increases continuously due to the heat input from the flame reaching a quasi-steady temperature (wet-bulb temperature, $T_{wb}$), after which the heat input to the droplet only results in phase change at the droplet surface. This quasi-steady stage of droplet evaporation follows $d^2$ law, where the square of the droplet diameter decreases linearly with time. Eq. \ref{Eq5: nanofuel} gives the droplet regression for droplet combustion in a quiescent environment with a spherical flame enveloping the droplet at a specific flame radius. The droplet temperature $T_d$ is considered to be at wet-bulb temperature ($T_{wb}$) and the surrounding temperature causing the evaporation ($T_a$) has to be considered to be the flame temperature ($T_f$). The thermal boundary thickness ($\delta$) can be considered to of the order of flame radius and the constant $C_1$ can be chosen accordingly. 

However, when an external flow is imposed on a burning droplet, the heat input (${q^{"}}_s$) to the droplet is altered due to convective heat transfer. Thus, the ${q^{"}}_s$ can be obtained as follows:

\begin{equation}\label{Eq7: nanofuel} 
     \left.{q^{"}}_s\right|_{conv}\sim \ \bar{h}\ (T_f - T_{wb})
\end{equation}

\begin{equation}\label{Eq8: nanofuel} 
     Nu\ =\ \frac{\bar{h}D}{k}\ =\ 2\ +\ 0.4\ {{\rm Re}_d}^{1/2}\ Pr^{1/3}
\end{equation}
Where, $\bar{h}$ is the convective heat transfer coefficient that can be obtained from Nusselt number correlation for flow over a sphere (shown in Eq. \ref{Eq8: nanofuel}). Thus, rewriting Eq. \ref{Eq1: nanofuel} considering forced convection for heat input to the droplet, we get:
\begin{equation}\label{Eq9: nanofuel} 
     -\frac{d}{dt}\left[\rho_l\ \frac{4}{3}\pi\left(\frac{d}{2}\right)^3 h_{fg}\right]\ \sim \pi d^2\ \dot \ \left\{\bar{h}\ (T_f -T_{wb})\right\}
\end{equation}
Following similar procedure, this equation reduces to 
\begin{equation}\label{Eq10: nanofuel} 
     d^2\sim {d_o}^2-\beta_ot -\beta_1t
\end{equation}
Where, $\beta_1 = \frac{1.6k\ {{Re}_d}^{1/2}\ Pr^{1/3}\ (T_f-T_{wb})}{\rho_l h_{fg}}$

\begin{figure}[t!]
    \centering
    \includegraphics[width=\textwidth,height=\textheight,keepaspectratio]{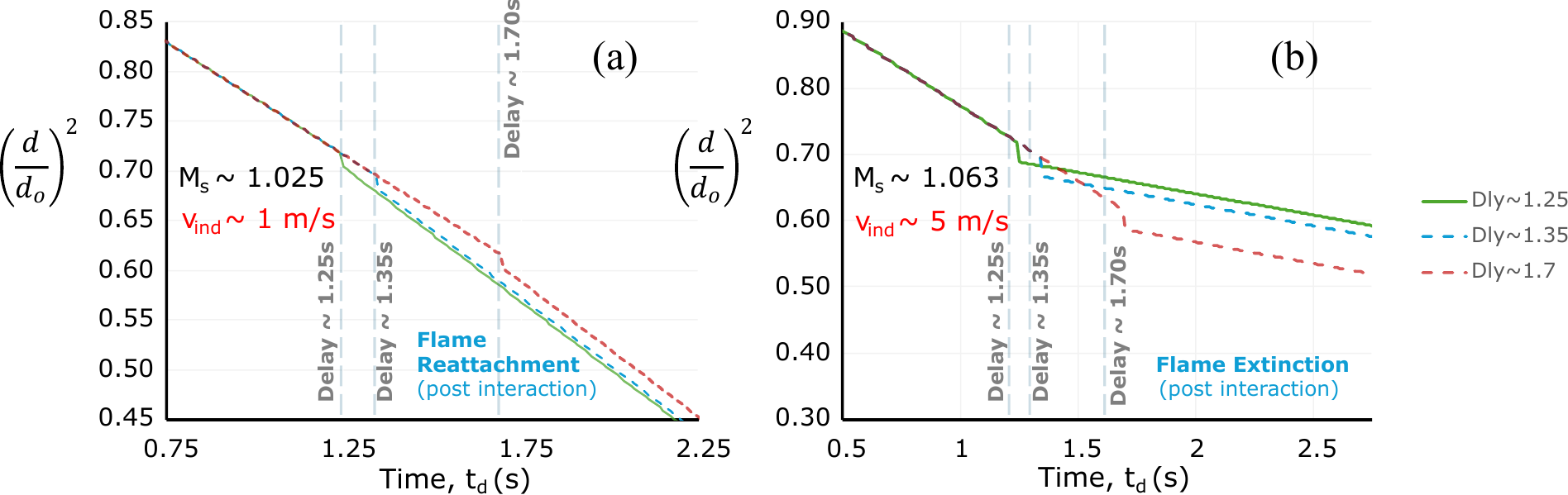}
    \caption{Theoretically obtained plots of droplet regression trends ($(d/d_o)^2$ vs droplet lifetime $t_d$) before and after the interaction with the flow (based on the energy balance scaling at the droplet) for (a) $M_s < 1.06$ (Flame reattachment) and (b) $M_s > 1.06$ (Flame extinction). The three different shock delays (time delay for shock incidence after the droplet ignition) i.e., 1.25s, 1.35s and 1.70s have been plotted with color code shown in the legend. The instance of shock incidence for different shock delays is indicated by vertical grey dotted lines.}
    \label{fig6p5:IRO:Th_DropReg}
\end{figure}
\par Comparing Eq. \ref{Eq6: nanofuel} and Eq. \ref{Eq10: nanofuel} show that the droplet vaporization enhancement occurs (additional term of $\beta_1t$) when external flow is imposed. In current experiments for $M_s < 1.1$, the induced flow ($\rm v_{ind}$) is responsible for the majority of the flame dynamics, and it has been established from the experiments that the entire interaction with the shock – flow for $M_s < 1.1$ occurs in a timescale of $\sim O(10^0-10^1)ms$, Thus, the effect of $\rm v_{ind}$ on droplet evaporation will only occur during the small time period of $\sim O(10^0-10^1)ms$ from the shock incidence, which is significantly smaller than the droplet lifetime $\sim O(10^0)s$. Thus, for current experiments, Eq. \ref{Eq6: nanofuel} is valid for majority of the droplet lifetime ($\sim O(10^0)s$), and Eq. \ref{Eq10: nanofuel} is only applicable during the interaction of shock-flow time period of $\sim O(10^{-2})s$, just after the shock incidence. Before the shock interaction, Eq. \ref{Eq6: nanofuel} is used for all the cases as the droplet combustion is quiescent. During the interaction, Eq. \ref{Eq10: nanofuel} is valid, and $Re_d$ is evaluated using the velocity of order of magnitude of $\rm v_{ind}$ for different cases i.e, $\rm v_{ind} \sim 1 m/s$ (for $M_s \sim 1.025$) and $\rm v_{ind} \sim 5 m/s$ (for $M_s \sim 1.063$). Additionally, in case of $M_s < 1.06$, the flame heat release rate enhancement (see Fig. \ref{fig5:IRO:Ms<1.06} d,f,h) can also be incorporated into the Eq. \ref{Eq6: nanofuel} model post-interaction. While for $M_s > 1.06$, as full flame extinction occurs post-interaction, the flame temperature ($T_f$) can be substituted to be lower value near $T_{wb}$ post-interaction. The droplet regression plots have been plotted in Fig. \ref{fig6p5:IRO:Th_DropReg}, where the $d/d_o$ is plotted against droplet lifetime ($t_d$) for $M_s < 1.06$ and $M_s > 1.06$ regimes similar to Fig. \ref{fig5:IRO:Ms<1.06} and Fig \ref{fig6:IRO:Ms>1.06} respectively. The theoretical plots of droplet regression plotted in Fig. \ref{fig6p5:IRO:Th_DropReg} for different cases also shows similar trend as the experimental observations of droplet regression for $M_s < 1.1$. Since these theoretical plots are obtained based on the order of magnitude analysis of energy balance at the droplet, the scope of this analysis is only limited to validate the trend of the droplet regression for different shock interaction cases. 


\subsubsection{Effect of experimental parameters on droplet and flame dynamics ($M_s<1.1$)}


The $d^2$ law is followed by the droplet regression during the droplet combustion in pure fuels (Dodecane and Ethanol) as shown in Fig. \ref{fig5:IRO:Ms<1.06}c, Fig. \ref{fig6:IRO:Ms>1.06}c and Fig. \ref{fig4p5:IRO:NP_Eth_effect}b. However, with addition of nanoparticles, the droplet regression rate is observed to slow down during the later stages of droplet lifetime ($t_d > 0.6 t_{d,max}$), as shown in Fig. \ref{fig5:IRO:Ms<1.06}c. This can be attributed to the accumulation of the NP aggregates at the droplet surface which hinders the fuel evaporation during unsteady evaporation stage as explained in \autoref{sec:FuelEffect}. Furthermore,the nanoparticle (NP) addition also results in significant oscillations and spikes in the droplet regression plot in Fig. \ref{fig4p5:IRO:NP_Eth_effect}b. These oscillations in droplet regression plots correspond to the droplet shape oscillation due to the phenomena such as internal boiling, bubble growth, bubble ejections, secondary atomization that occurs during the droplet lifetime ($t_d$). This is because of the increased heterogeneous nucleation due to the aggregation and agglomeration of the nanoparticle (NP) inside the droplet. Moreover, as shown in supplementary Fig. S3, high particle loading rates (PLR) of nanoparticles i.e., $5\%$ wt NPs + Dodecane showed minimal droplet oscillations and bubble ejections compared to intermediate NP PLR of ($0.5\%$ wt NPs). This can be attributed to the shell formation by NP aggregates at the droplet surface, due to very high nanoparticle (NP) concentration, resulting in suppression of fuel evaporation rate. This can be reflected in the lower heat release rate (HRR) of the unforced flame (without shock interaction) in case of high PLR ($5\%$wt) compared to intermediate and low PLR ($0.5\%$wt) and pure dodecane (see supplementary Fig. S3). 

The lower heat release rate (HRR) results in lower heat input to the droplet, which results in the reduction in the heat available for the heterogeneous nucleation inside the droplet that result in bubble ejection events. The same can be observed in Fig. \ref{fig5:IRO:Ms<1.06}d and supplementary Fig. S6(b,d), where the bubble ejections correspond to the local peaks in the flame HRR time series. The significant oscillations in droplet regression plot during the final stages in $5\%$ wt case shown in Fig. S6 (a), corresponds to the larger bubble formation whose bubble diameter is comparable to or more than the instantaneous droplet diameter (d). These larger bubble formation occurs because of the elastic behavior of the nanofuel due to higher NP concentration and skeletal structure formation at the surface during the later stages of the droplet lifetime ($t_d > 0.6 t_{d,max}$). Furthermore, the periodic oscillations in droplet diameter regression for Pure Dodecane in Fig. \ref{fig5:IRO:Ms<1.06}c can be attributed to the lower viscosity of the pure Dodecane compared to nanofuels during the later stages of droplet life time due to increased viscosity  because of higher particle concentration. This results in the reduction in the dampening effects of the droplet oscillations due to interaction with the shock-flow in case of pure fuel compared to nanofuels.

In regime-I ($M_s<1.06$), even though the partial extinciton followed by the reattachment of the flame is observed due to the interaction with the shock-flow (see Fig. \ref{fig5:IRO:Ms<1.06}), the general phenomena of droplet dynamics broadly remained same with and without the interaction of the shock-flow (see Fig. \ref{fig4p5:IRO:NP_Eth_effect}b and Fig. \ref{fig5:IRO:Ms<1.06}c). This is majorly because of the timescale of the interaction of shock-flow ($\sim O(10^1)ms$) is far less than the overall timescale of the droplet lifetime, $t_d$ ($\sim O(1)s$). The droplet regression plot showed quiescent regression for pure fuel for both unforced (quiescent burning) and forced (with interaction) cases. Whereas with nanoparticle addition, both the forced and unforced cases showed bubble ejections and droplet shape deformations which correspond to the oscillations in the droplet regression plots shown in Fig. \ref{fig4p5:IRO:NP_Eth_effect}b and Fig. \ref{fig5:IRO:Ms<1.06}c. Thus, effect of the nanoparticle addition affected the droplet regression in similar way with and without the external forcing imposed due to the interaction with shock-flow.

However, as discussed in the analysis about the regime-I, even though the broad trend observed in the droplet regression plots are similar with and without the interaction of the shock-flow, the droplet regression rate is observed to alter due to the interaction with shock-flow (see Fig. \ref{fig5:IRO:Ms<1.06}c). It has been established in \autoref{sec_regressionmodel}, that this alteration in droplet regression rate (slope) after the interaction with the shock-flow is due to the combined effect of the fuel evaporation enhancement due to convective effects (because of imposed flow) as well as the increase in the heat input from the flame post-interaction. The flame heat release rate enhancement post-interaction is in turn caused by the higher fuel vapor availability at the flame due to the fuel evaporation enhancement due to convective effects. On the other hand, in regime-II, where full extinction of the droplet flame is observed due to higher velocity scales of the imposed flow, the droplet regression rate reduces post-interaction, as discussed previously. Interestingly, due to the absence of the flame post-interaction, the droplet dynamics for regime-II (see Fig. \ref{fig6:IRO:Ms>1.06}c) are found to be completely altered compared to the unforced cases (see Fig. \ref{fig4p5:IRO:NP_Eth_effect}b) for different fuels. The alteration in droplet regression plot behavior post-interaction in regime-II is primarily because of two factors. The absence of the heat source which reduces results in the drastic reduction in the droplet evaporation rate. Secondly, the sudden imposition of higher velocity flow ($\rm v_{ind} \sim 5 - 7 m/s$) corresponding to  Weber numbers of order $We \sim 5$ induces significant droplet shape oscillations along with slight enhancement in the droplet evaporation rate due to convective effects (as discussed in \autoref{sec_regressionmodel}). 

On the other hand, in case of nanofuels, the full extinction of the flame post-interaction resulted in reduction and suppression of heterogeneous nucleation inside the droplet due to the lack of sufficient heat flux, i.e., $\dot{Q}_{input} \sim 4/3\pi \rho_l (r_{bubble}^3 - r_{agg}^3)h_{fg} \Delta t$. The excess heat flux '$\dot{Q}_{input}$' from the flame is responsible for the internal boiling, i.e, phase change at the NP aggregates inside the droplet that results in the formation of a vapor bubble within a residence time of $\Delta t$, during nanofuel droplet combustion (without flame extinction). However, as discussed before, in regime-II the flame extinction itself occurs at significantly faster timescales of $\sim O(10^0)ms$ compared to droplet lifetime ($t_d$), however the extinction occurs at relatively slower timescales in comparison to the blast wave propagation ($\sim O(10^{-1})ms$). As shown in Fig. \ref{fig6:IRO:Ms>1.06}d, the flame extinction in regime-II occurs in two stages: initial interaction with $\rm v_s$ (green background) and subsequent interaction with $\rm v_{ind}$ (yellow background). As discussed before in \autoref{sec:HRR_Effect}, the flame initially responds to the blast wave velocity profile ($\rm v_s$) exhibiting a momentary flame stretching, followed by a continuous advection of the flame-base downstream due to the subsequent induced flow interaction ($\rm v_{ind}$). The droplet flame undergoes momentary stretching during the interaction with $\rm v_s$ (green), and subsequently undergoes flame shedding due to the interaction of $\rm v_{ind}$. Corresponding to these two stages, the normalized flame HRR shows momentary peaks in green and yellow region indicating the aforementioned HRR enhancement in regime-II, as shown in Fig. \ref{fig6:IRO:Ms>1.06}d before the imminent extinction. The heat release rate (HRR) enhancement of roughly twice the base HRR of unforced flame is observed for all the fuels. Furthermore, it can also be concluded that the HRR enhancement during the interaction with $\rm v_{ind}$ is more prominent in regime-II as it lasts for longer timescales (yellow region, in Fig. \ref{fig6:IRO:Ms>1.06}d) compared to $\rm v_s$ interaction (green region). This momentary HRR enhancement is hypothesized to be due to the fuel vapor accumulation at the flame location as the flame base is advected towards the flame tip due to the externally imposed flow.

However, it can be also be observed from Fig. \ref{fig6:IRO:Ms>1.06}d that the extent of HRR enhancement is higher for Dodecane based fuels (both pure and nanofuels) compared to pure Ethanol. This can be attributed to the lower fuel vapor availability in case of Ethanol compared to Dodecane. The lower fuel vapor availability of Ethanol corresponds to lower droplet regression rate in case of Ethanol (see Fig. \ref{fig4p5:IRO:NP_Eth_effect}c) due to lower heat input from the flame (see supplementary Fig. S3) compared to Dodecane. This also can explain the faster extinction of the Ethanol flame (shorter timescales) compared to Dodecane based fuels, as shown in Fig. \ref{fig6:IRO:Ms>1.06}d. On the other hand, as shown in Fig. \ref{fig6:IRO:Ms>1.06}d, the nanoparticle addition to Dodecane showed minimal variation in the flame HRR enhancement during interaction ($\dot{Q} / \dot{Q}_{o,peak} \sim 2.3$ for different nanofuels), as the base fuel vapor is same in both cases (n-Dodecane) with similar order of fuel availability due to similar droplet regression slope for different PLR of nanofuels (see Fig. \ref{fig5:IRO:Ms<1.06}a and Fig. \ref{fig6:IRO:Ms>1.06}a). However, there is another experimental parameter of shock delay (the delay after which shock is incident on droplet flame from droplet ignition, $t_d =0$) that can alter this behavior. 

The fuel vapor availability at the instant of interaction of shock-flow may also vary due to the stochastic nature of the bubble ejection events which momentarily may increase the fuel vapor availability at different stages of droplet lifetime ($t_d$). This argument is supported by the relatively higher slope of droplet regression observed in unforced nanofuel droplet combustion compared to pure Dodecane especially during the initial stages of droplet lifetime ($t_d$). This is evident from the droplet oscillations and bubble ejection events present throughout the droplet lifetime, as shown in Fig. \ref{fig4p5:IRO:NP_Eth_effect}, which is also reflected in Fig. \ref{fig5:IRO:Ms<1.06}a and Fig. \ref{fig6:IRO:Ms>1.06}a as well. However, this enhancement of net fuel vaporization rate is only observed during initial stages of droplet regression ($t_d \sim (0-0.6)t_{d,max}$), after which the effects of porous skeletal structure formation become dominant at higher PLR which hinders the fuel evaporation rate at the droplet surface (as discussed before in 
\autoref{sec:FuelEffect}). 

Thus, the extent of flame HRR enhancement is observed to be marginally higher for nanofuels when compared to pure Dodecane when the interaction of shock-flow occurs during earlier stages of droplet lifetime ($t_d < 0.6 t_{d,max}$), as shown in Fig. \ref{fig6:IRO:Ms>1.06}d. On the contrary, the HRR enhancement is marginally lower compared in case of nanofuels compared to pure Dodecane when the interaction of shock-flow occurs during later stages of droplet lifetime ($t_d > 0.6 t_{d,max}$), i.e., shock delay of $t_d \sim 1.70s$. This can be seen in the normalized HRR plots presented in supplementary Fig. S7 (b,e,h), where nanofuels and pure fuels showed higher peak value ($\dot{Q}/\dot{Q}_{o,peak,Dly \sim 1.3s} \sim 2.3$) for lower shock delay of $t_d \sim 1.3$ (Fig. S7 b,h) compared to the peak value (Fig. S7 e) obtained for higher shock delay ($\dot{Q}/\dot{Q}_{o,peak,Dly \sim 1.7s} \sim 1.5$). It is to be noted that the last stage of droplet lifetime ($t_d \sim 0.9t_{d,max}$) where shell formation occurs in nanofuel droplet combustion is out of scope of current study as the droplet size becomes comparable to the quartz filament holding it. Hence, the effect of bubble ejections during this final stages of nanofuel droplet combustion after shell formation, as shown in Fig. \ref{fig4p5:IRO:NP_Eth_effect}b is also out of scope of current experiments. Thus, the current experiments were only performed with shock delays in the range of $t_d < 0.7 t_{d,max}$, where the instantaneous droplet size is considerably higher than the quart filament holding it.

So far, the effect of different parameters such as shock delay, PLR of NP addition to the fuel and fuel type on the response of the droplet flame to the shock-flow (at a given $M_s$) has been discussed in detail. The general trend of droplet and flame response to the imposed flow is observed to remain similar for different PLRs of NPs as well as for both Dodecane and Ethanol. Ethanol showed relatively lower HRR enhancement of the flame during the interaction compared to Dodecane and the NP addition showed marginally higher flame HRR enhancement. Interestingly, the effect of shock delay is only found to be noticeable in case of nanofuels. The extent of the enhancement of the flame HRR during interaction is observed to be marginally higher for nanofuels compared to pure Dodecane when the interaction of shock-flow occurs during the earlier stages of droplet lifetime, i.e., lower shock delay ($t_d < 0.6t_{d,max}$). On the other hand, the extent of HRR enhancement is found to be marginally lower for nanofuels compared to pure Dodecane, when the blast is incident on the droplet at a later point of droplet lifetime, i.e., $t_d \sim 0.7t_{d,max}$ (higher shock delay). However, these differences due to shock delay are marginal and don't significantly alter the flame HRR trends. Interestingly, even though all these parameters influenced the droplet flame response during the interaction, all the cases were found to categorically follow the same set of regimes irrespective of the base fuel, nanoparticle PLR and shock delay. 

The absolute values of droplet regression rate, flame HRR, etc during the flame and droplet response may vary when the experimental parameters are varied, however, the effect of these parameters is not as significant to completely alter the general behavior of the droplet flame during the interaction. This can be attributed to the timescale of the interaction of the shock-flow ($\sim O(10^0-10^1)ms$) being significantly lower compared to the droplet lifetime for different fuels ($t_d \sim O(1)s$). Hence, even though, the fuel type, NP PLR, shock delay can vary the local fuel vaporization rate, the shock interaction is a significantly high-speed process. Therefore, the interaction of the shock-flow with the droplet flame mainly imposes significantly high velocity scales ($\gg \rm v_{NC}$) involving timescales, and the ratio of $\rm v_{imposed}/\rm v_{NC}$ primarily dictates the different flame regimes irrespective of the local variations in fuel vaporization due to the fuel type, NP PLR and shock delay. Furthermore, the high convective effects due to high velocity scales during the interaction in current experiments also masks the minor variations in the droplet flame behavior in terms of HRR and droplet regression rate.

\subsubsection{Mechanisms involved in Flame Extinction in different regimes}

\begin{figure}[t!]
	\centering
	\includegraphics[width=\textwidth,height=\textheight,keepaspectratio]{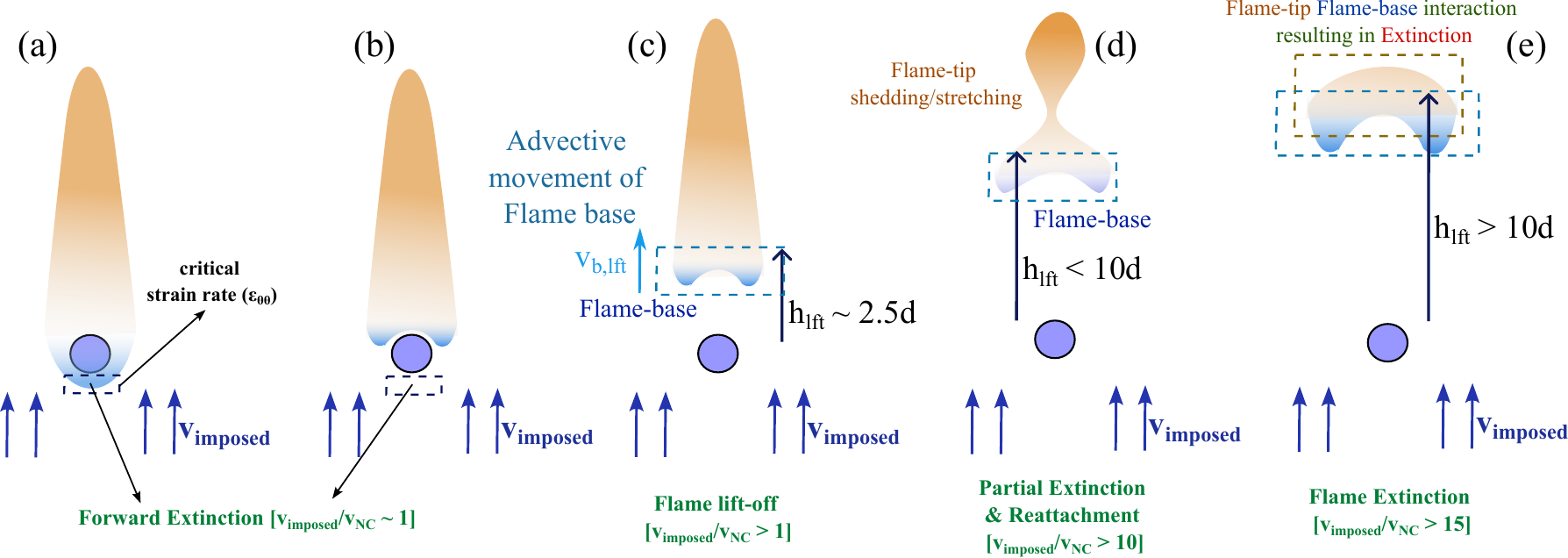}
	\caption{(a-e) Schematics of different phenomena observed during the droplet flame interaction with shock-flow: Forward Extinction, Flame lift-off, Partial Extinction and Reattachment, Complete Flame extinction. The velocity scales of the imposed flow ($\rm v_{imposed}$) at which different phenomena occur and the corresponding flame liftoff heights associated with each of these behaviors are also indicated in the figure.}
	\label{fig9p5:Flame_Ext_phenomena}
\end{figure}

The flame extinction and reattachment behavior in current experiments can be broadly divided into three regimes: 1. Flame extinction and reattachment (for $M_s < 1.06$ regime, due to $\rm v_{ind}\sim 1m/s$), 2. Complete flame extinction (for $M_s > 1.06$ regime, due to $\rm v_{ind}\sim 5m/s$), 3. High-speed flame extinction (for $M_s > 1.1$ regime, due to $\rm v_s \sim 50-250m/s$). The flame extinction in all these three regimes is observed to involve four phenomena: a) Forward extinction of droplet flame, b) Flame lift-off, c) Flame reattachment, d) Flame extinction or blowout. Fig. \ref{fig9p5:Flame_Ext_phenomena} shows these different phenomena observed as the droplet flame responds to the externally imposed transient flow for different velocity scales ($\rm v_{imposed}$). 

Among these phenomena, the most basic event that occurs during the interaction of a droplet flame with an external flow is 'Forward Extinction'. In droplet combustion in quiescent environment, the droplet flame is in fully enveloped configuration, however, as external flow is imposed, local extinction occurs at the forward stagnation point of the enveloped flame. This occurs when the local strain rate at the forward stagnation point reaches a critical value, as shown in Fig. \ref{fig9p5:Flame_Ext_phenomena}a and this phenomenon is termed as 'Forward Extinction'. The local strain rate ($\epsilon_{\theta \theta}$) near the forward stagnation point can be obtained using the following relation (using cylindrical coordinate system) \cite{pandey_self-tuning_2020}: 
\begin{equation}\label{Eq0ex1_: imposed flow}
	\epsilon_{\theta \theta} = \frac{u_r}{r}+\frac{\partial u_{\theta}}{r\partial \theta}
\end{equation}

where, $u_r$ is radial velocity component, $u_{\theta}$ is the tangential velocity component. The stream-function ($\psi$) Onseen's solution for flow over the sphere can be used to evaluate the velocity components to determine the critical value of $\epsilon_{\theta \theta}$ corresponding to a velocity scale of $\rm v_{imposed}$ \cite{pandey_self-tuning_2020}. Since, the primary velocity scale associated with a buoyant droplet diffusion flame is buoyancy-driven, the natural convection velocity scale associated with the droplet flame is evaluated as $\rm v_{NC} \sim \sqrt{g\beta (T_{flame}-T_{\infty})}$. Thus, the externally imposed velocity ($\rm v_{imposed}$) can be normalized using the natural convection velocity scale, which reflects the underlying dynamics of forced and natural convection in an externally forced buoyant droplet flame. The envelope-to-wake flame transition is observed to occur at Reynolds number, $Re > 5$ \cite{pandey_self-tuning_2020}, which corresponds to the normalized the velocity scale of $\rm v_{imposed}/\rm v_{NC} \sim 1$. This also is in agreement with the partial extinction regimes reported in the literature \cite{pandey_dynamic_2021,thirumalaikumaran2022insight}, which exhibits forward extinction for $\rm v_{imposed}/\rm v_{NC} > 1$. Thus, for $\rm v_{imposed}/\rm v_{NC} \ge 1$, the droplet flame undergoes forward extinction and stabilizes at the droplet wake, as shown in Fig. \ref{fig9p5:Flame_Ext_phenomena}a,b. The velocity scales imposed by the shock-flow ($\rm v_s$ and $\rm v_{ind}$) in current experiments is found to be always greater than $\rm v_{NC}$, thus, forward extinction is observed in all the cases. This is also evident from Fig. \ref{fig3:IRO:CH_Timeseries}a (at $t\sim 1.15ms$), where flame has undergone forward extinction at the forward stagnation point, even for the lowest $M_s$ case in current experiments.

Furthermore, when $\rm v_{imposed}$ is increased further, the flame in the droplet wake lifts off towards downstream, as shown in Fig. \ref{fig9p5:Flame_Ext_phenomena}c. The liftoff height of the droplet flame ($h_{lft}$) increases with increase in $\rm v_{imposed}$ as the flame base is swept downstream due to the imposed flow (see Fig. \ref{fig9p5:Flame_Ext_phenomena}c,d). Along with liftoff, the droplet flame exhibits shedding behavior due to buoyancy-induced instability at the flame for $\rm v_{imposed}/\rm v_{NC}> 10$, corresponding to regime-I in current experiments ($M_s<1.06$, with $\rm v_{imposed} = \rm v_{ind} \sim 1-3m/s$). This regime also shows flame reattachment at the droplet, after the initial liftoff, as shown in Fig. \ref{fig9p5:Flame_Ext_phenomena}(c,d). The vortex-droplet flame interaction experiments by Thirumalaikumaran et al. \cite{thirumalaikumaran2022insight} also showed that the partial extinction and reattachment regime occurs for  $\rm v_{vortex}/\rm v_{NC} > 8$, where flame liftoff becomes significantly higher ($\sim (2-10)d$), before the reattachment. The bright yellow tip was reported to disappear during shedding and subsequently the flame base recedes towards the droplet leading to enveloped flame after reattachment. This flame behavior reported in the literature is in very good agreement with the current experiments as shown in Fig. \ref{fig2:IRO:Timeseries}a and Fig. \ref{fig3:IRO:CH_Timeseries}a.

On the other hand, in regime-II, the flame-base lifts off significantly reaching the flame tip due to the imposed velocity ($\rm v_{imposed} = \rm v_{ind} \sim 5 m/s$), resulting in complete extinction of the flame due to the interaction of flame base and flame tip, as shown in Fig. \ref{fig2:IRO:Timeseries}b and Fig. \ref{fig9p5:Flame_Ext_phenomena}(e). Based on the experimental observations, a flame extinction criteria had been proposed in our previous experiments on shock-droplet flame interaction \cite{vadlamudi2024insights}. This implies that the flame extinction occurs when the flame base reaches and interacts with the flame tip during the liftoff, as shown in Fig. \ref{fig9p5:Flame_Ext_phenomena}e. According to the criteria, if the distance traveled by the flame-base ($h_{lft}$) during the interaction is greater than the flame tip distance, then flame extinction is possible. Based on this criteria, it has been reported that flame reattachment is only possible for low $M_s$, i.e., regime-I, which is in agreement with current experiments.


It has been reported in the literature \cite{thevenin1998extinction,thirumalaikumaran2022insight} that complete extinction of the flame is observed when the imposed flow velocity is around $3m/s$, which roughly corresponds to $\rm v_{imposed}/\rm v_{NC}> 15$. This is in good agreement with the velocity scales associated with the interaction of shock-flow in regime-II ($M_s>1.06$) of current experiments. The flame liftoff is significantly higher ($> 10d$) in this regime-II during the interaction with shock-flow. Thus, the flame extinction behavior during the interaction with a highly transient flow imposed by blast wave in current experiments showed similar fundamental characteristics as that of different kinds of experiments in the literature such as flame interaction with acoustics, coaxial vortex, continuously varying flow, etc \cite{pandey_dynamic_2021,thirumalaikumaran2022insight,pandey_self-tuning_2020}. Thus, it can be concluded the fundamental mechanism of the flame dynamics such as forward extinction, flame shedding, partial and complete extinction is same in all the different interaction settings. Hence, all the experiments showed the same regimes of the droplet flame response behavior even though the nature of the imposed flow itself is drastically changed from experiment-to-experiment. 

Thus, it has been established that complete extinction occurs for higher Mach numbers of $M_s>1.06$ in current experiments. However, due to the nature of the flow imposed by the shock generator in current experiments, the flame extinction is observed to occur at different rates for $1.06<M_s<1.1$ regime (regime-II) and the high Mach number regime ($M_s>1.1$). The flame extinction occurs almost one order faster for $M_s>1.1$ regime compared to regime-II ($1.06<M_s<1.1$). This will be discussed in detail in the following sections.


\subsection{Droplet flame response: High Mach number regime ($M_s > 1.1$)}

As the Mach number is increased beyond $M_s > 1.1$, the flame dynamics alter significantly as shown in Fig. \ref{fig2:IRO:Timeseries}c,d. At high Mach numbers, the droplet flame is observed to fully extinguish during the initial interaction with $\rm v_s$ (green region). Fig. \ref{fig7:IRO:Ms>1.1} shows the normalized heat release rate (HRR) variation of the droplet flame during this interaction, where, the red vertical solid line indicates the instant at which shock wave begins to interact with the droplet flame, the violet vertical dotted line indicates the instant where the flame HRR peak occurs and the time period of the interaction  with $\rm v_s$ is represented by green background. 

\subsubsection{Momentary Rapid increment in flame Heat release rate}

\begin{figure}[b!]
	\centering
	\includegraphics[width=1\textwidth,height=\textheight,keepaspectratio]{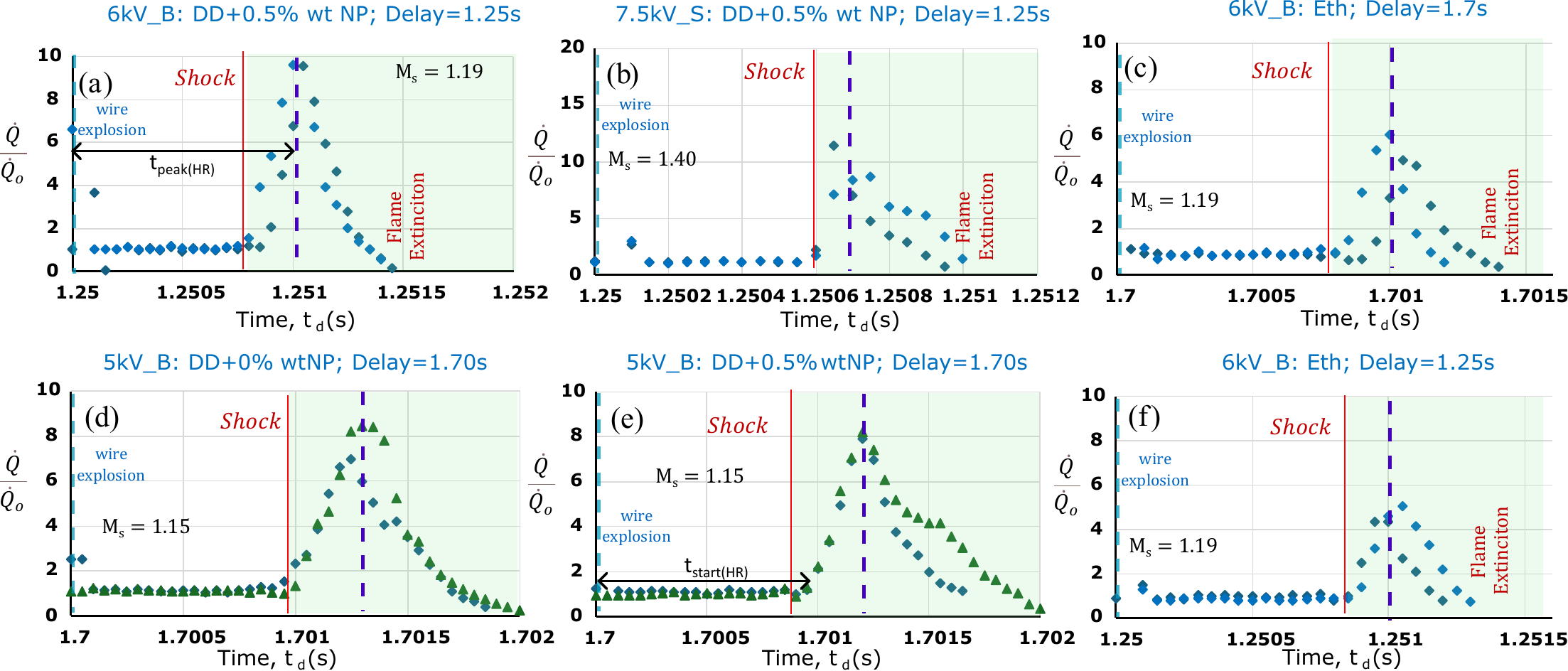}
	\caption{(a-f)The normalized flame HHR ($\dot{Q}/\dot{Q}_o$) variation before and after interaction with the shock-flow plotted against droplet life time ($t_d$) for (a) $6kV\_B$: $0.5\%$ wt NP + DD, shock delay $\sim 1.25s$, (b) $7.5kV\_S$: $0.5\%$ wt NP + DD, shock delay $\sim 1.25s$, (c) $6kV\_B$: Eth, shock delay $\sim 1.70s$, (d) $5kV\_B$: $0\%$ wt NP + DD, shock delay $\sim 1.70s$, (e) $5kV\_B$: $0.5\%$ wt NP + DD, shock delay $\sim 1.70s$, (f) $6kV\_B$: Eth, shock delay $\sim 1.25s$. The blue vertical dotted line indicate the instant of wire-explosion, red vertical solid line indicates the time instant of shock incidence, vertical violet dotted line represents the instant at which the flame HRR peaks and the Green background indicates the initial blastwave-imposed velocity profile ($\rm v_s$) interaction. The data from two different experimental runs were plotted in each of the plots in subfigures(a-f).}
	\label{fig7:IRO:Ms>1.1}
\end{figure}

The flame HRR is observed to increase as the flame interacts with $\rm v_s$, which is evident from the peak in the plot shown in Fig. \ref{fig7:IRO:Ms>1.1} for $\tau < 1$. The violet vertical dotted line are used to indicate the instant where the flame HRR peak occurs in Fig. \ref{fig7:IRO:Ms>1.1}, which correspond to the red circles used in Fig. \ref{fig4:IRO:HRR_Global}c,d that indicate the momentary HRR peak that occurs just before extinction. Similar to lower Mach number ($1.06 < M_s < 1.1$) the flame undergoes continuous liftoff away from the droplet during the interaction with $\rm v_s$ for $M_s > 1.1$. This flame-base advection towards the flame tip results in flame extinction when flame base interacts with flame tip. As discussed before, during this process, the flame HRR undergoes significant enhancement before the imminent flame extinction. However, all these flame dynamics occur during the initial interaction with $\rm v_s$ for $M_s > 1.1$ case ($\tau<1$), which is the main deviation from the flame dynamics observed in $M_s < 1.1$ regimes. It has been established in previous experiments \cite{vadlamudi2024insights} that the flame base advection during the interaction for $M_s > 1.1$ occurs as the direct consequence of the velocity imposed by the propagating blast wave at the droplet location ($\rm v_s$), as shown in supplementary Fig. S5, as discussed at the end of \autopageref{v_Th} in \autoref{sec:Global}. Thus, due to the higher velocity scales of $\rm v_s$ ($\sim 50-250 m/s$), this whole interaction involving flame liftoff and extinction occurs at faster timescales ($\sim O(10^{-1})ms$) for $M_s>1.1$ regime. Furthermore, the timescale of the flame extinction process for $M_s>1.1$ regime is observed to decrease with increase in Mach number ($M_s$).

Fig. \ref{fig7:IRO:Ms>1.1}a-f show the temporal variation of the flame HRR for different fuels i.e., pure Dodecane, Ethanol and Alumina NP + Dodecane nanofuels, undergoing interaction with blast wave at different Mach numbers and different delays for wire explosion from ignition (shock delay). The normalized flame heat release rate plots in Fig. \ref{fig7:IRO:Ms>1.1} are plotted with respect to the droplet lifetime (td). Fig. \ref{fig7:IRO:Ms>1.1}a shows the flame HRR variation for $0.5\%$ wt NP + DD nanofuel, during the flame interaction with a blast wave ($M_s \sim 1.19$) with a shock delay set to be $t_d \sim 1.25s$ (from the droplet ignition). The shock interaction was observed to begin at $t_d \sim 1.25075s$ and the peak of flame HRR enhancement is reached at $t_d \sim 1.251s$ (i.e., $0.25ms$ from shock incidence) and flame extinction occurs at $t_d \sim 1.2515s$ (i.e., $0.40ms$ from shock incidence). Using the same nanofuel, and maintaining the same shock delay, the shock Mach number is increased to $M_s \sim 1.26$ and the resulting flame HRR variation obtained is plotted in Fig. \ref{fig7:IRO:Ms>1.1}b. The peak of flame HRR is found to occur after $0.20ms$ from shock incidence (red solid vertical line) and the flame extinction is observed to occur after $0.50ms$ from the vertical red line (shock incidence). It can be observed from Fig. \ref{fig7:IRO:Ms>1.1}a,b that increasing the Mach number ($M_s$) at the same shock delay and for the same fuel, reduced the time scale of the flame dynamics, yet the characteristics of flame dynamics remained consistent. 

Fig. \ref{fig7:IRO:Ms>1.1}c,f show the plots of flame HRR variation obtained at same Mach number ($M_s \sim 1.19$), and and for the same fuel (Ethanol) but at different shock delays. It is evident form the plots that in both the cases the time scale of the HRR peak and the flame extinction remained similar ($\sim 0.25ms$ and $\sim 0.40ms$ respectively) even when the shock delay is varied. This shows that the underlying phenomena responsible for the HRR enhancement is primarily flow driven (i.e., decaying velocity profile due to blast wave, $\rm v_s$). Furthermore, Fig. \ref{fig7:IRO:Ms>1.1}d,e show the effect of the nanoparticle particle loading where, NP concentration is increased from $0\%$ (by weight) in Fig. \ref{fig7:IRO:Ms>1.1}d to $0.5\%$ (by weight) in Fig. \ref{fig7:IRO:Ms>1.1}e, and the time scales are observed to remain unchanged with variation in NP addition (violet dotted vertical lines). This is in agreement with the previous observation from Fig. \ref{fig7:IRO:Ms>1.1}c,f. In all these cases, the flame HRR enhancement is observed to be significantly higher ($> 8$ times the unforced HRR for dodecane and nanofuels) compared to the enhancement that was observed in lower Mach number regime i.e., $1.06 < M_s < 1.1$. The major difference between these two regimes $M_s > 1.1$ and $1.06 < M_s < 1.1$ is that the flame extinction occurs only due to $\rm v_s$ with significantly higher magnitude ($30 m/s < \rm v_s < 250 m/s$) compared to the velocity scales of $\rm v_{ind}$ ($2 m/s < \rm v_{ind} < 10 m/s$) responsible for flame extinction in case of $1.06 < M_s < 1.1$ regime. This significant difference in the velocity scale of imposed flow in each regime influences the fuel vapor transport and accumulation rate at the high temperature zone (flame tip). Thus, the velocity scale responsible for the advection or transport of the fuel vapor towards the flame tip for $M_s>1.1$ regime is almost one order higher than $M_s<1.1$ regime, which drastically increases the fuel accumulation rate at the flame tip (high temperature zone). It can be hypothesized that the fuel vapor thus accumulated near the flame tip auto-ignites spontaneously which is responsible for the drastic increase in the flame heat release rate (HRR).

\subsubsection{Different stages of Flame interaction with blast wave velocity profile ($\rm v_s$)}

\begin{figure}[t!]
    \centering
    \includegraphics[width=1\textwidth,height=1.1\textheight,keepaspectratio]{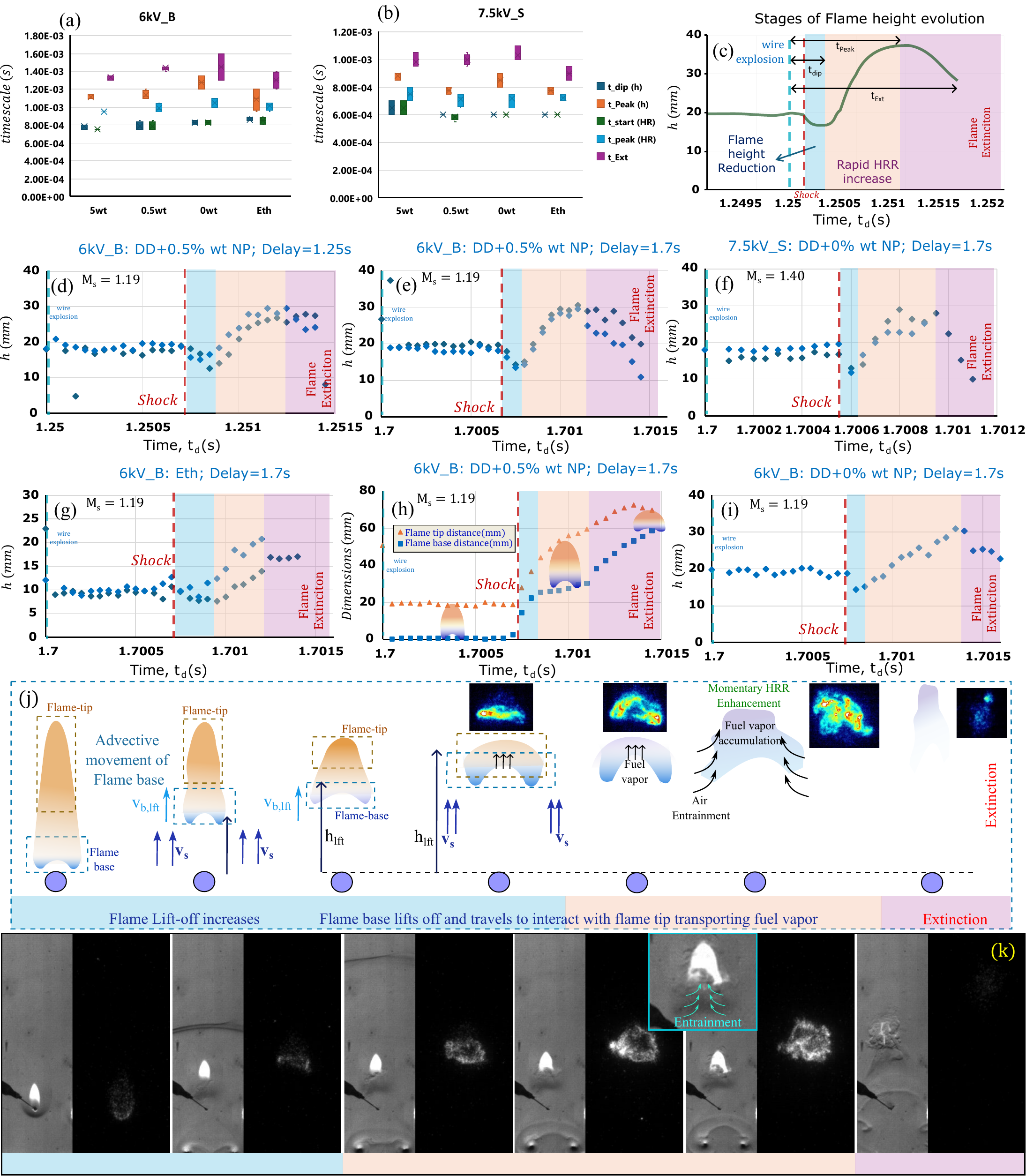}
    \caption{(a,b) Box and whisker plots of different time scales involved in droplet flame extinction for $M_s > 1.1$ for $6kV\_B$ ($M_s \sim 1.19$) and $7.5kV\_S$ ($M_s \sim 1.40$) respectively. (c) Schematic of the temporal variation of the instantaneous flame height, '$h$' (flame tip-to-base distance) plotted with respect to droplet lifetime ($t_d$), showing different stages involved in high-speed droplet flame extinction. (d,e,f,g,i) Temporal variation of the instantaneous flame height (‘$h$’ in mm) during shock droplet interaction for  (d) $6kV_B$, $0.5\%$wt NP + DD (Delay $\sim 1.25s$), (e) $6kV\_B$, $0.5\%$wt NP + DD (Delay $\sim 1.70s$), (f) $7.5kV\_S$, $0\%$wt NP + DD (Delay $\sim 1.70s$), (g) $6kV\_B$, Eth (Delay $\sim 1.70s$), (i) $6kV\_B$, $0\%$wt NP + DD (Delay $\sim 1.70s$). (h) Plot of temporal variation of the flame dimensions i.e., distance in ‘mm’ (from droplet) to flame base (blue data points) and flame height (orange data points), corresponding to the HRR plot shown in sub-figure ‘e’ i.e., $6kV\_B$, $0.5\%$wt NP + DD (Delay $\sim 1.70s$). (j) Schematic of different stages of the high-speed flame extinction (observed in case of $M_s > 1.1$), (k) Time series of simultaneous Schlieren and OH* chemiluminescence depicting different stages during high-speed flame extinction observed in current experiments (for $5kV\_B$, $5\%$wt NP + DD case). Data from multiple experimental runs corresponding to were plotted in the each of the plots in sub-figures(d-g) that correspond to different cases.}
    \label{fig8:IRO:height_Ms>1.1}
\end{figure}

To get a better understanding of this high-speed flame extinction phenomena, the temporal variation of instantaneous flame height ($h$) in ‘mm’ (i.e., the instantaneous distance between the flame-tip and flame-base) is evaluated and it is plotted against the droplet lifetime ($t_d$) in Fig. \ref{fig8:IRO:height_Ms>1.1}(d-i) for different cases. Similar to Fig. \ref{fig7:IRO:Ms>1.1}, the blue vertical dotted line indicates the instant of wire-explosion, and the instant of shock arrival at the droplet is indicated using red vertical dotted lines in Fig. \ref{fig8:IRO:height_Ms>1.1}(c-i). The trend of the flame height variation ($h$) is found to follow similar trend in all the cases, and schematic of this trend is shown in Fig. \ref{fig8:IRO:height_Ms>1.1}c. Similar to the flame heat release rate shown in Fig. \ref{fig7:IRO:Ms>1.1}, the variation of the flame height also showed a significant peak during the interaction with $\rm v_s$ (see Fig. \ref{fig8:IRO:height_Ms>1.1}c-i). Based on the observation of the trend, the temporal variation of the flame height ($h$) can be divided into three successive stages as shown in Fig. \ref{fig8:IRO:height_Ms>1.1}c, which are indicated by the blue, orange and pink background respectively. Fig. \ref{fig8:IRO:height_Ms>1.1}h shows the temporal evolution of the flame dimension (flame-base and flame-tip distance from the droplet) for $6kV\_B$ $0.5\%$ wt NP + DD case corresponding to Fig. \ref{fig8:IRO:height_Ms>1.1}e which shows flame HRR variation for the same case. It can be observed in Fig. \ref{fig8:IRO:height_Ms>1.1}h that initially the flame-base lifts off from the droplet rapidly resulting in reduction in the instantaneous flame height ($h$), as shown in Fig. \ref{fig8:IRO:height_Ms>1.1}e (blue region). Blue region is the initial stage during the flame interaction with $\rm v_s$ where the instantaneous flame height decreases momentarily just after shock interaction. Later, the flame height starts to increase subsequently (represented by orange region in Fig. \ref{fig8:IRO:height_Ms>1.1}e,h) reaching a maxima. Following this the flame HRR starts to decrease corresponding to a marginal reduction in flame height before imminent extinction (represented by pink background in Fig. \ref{fig8:IRO:height_Ms>1.1}e,h). 

\par Fig. \ref{fig8:IRO:height_Ms>1.1}a,b show the box and whisker plots for different timescales involved during this process for $6kV\_B$ ($M_s \sim 1.19$) and $7.5kV\_S$ ($M_s \sim 1.40$) for different fuels. The timescales $t = t_{dip}$ and $t = t_{Peak}(h)$ are the time taken from explosion for the flame height ($h$) to reach minima in blue region and to attain maxima in orange region (see Fig. \ref{fig8:IRO:height_Ms>1.1}c-i). On the other hand, as shown in Fig. \ref{fig7:IRO:Ms>1.1}a,e, the timescales $t = t_{start(HR)}$ and $t = t_{Peak (HR)}$ are the time taken from explosion till the flame HRR starts to increase and it reaches maxima (violet dotted line in Fig. \ref{fig7:IRO:Ms>1.1}) during shock interaction. The timescale $t = t_{Ext}$ (from explosion) is the timescale after which complete flame extinction occurs (see Fig. \ref{fig8:IRO:height_Ms>1.1}c). 

Fig. \ref{fig8:IRO:height_Ms>1.1}a,b show that, the end of first stage i.e., at $t=t_{dip}$ the flame height is minimum, which corresponds to the beginning of the interaction between flame base and flame height. Fig. \ref{fig8:IRO:height_Ms>1.1}a,b also show that the time $t=t_{start(HR)}$ where the flame heat release rate (HRR) begins to increase coincides with the $t=t_{dip}$ timescale (end of blue region in \ref{fig8:IRO:height_Ms>1.1}c). Thus, as shown in Fig. \ref{fig8:IRO:height_Ms>1.1}a,b that the beginning of the increment of the flame HRR ($t = t_{start (HR)}$) and minima of flame height (at $t = t_{dip}$) occurs in a narrow timespan, at which the second stage begins (increment in ‘$h$’ i.e., orange region). Hereafter, the increasing flame HRR reaches a maxima at $t = t_{Peak(HR)}$ where the instantaneous flame heat release rate (HRR) is more than 8 times that of the nominal unforced flame HRR. This flame HRR enhancement can be attributed to the fuel vapor accumulation at the flame tip as the flame base advects downstream along with the fuel vapor due to the interaction with the shock-flow. The timescales at the beginning of HRR enhancement i.e., $t_{dip(h)}$ (flame height minima) and $t_{start(HR)}$ (HRR enhancement beginning) coincide, which supports the fuel vapor accumulation hypothesis, as the flame HRR enhancement occurs only after the flame base interacts with flame tip (fuel vapor accumulated near the flame tip in small volume).

The flame HRR enhancement is accompanied by rapid enhancement in the instantaneous line-of-sight area of the flame (in OH* Chemiluminescence imaging), which corresponds to increase in flame dimensions such as flame height. The flame height ($h$) reaches a maxima subsequently at $t = t_{Peak}(h)$, entering the extinction phase leading to imminent extinction at $t = t_{Ext}$. This is also shown in the supplementary Figure S4, where normalized flame HRR and instantaneous flame height ($h$) are plotted simultaneously for different cases. It can also be observed that the timescale of these different stages remain similar irrespective of the shock delay or fuel type. However, the timescales become shorter as the Mach number is increased (see Fig. \ref{fig8:IRO:height_Ms>1.1}d-i).


Fig. \ref{fig8:IRO:height_Ms>1.1}j and Fig. \ref{fig8:IRO:height_Ms>1.1}k show the schematic of the high-speed flame extinction for $M_s > 1.1$ regime and simultaneous Schlieren, OH* chemiluminescence image snapshots of the phenomena respectively. The corresponding colors of the different stages of flame extinction (blue, orange and pink) are shown along with the schematic (Fig. \ref{fig8:IRO:height_Ms>1.1}j) and the snapshots of Schlieren and flame images (Fig. \ref{fig8:IRO:height_Ms>1.1}k). As shown in the schematic in Fig. \ref{fig8:IRO:height_Ms>1.1}j, the flame-base lifts off away from the droplet due to the velocity associated with blast wave ($\rm v_s$) and flame base along with fuel vapor gets swept downstream towards the flame tip in the process. During the second stage, as the flame base approaches the flame tip, more fuel vapor is fed into the flame. The RM instability at the location of the advecting flame-base is shown to occur due to the imposition of blast wave in previous work \cite{vadlamudi2024insights}, which results in vorticity generation at the flame base. The higher velocity scales ($\rm v_s\sim O(10^1-10^2)m/s$) in case of $M_s>1.1$ also contribute to the faster rate of fuel vapor addition to the high temperature zone near the flame tip, and this vapor auto-ignites as it heats up to auto-ignition temperature at the flame. Thus, the combined effect of the higher fuel vapor accumulation rate coupled with the air entrainment occurring at the flame-base (shown in the sub-figure in Fig. \ref{fig8:IRO:height_Ms>1.1}k) and enhancement in mixing due to vorticity generation because of RM instability, can be attributed to the enhancement in the flame HRR to significant levels, in case $M_s > 1.1$ regime. It is interesting to note that all these phenomena during the high-speed flame extinction for $M_s > 1.1$ regime causing significant flame HRR enhancement (momentarily), occur at significantly small timescales of $\sim O(10^{-1})ms$ from shock arrival at the droplet due to the imposition of the velocity scale $\rm v_s$.

\begin{figure}[t!]
    \centering
    \includegraphics[width=\textwidth,height=\textheight,keepaspectratio]{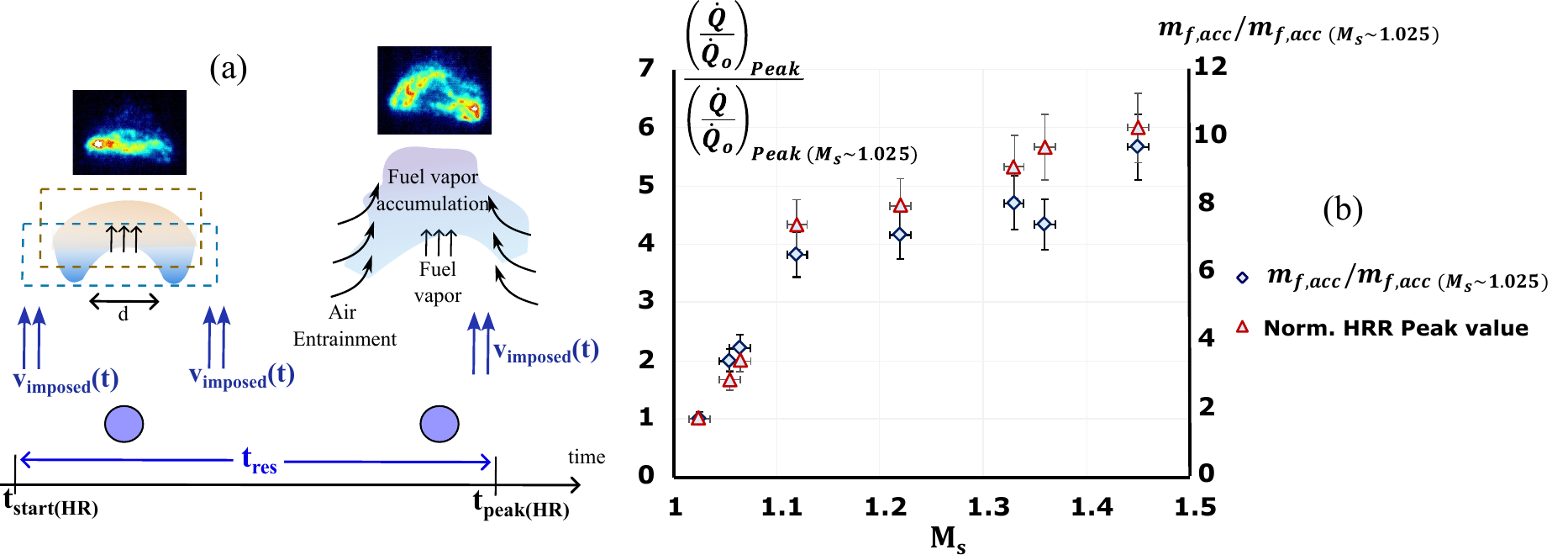}
    \caption{(a) Schematic of the droplet flame and fuel vapor advecting downstream towards flame tip, (b) Normalized flame HRR peak value ($(\dot{Q}/\dot{Q}_o)/(\dot{Q}/\dot{Q}_o)_{M_s\sim 1.025}$) and the normalized fuel vapor accumulation (${m}_{f,accumulated}/{m}_{f,accumulated(M_s\sim 1.025)}$) during $t_{res}$ which is given by Eq. \ref{Eq0p3_: imposed flow}.}
    \label{fig11:Shock:Flame}
\end{figure}

Thus, the HRR enhancement can be conjectured to be directly influenced by the fuel vapor consumption at the flame due to the accumulation of fuel vapor at the flame tip region, which subsequently undergoes auto-ignition. Thus, during the interaction with the shock-flow, the fuel vapor around the droplet gets swept downstream that gets accumulated towards the flame tip region. The time period of $t_{peak(HR)}-t_{start(HR)}$ i.e., time taken for the flame HRR enhancement to reach its peak value (from the beginning of flame base - flame tip interaction) obtained experimentally can be assumed to be the residence time ($t_{res}$) required for the pre-heating of the accumulated fuel vapor at the flame-tip for achieving auto-ignition. Thus, the amount of fuel accumulated at the flame tip during this residence time ($t_{res}$) can be considered to be of same order as the excess fuel that contributes to the HRR enhancement during interaction. Hence, evaluating the rate of fuel vapor transport towards the flame tip assuming the advection velocity of the vapor to be of same order as that of the externally imposed flow ($\rm v_{imposed}$) which is given by:
\begin{equation}\label{Eq0p1_: imposed flow}
   \rm v_{imposed,Open}(t)=\rm v_{s,Th}(t)+\rm v_{ind}(t)
\end{equation}

\begin{equation}\label{Eq0p2_: imposed flow}
   \rm v_{imposed,focused}(t)=\rm v_{s,Th}(t)
\end{equation}
Since, the flow-flame interaction is of interest, the imposed flow on the flame is assumed to be equal to the sum of instantaneous blast wave imposed velocity ($\rm v_s$) and the instantaneous induced flow ($\rm v_{ind}$) for open-field configuration cases (see Eq. \ref{Eq0p1_: imposed flow}), where as it is assumed to be equal to the blast wave imposed velocity ($\rm v_s$) for shocktube focused cases (see Eq. \ref{Eq0p2_: imposed flow}). Assuming the cross-sectional area of the fuel stream in the droplet wake (which is swept downstream) is a circle with diameter of the order of droplet diameter ($d$), and the fuel vapor transport velocity is of the order of the externally imposed flow ($\rm v_{imposed}(t)$), and the residence time, $t_{res} = t_{peak(HR)}-t_{start(HR)}$ is the timescale for accumulation, the net fuel accumulation (${m}_{f,accumulated}$) at the flame tip is obtained (see Fig. \ref{fig11:Shock:Flame}a) as follows:
\begin{equation}\label{Eq0p3_: imposed flow}
    {m}_{f,accumulated} \sim \int_{t_{start(HR)}}^{t_{peak(HR)}} \rho \frac{\pi d^2}{4}\rm v_{imposed}(t)  \,dt
\end{equation}

Assuming sufficient mixing and air entrainment due to RM instability, the obtained the fuel vapor accumulation is normalized using the value corresponding to lowest Mach number, i.e., $(\dot{Q}/\dot{Q}_o)_{M_s\sim 1.025}$ and this ratio $(\dot{Q}/\dot{Q}_o)/(\dot{Q}/\dot{Q}_o)_{M_s\sim 1.025}$ is plotted in Fig. \ref{fig11:Shock:Flame}b. Along side, the ratio of the peak value of the normalized flame HRR with peak value of normalized HRR for lowest $M_s$ case i.e., $(\dot{Q}/\dot{Q}_o)/(\dot{Q}/\dot{Q}_o)_{M_s\sim 1.025}$ is also plotted in Fig. \ref{fig11:Shock:Flame}b. Since, the values of HRR peak during enhancement ($\dot{Q}/\dot{Q}_o$) and fuel vapor accumulation at the flame (${m}_{f,accumulated}$) are plotted relative to the reference value corresponding to $M_s\sim 1.025$ case, and the two ratios plotted in Fig. \ref{fig11:Shock:Flame}b can be compared with each other on relative basis. Thus, it can be observed from Fig. \ref{fig11:Shock:Flame} that the trend of the normalized fuel mass accumulation (${m}_{f,accumulated}/{m}_{f,accumulated(M_s\sim 1.025)}$) at the flame during the residence time ($t_{res}$) is found to be in good agreement with the peak value achieved by the normalized flame HRR ($(\dot{Q}/\dot{Q}_o)/(\dot{Q}/\dot{Q}_o)_{M_s\sim 1.025}$) during interaction, for different Mach numbers.

\subsection{Droplet breakup during interaction ($M_s>1.1$)}
The shock flow interacting with the droplet flame also influences the droplet behavior as shown in $M_s < 1.1$ regime. However, to achieve higher Mach numbers ($M_s>1.1$) due to experimental limitations on the charging voltage of the shock generator, the shock tube focusing is employed. Moreover, the usage of shock tube focusing resulted in the significantly higher velocity scales associated with the induced flow following the blast wave. This high velocity scales resulted in droplet breakup and atomization during the interaction unlike $M_s<1.1$ regime.

\subsubsection{Flow - Droplet interaction ($\tau > 1$)}\label{sec:shock_tau>1}
\begin{figure}[t!]
    \centering
    \includegraphics[width=0.8\textwidth,height=\textheight,keepaspectratio]{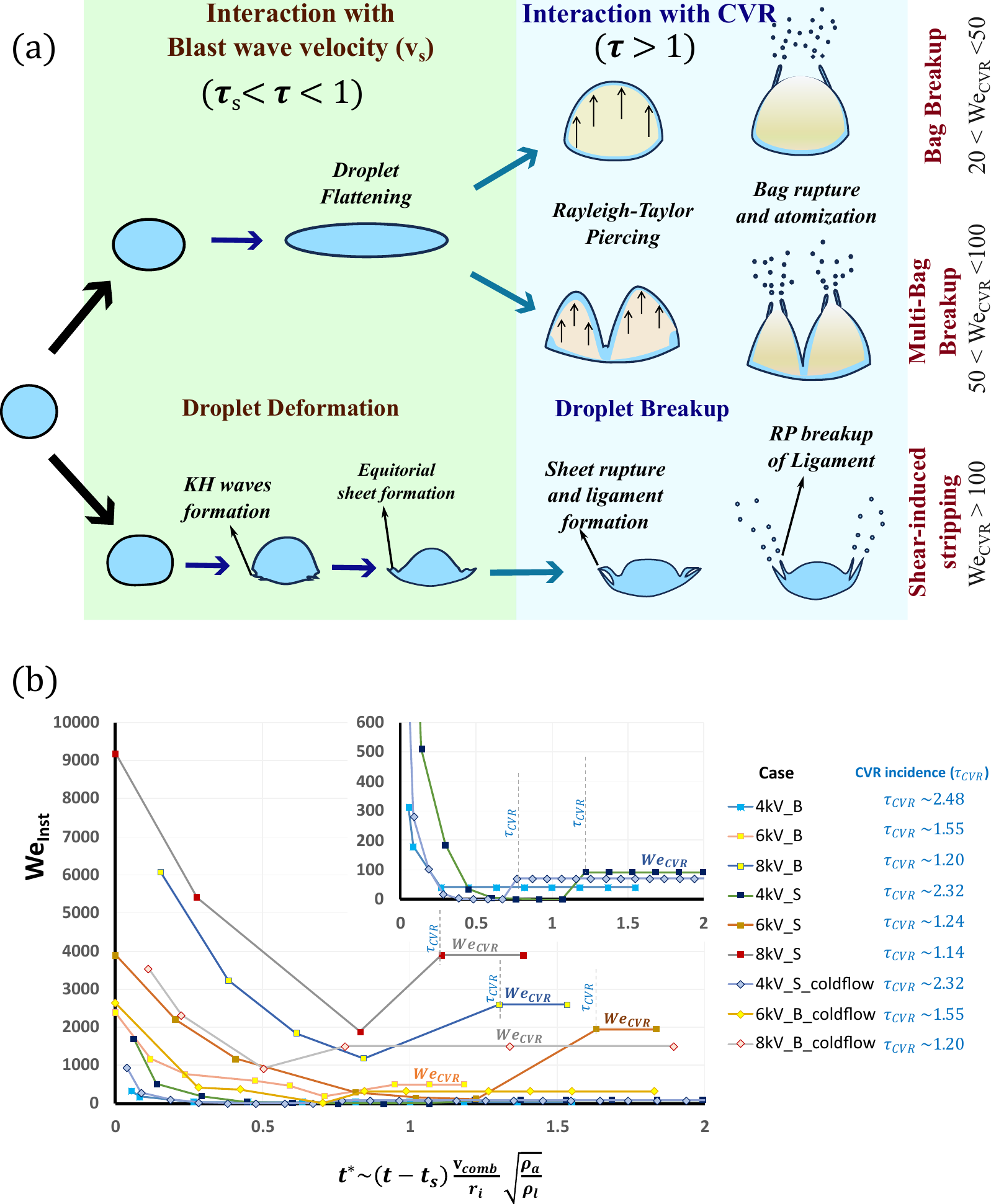}
    \caption{(a) Schematic of the droplet breakup modes at different Weber number range. Green region represents the blastwave decay profile ($\rm v_s$) interaction and blue region represents the induced flow interaction ($\rm v_{ind}$). (b) Plot showing temporal variation of instantaneous Weber number ($We_{inst}$) at the droplet location due to $\rm v_s$ and $\rm v_{CVR}$, plotted against normalized time, $t^*$ (normalized with inertia timescale of the liquid phase). The initial decaying portion of $We_{inst}$ is due to $\rm v_s$ and the jump in later portion is due to the incidence of CVR ($\rm v_{CVR}$) at the droplet location. This instant of CVR incidence ($\tau = \tau_{CVR}$) is denoted by grey vertical dotted lines in the plot for different cases and the values of $\tau_{CVR}$ corresponding to different cases are displayed alongside the legend.}
    \label{fig10:Shock:Droplet}
\end{figure}
During this initial stage of droplet dynamics, due to $\rm v_s$, the droplet starts to show deformation and KH instabilities (at high $M_s$). However, droplet breakup is not observed till the droplet starts to interact with the oncoming compressible vortex ring (CVR) formed from the induced flow behind the blast wave, which reaches the droplet after some delay (Fig. \ref{fig10:Shock:Droplet}a, blue region). The Schlieren visualisation of the induced flow CVR has been used to estimate the translational velocity scales associated with the CVR and it has been found that the velocity scale of the CVR ($\rm v_{CVR}$) is smaller than the shock propagation velocity, however, it is sufficient to disintegrate the droplet as shown in Fig. \ref{fig10:Shock:Droplet}a. 

In the previous experiments investigating droplet - blast wave (wire explosion technique) interaction \cite{sharma2021shock}, normal shock assumption has been considered for the planar blast wave (with decaying profile) for evaluating the Weber number. Even though the order of magnitude of the velocity scales matched with the Weber number range, the temporal variation in blastwave-imposed velocity and the subsequent induced flow has not been discussed in previous studies. Thus, the time scale of the two stages of the droplet interaction i.e., initial deformation ($\tau_s<\tau<1$) and subsequent breakup ($\tau>1$) has not been addressed. Thus, an attempt has been made here to incorporate the temporal variation of the velocity to evaluate the instantaneous Weber number at the droplet.The Fig. \ref{fig10:Shock:Droplet}a shows the two stages of the flow - droplet interaction. Even though the droplet doesn’t respond to the imposed blast wave, the velocity profile behind the blast wave ($\rm v_s$) ensues droplet deformation and initiation of KH instabilities ($\tau_s<\tau<1$). Subsequently, when the compressible vortex (CVR) reaches the droplet ($\tau>1$), the instabilities intensify leading to the droplet fragmentation. The green region in Fig. \ref{fig10:Shock:Droplet}a represents the initial interaction between the decaying velocity profile behind the blast wave ($\rm v_s$) and the droplet. The blue region represents the interaction between the CVR and the droplet. Thus, the velocity at the droplet location can be written as  
\begin{equation}\label{Eq1_: Shock_Droplet}
   \rm v_{comb}(t)=\rm v_{s,Th}(t)+\rm v_{CVR}(t)
\end{equation}
where, $\rm v_{comb}$ is the total combined velocity at the droplet due to the blast wave profile and CVR. $\rm v_{s,Th}$ is the theoretical estimate of the instantaneous local velocity at the droplet imposed by the blast wave and $\rm v_{CVR}$ is the velocity scale associated with the CVR. $\rm v_{s,Th}$ can be obtained using the blast wave formulation based on power-law density profile\cite{vadlamudi2024insights} using the instantaneous Mach number, as shown in our previous work. The instantaneous Weber number (${\rm We}_{inst}$) at the droplet location based on the temporally varying $\rm v_{comb}$ is evaluated as follows:
\begin{equation}\label{Eq1p5_: Shock_Droplet}
   {\rm We}_{inst} \sim \frac{\rho d_i\rm v_{comb}^2}{\sigma}
\end{equation}
Fig. \ref{fig10:Shock:Droplet}b shows the plot of the instantaneous Weber number plotted against normalized time ($t^*$), which is the time from shock incidence ($t-t_s$) normalized using the droplet inertial timescale, $\frac{r_i}{\rm v_{comb}} \sqrt{\frac{\rho_l}{\rho_a}}$ (where, $\rho_l, \rho_a$ are densities of liquid and gas phase, $r_i$ is the order of magnitude of the initial droplet radius and $t_s$ is the time instant at which blast wave reaches the droplet). The plot shows an initial temporally decaying Weber number ($We_{inst}$) which shows a jump in magnitude during the later stage at $t^* \sim O(10^0)$ due to the incidence of CVR at the droplet. The grey vertical dotted lines are used to indicate the instant where CVR starts to interact with the droplet and the value of the non-dimensional time w.r.t. blast wave profile ($\tau$) at the CVR incidence instant is mentioned in Fig. \ref{fig10:Shock:Droplet}b for different cases. As mentioned before, during the initial stage of interaction with $\rm v_s$, the droplet only exhibits deformation (green region), where as droplet breakup is only observed after the CVR reaches the droplet. This is evident from the plot in Fig. \ref{fig10:Shock:Droplet}b, where even though high Weber numbers are imposed initially, there is no sufficient time for the droplet to respond to the high $We_{inst}$ for $t^* \ll 1$. Furthermore, the droplet breakup during the interaction with CVR corresponds to the Weber number jump to $We_{CVR}$ shown in Fig. \ref{fig10:Shock:Droplet}b. 

It is to be noted that, for the Eq. \ref{Eq1_: Shock_Droplet}, before the vortex reaches the droplet, $\rm v_{CVR} = 0$, hence $\rm v_{comb} \sim \rm v_{s,Th}(t)$. As shown in Fig. \ref{fig10:Shock:Droplet}b, for all the cases, $\tau$ becomes more than unity by the time CVR reaches the droplet at $\tau \sim \tau_{CVR}$ which is greater than '1'. This indicates that the $\rm v_s$ decays significantly and approaches zero during CVR interaction and $\rm v_{CVR} \gg \rm v_{s,Th}$, thus, $\rm v_{comb} \sim \rm v_{CVR}$ during the CVR - droplet interaction. Since, the droplet breakup begins during the second stage (interaction with CVR) i.e., blue region in Fig. \ref{fig10:Shock:Droplet}a, Weber number based on the droplet diameter (d) and $\rm v_{CVR}$ (velocity scale associated with CVR) i.e., $We_{CVR}$ is more relevant to understand the droplet breakup phenomena (see Eq. \ref{Eq2_: Shock_Droplet}).
\begin{equation}\label{Eq2_: Shock_Droplet}
   {\rm We}_{CVR} \sim \frac{\rho d_i\rm v_{CVR}^2}{\sigma}
\end{equation}
 
\begin{figure}
    \centering
    \includegraphics[width=\textwidth,height=\textheight,keepaspectratio]{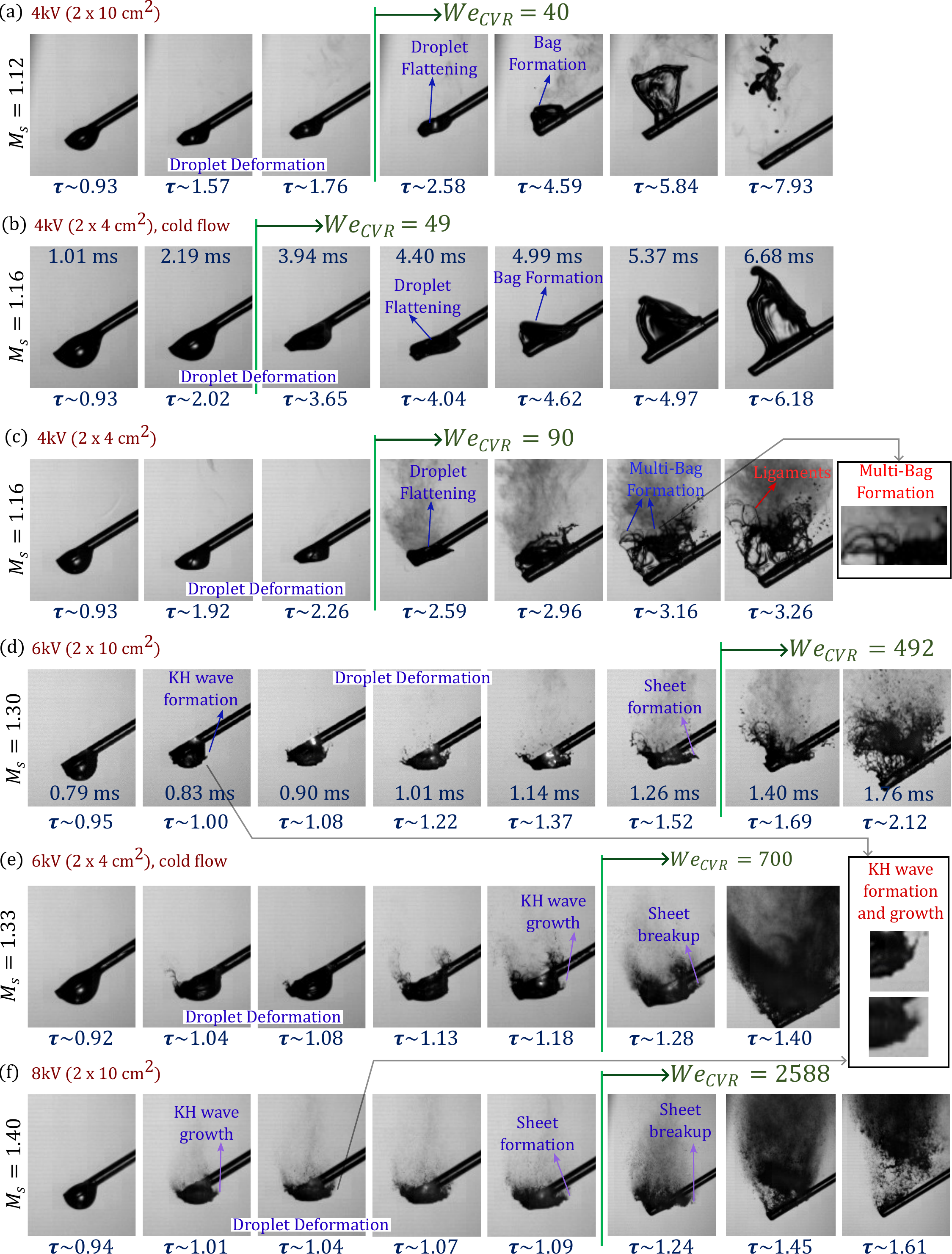}
    \caption{Shadowgraphy time series images of the droplet breakup dynamics during interaction with blast wave and induced flow for Combusting droplet with respect to time from explosion (t): (a) $4kV\_B$, (c) $4kV\_S$, (d) $6kV\_B$, (f) $8kV\_B$ and for Non-combusting droplet (cold flow): (b) $4kV\_S$, (e) $6kV\_S$. The CVR interaction occurs right side of the vertical green line and the corresponding $We_{CVR}$ is mentioned.}
    \label{fig13_IRO_shock}
\end{figure}

As shown in the schematic in Fig. \ref{fig10:Shock:Droplet}a, the droplet breakup behaviour significantly changes as the Mach number is varied, where bag breakup mode is observed at lower Mach numbers ($M_s<1.3$), whereas, as Mach number is increased ($M_s>1.3$), which corresponds to increase in Weber number ($We_{CVR}$), the shear stripping of the droplet is observed. The following Fig. \ref{fig13_IRO_shock} shows the time series of the droplet breakup (with and without ignition) at different flow conditions obtained by varying the charging voltages in both the shock tube channels. The cases without droplet ignition are labelled as ‘cold flow’ in the figure. Fig. \ref{fig13_IRO_shock}a-c are the low Mach number cases where the velocity scales are slower corresponding to low Weber number CVR i.e., $We_{CVR} < 100$, and Fig. \ref{fig13_IRO_shock}d-f correspond to high $We_{CVR}$ cases. As discussed before, the droplet deformation is observed in all the cases. Fig. \ref{fig13_IRO_shock} shows the two-stage droplet response to the imposed flow and the time instant from where CVR interaction (second stage, $\tau>1$) starts is mentioned. It is to be noted that the increase in shock strength results in higher $\rm v_{CVR}$ and this corresponds to a smaller time delay between blast wave incidence ($\tau \sim \tau_s$) and CVR interaction ($\tau>1$). Thus, the induced flow reaches the droplet quicker i.e., closer to $\tau\sim 1^+$ as shock strength ($M_s$) is increased which is also seen from Fig. \ref{fig13_IRO_shock} where CVR interacts with the droplet for $\tau>1$ at smaller values of $\tau$ as $M_s$ is increased.

For low $We_{CVR}$ cases, the droplet deformation at $\tau_s<\tau<1$ continues leading to the flattening of the droplet into a disc-like shape whose thickness continuously decreases due to liquid transport towards the periphery as the droplet is elongated in the equatorial direction. Subsequently, for $\tau>1$, Rayleigh-Taylor piercing is observed which leads to blowup of the membrane into a bag-like structure, which eventually ruptures due to nucleation of holes in the thin membrane, leading to droplet disintegration and atomization. For $We_{CVR} > 50$, the multi-bag Rayleigh-Taylor piercing is observed, and these droplet breakup behaviours are in good agreement with the literature \cite{sharma2023aadvances}. Fig. \ref{fig13_IRO_shock}a,c shows the droplet breakup for bigger and smaller channel for the same charging voltage of 4kV, which shows bag breakup in case of the bigger channel, multi-bag breakup in case of the smaller channel. This can be attributed to higher velocity scales achieved while focusing using smaller channel, which results in higher $We_{CVR}$.

In addition to this, at higher flow velocities (higher Weber number, i.e., $We_{CVR}$), along with the initial droplet deformation ($\tau_s<\tau<1$), surface waves due to KH instability also starts to form on the windward surface of the droplet near the high shear forward stagnation region, as shown in Fig. \ref{fig10:Shock:Droplet}a. As the KH waves travel radially outward and grow in amplitude, the fluid is carried outward which gets accumulated towards the periphery. When the KH waves reach the edge, they get deflected along the flow direction, undergoing flow entrainment leading to the formation of a thin sheet of the accumulated fluid at the periphery (see Fig. \ref{fig13_IRO_shock}d-f). The thin sheet ruptures due to the nucleation of multiple holes which grow temporally leading to the breakup of the sheet into multiple ligaments. These ligaments undergo secondary atomization due to Rayleigh - Plateau instability forming numerous daughter droplets. This shear-induced breakup mode is observed to be dominant at higher Weber numbers as shown in Fig. \ref{fig13_IRO_shock}d-f. At very high Weber numbers, the droplet breakup through shear stripping becomes more intense, forming a dense cloud of droplet spray. The shear induced stripping is dominant at high Weber numbers ($We_{CVR}$) due to faster growth rates of KH instability compared to droplet deformation, which leads to faster disintegration of the droplet through shear stripping at high $We_{CVR}$. The KH instability growth is observed to be occurring at shorter time scales ($\sim O(10^{-1}) ms$) compared to the droplet deformation and Rayleigh-Taylor piercing mode  time scale ($\sim O(10^0) ms$), see Fig. \ref{fig13_IRO_shock}b,d. Thus, as pointed out by Sharma et al. \cite{sharma2021shock}, at lower $M_s$, i.e., lower Weber number, when velocity scales are low, the KH instability does not occur, and hence, the slower Rayleigh-Piercing mode of breakup due to continuous droplet deformation is observed. However, when $M_s$ is increased, due to higher velocity scales, KH waves start to form, leading to a shear stripping mode of breakup that occur at shorter time scales ($\sim O(10^0) ms$), see Fig. \ref{fig13_IRO_shock}b,d.

Interestingly, the droplet breakup phenomena are observed to change for the same flow conditions (shock tube and charging voltage) when the droplet is ignited (see Fig. \ref{fig13_IRO_shock}b,c). This can be attributed to the higher droplet temperatures during combustion. In the absence of the flame, the droplet temperature is same as the ambient temperature, however, when the droplet is ignited, the droplet surface temperature will become wet bulb temperature due to the phase change occurring at the surface. Since sufficient time is given after the ignition before the shock interaction, it is reasonable to assume the droplet heat-up phase is completed and the bulk liquid in the droplet is at the wet bulb temperature, for simplicity. Thus, the fluid properties during the interaction with the vortex are at an elevated temperature which resulted in alteration of the droplet breakup phenomena at the same input flow conditions. Thus, the surface tension of the liquid droplet at its wet bulb temperature will be lower leading to higher Weber number values in the presence of the flame for the same flow conditions. The droplet breakup behavior is also observed to change when the fuel properties are varied, as Weber number ($We_{CVR}$) changes with surface tension of the fuel.

All these observations are in good agreement with the Weber number range ($We_{CVR}$) corresponding to classical breakup modes in the literature, which is valid for the droplet size that is far greater than the order of magnitude of the wavelength of the KH wave, i.e., $d\gg\lambda$ \cite{sharma2021shock}. Since, the droplet is in pendant mode, the effect of quartz rod is significant on droplet breakup phenomena, thus, the droplet dynamics on the windward side of the quartz rod were used to deduce conclusions regarding droplet breakup characteristics. Furthermore, this section primarily sheds light on the time-scales involved in droplet breakup during shock - droplet interaction which has not been addressed in the literature previously. 

\subsubsection{Effect of Nanoparticle addition on droplet breakup}
\begin{figure}[t!]
    \includegraphics[width=1\textwidth,height=\textheight,keepaspectratio]{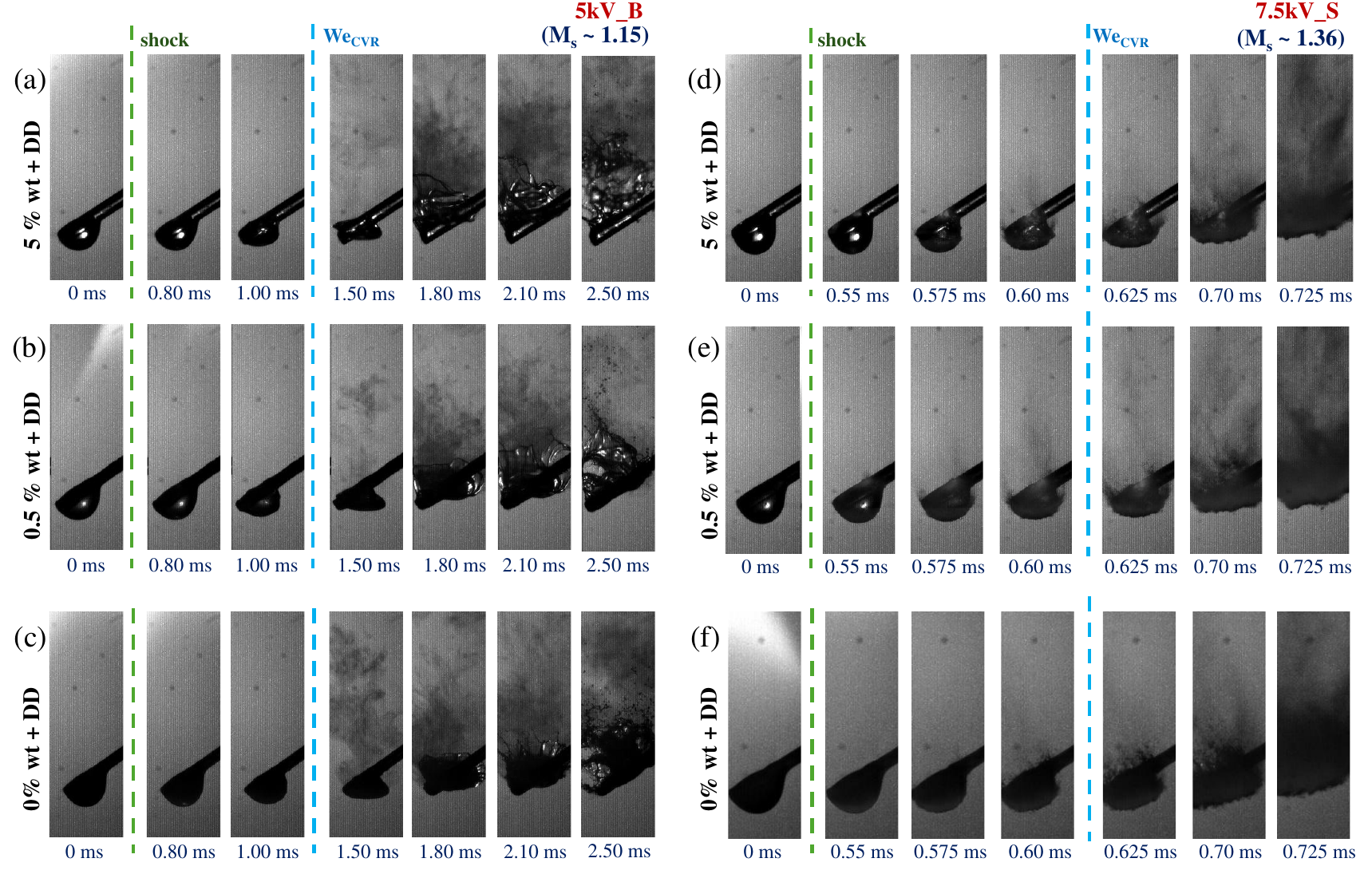}
    \caption{Time series of ignited fuel droplet interaction with shock flow, resulting in droplet breakup for three nanoparticle concentrations with respect to time from explosion (t): (a,d) Pure Dodecane, (b,e) $0.5\%$ wt $Al_2O_3$ + DD, (c,f) $5\%$ wt $Al_2O_3$ + DD; at two different shock strengths: (a,b,c) $M_s \sim 1.15$ ($5kV\_B$), (d,e,f) $M_s \sim 1.36$ ($7.5kV\_S$).}
    \label{fig10_IRO_shock}
\end{figure}

Due to experimental limitations, the shock incidence delay could not be set to a higher value corresponding to the later stages of the nanofuel droplet lifetime where shell formation occurs that can potentially alter the droplet breakup dynamics or flame dynamics drastically from pure fuel, which has to be investigated in the future. However, interesting observations were found in droplet breakup phenomena during interaction with the shock flow for different nanofuel concentrations at lower Mach numbers ($M_s < 1.20$). Fig. \ref{fig10_IRO_shock} shows the time series of the droplet breakup phenomena observed for different NP particle loading rates and shock strengths ($M_s$). The vertical green dotted lines are shown in Fig. \ref{fig10_IRO_shock} to indicate the instant of shock incidence on droplet and the vertical blue dotted lines in Fig. \ref{fig10_IRO_shock} indicate the instant when CVR reaches the droplet. The image on the right side of the green vertical dotted line in all the time series provided in Fig. \ref{fig10_IRO_shock} correspond to the instant when flame liftoff occurs that subsequently results in extinction. Similar to the shock flow - flame interaction, the droplet interaction with shock flow also occurs in two stages: interaction with decaying $\rm v_s$ imposed by blast wave and the interaction with the compressible vortex (CVR) exiting the shock tube. Thus, as explained in Fig. \ref{fig10:Shock:Droplet}a, during the first stage, the droplet undergoes deformation, but the droplet breakup only occurs after the interaction with CVR.

The time series in Fig.\ref{fig10_IRO_shock} a.b.c (left side) correspond to the lower Mach number ($M_s < 1.20$) and the droplet dynamics are observed to be similar in all the three cases with different NP concentrations. However, there is a marginal mismatch in the timescale of the droplet dynamics among different NP particle loading rates for low $M_s$ ($< 1.20$). This is clearly evident from Fig. \ref{fig10_IRO_shock}a,b,c where, the droplet deformation in Fig. \ref{fig10_IRO_shock}a is relatively faster compared to Fig. \ref{fig10_IRO_shock}b, which is marginally faster than Fig. \ref{fig10_IRO_shock}c. That means, at the same time stamp, Fig. \ref{fig10_IRO_shock}a showed more droplet deformation compared to Fig. \ref{fig10_IRO_shock}b and Fig. \ref{fig10_IRO_shock}c. Thus, it can be inferred that the droplet deformation occurs at slower rate as the NP concentration is increased. This result can be attributed to the increase in droplet viscosity due to the agglomeration of Nanoparticles (inside the droplet), that contribute to the resistance to the droplet deformation (i.e., inertial forces due external forcing). The velocity scale corresponding to this deforming inertial forces i.e., induced flow velocity ($\rm v_{ind}$) is $\sim 30-50 m/s$ for $M_s < 1.20$. The similar effect of viscosity can also be seen in the droplet regression plots in Fig. \ref{fig5:IRO:Ms<1.06}(c,e,g), where the droplet oscillations due to the interaction with shock flow are dampened immediately in higher PLR cases (Fig. \ref{fig5:IRO:Ms<1.06}c,e) compared to pure Dodecane, which exhibited significant oscillations throughout the droplet lifetime (Fig. \ref{fig5:IRO:Ms<1.06}g). However, as the shock Mach number is increased, the induced flow velocity scale ($\rm v_{ind} \sim 140 m/s$) is also increased, resulting in more dominant inertial forces. Thus, for $M_s > 1.20$, the effect of viscous dampening of the inertial forces is less pronounced due to the relative dominance of inertial forces corresponding to higher $\rm v_{ind}$. This can also be observed in Fig. \ref{fig10_IRO_shock}(d,e,f), where negligible difference is observed in the droplet breakup and deformation phenomena, when the droplet snapshots corresponding to same time stamps are compared for different NP concentrations. Thus, the effect of Nanoparticle addition primarily affects the droplet dynamics at lower velocity scales of the externally imposed flow. However, the extreme case of shell formation and its effects on droplet breakup needs to be studied in future by investigating the droplet breakup behavior due to shock interaction at very high shock delay ($> 0.8 \cdot t_d$) in contactless environment.

    \section{Conclusion}
In current study, the droplet flame interaction with the shock flow has been investigated with primary focus on the flame topology, flame heat release rate (HRR) and droplet dynamics. The experiments were conducted using different fuels such as Pure Dodecane, Pure Ethanol and nanofuels (Dodecane + Alumina nanoparticles) and the phenomena is recorded using high-speed IRO for OH* chemiluminescence along with simultaneous high-speed Schlieren. The experiments gave deeper insights into flame and droplet dynamics during the interaction with the shock flow adding upon our previous work \cite{vadlamudi2024insights}. The flame imaging obtained using CH* chemiluminescence and OH* chemiluminescence gave further information about the flame topology and flame heat release rate variation respectively. The flame shedding, partial extinction followed by reattachment occured for low Mach number regime ($M_s < 1.06$) and flame extinction is observed for regime-I ($M_s > 1.06$). The flame extinction is observed to occur due to the interaction with induced flow ($\rm v_{ind}$) for regime-II ($1.06 < M_s < 1.1 $). Furthermore, the flame extinction occurs at significantly smaller timescales for $M_s>1.1$ where the flame extinction occurs solely due to the interaction with the decaying velocity profile ($\rm v_s$) imposed by the blast wave. Interestingly, during the extinction, the flame heat release rate (HRR) showed significant enhancement compared to unforced nominal HRR of the flame. It has been hypothesized that the fuel vapor transport during the flame base liftoff towards the flame tip contributes to the HRR enhancement. However, the flame HRR enhancement is significantly higher in case of $M_s > 1.1$ which can be attributed to higher velocity scales of $\rm v_s$ in this regime compared to $\rm v_{ind}$ (for $M_s <1.1$), which contribute to air entrainment, higher fuel accumulation rate and better mixing. Different mechanisms involved in the flame extinction in different regimes have been discussed. Later, the different stages during the flame evolution during HRR enhancement has also been discussed.

Along with the flame response, the droplet regression alteration and droplet breakup were also observed during the interaction for different $M_s$. At lower Mach number ($M_s < 1.1$), the droplet regression rate is altered due to the interaction with the shock flow. Interestingly, the droplet regression rate is enhanced after the interaction with shock flow for lower $M_s<1.06$ where droplet flame reattaches post-interaction. This enhancement in droplet regression is shown to be due to the increased fuel vaporization rate under externally imposed flow. On the other hand, when $M_s > 1.06$, the droplet regression rate is slowed due to the full extinction of the flame which deprives a heat source necessary for evaporation. In addition to this, the increased velocity scales ($\rm v_{ind}$) for $M_s > 1.06$ was shown to not contribute significantly towards the enhancement in fuel vaporization rate. As the Mach number is increased ($M_s >1.1$), droplet exhibited breakup and atomization due to interaction with flows at higher Weber numbers. The droplet exhibited temporal deformation into an oblate shape for all these cases and the droplet dynamics were observed to occur in two stages: Initial interaction with the blast wave and subsequent interaction with the CVR. At lower $M_s$, the droplet elongates equatorially and undergoes RT piercing exhibiting bag breakup mode of atomization during the eventual interaction with induced flow vortex (CVR). Unlike lower $M_s$, alongside continuous deformation, the droplet exhibited KH instability-induced perturbation growth on the windward surface. This leads to shear induced stripping as the perturbations further grow leading to sheet formation and rupture at the equator resulting in secondary atomization. The Weber number ranges based on the induced flow velocity scale ($\rm v_{ind}$) for different modes were observed to be in good agreement with the literature. 

Subsequently, the effect of fuel type, nanoparticle addition and shock delay on droplet flame response during the interaction has also been investigated in current experiments. The nanoparticle addition showed marginal effects on the flame dynamics during the interaction because of the alteration of the vaporization rate locally at the droplet level based on the NP particle loading rates (PLR) and shock delay (the point of time at which blast is incident during droplet lifetime). Ethanol showed relatively lower levels of HRR enhancement during interaction compared to Dodecane, due to lower vaporization rate of the droplet. However, the flame HRR enhancement (before extinction) during the interaction for different shock delays showed marginal variation for nanofuels compared to pure fuel. Nevertheless, irrespective of the different experimental parameters the droplet flame response to the interaction with the shock-flow is observed to categorically follow the same set of regimes irrespective of the fuel type, NP PLR, shock delay, etc. This can be attributed to the significantly high-speed nature of the interaction ($\sim O(10^0-10^1)ms$) and high velocity scales involved ($\gg \rm v_{NC}$), and only the ratio $\rm v_{imposed}/\rm v_{NC}$ majorly dictates the different flame behaviors.

Nanofuels showed higher viscous effects due to agglomeration resulting in slower rate of droplet deformation with increase in nanoparticle loading. However, this effect is found to become less prominent with increase in external forcing i.e., shock Mach number, which increases the dominance of the inertial forces over the viscous effects. Due to the experimental limitation, the upper limit of the delay of shock interaction with nanofuels is restricted, resulting in negligible shell formation effects in the scope of current study, which may potentially alter flame and droplet dynamics during the interaction, and needs further investigation.

\section*{Declaration of Interests}
{
The authors report no conflict of interest.
}

\section*{Acknowledgements}
{
The authors are thankful to SERB (Science and Engineering Research Board) - CRG: CRG/2020/000055 for financial support. S.B. acknowledges funding through the Pratt and Whitney Chair Professorship. The authors acknowledge Prof. Pratikash Panda and his lab for providing high-speed IRO for performing OH* Chemiluminescence.
}

    \bibliography{mybibfile}

    \end{sloppypar}

\end{document}